\documentclass[english]{article}
\usepackage[T1]{fontenc}
\usepackage[latin9]{inputenc}
\usepackage{geometry}
\usepackage{float}
\usepackage{booktabs}
\usepackage{amsmath}

\makeatletter

\providecommand{\tabularnewline}{\\}

\makeatother

\usepackage{babel}
\begin{document}
\title{Quantifying and Detecting Individual Level `Always Survivor' Causal
Effects Under `Truncation by Death' and Censoring Through Time}
\author{Jaffer M. Zaidi, Eric J. Tchetgen Tchetgen, Tyler J. VanderWeele}
\maketitle

\section*{Abstract}

The analysis of causal effects when the outcome of interest is possibly
truncated by death has a long history in statistics and causal inference.
The survivor average causal effect is commonly identified with more
assumptions than those guaranteed by the design of a randomized clinical
trial or using sensitivity analysis. This paper demonstrates that
individual level causal effects in the `always survivor' principal
stratum can be identified with no stronger identification assumptions
than randomization even in the presence of informative censoring.
We illustrate the practical utility of our methods using data from
a randomized clinical trial on patients with prostate cancer. Our
methodology is the first and, as of yet, only proposed procedure that
enables detecting causal effects in the presence of truncation by
death using only the assumptions that are guaranteed by design of
the clinical trial. This methodology is applicable to all distributions
concerning the outcome variable.

\section{Introduction}

Researchers would often like to evaluate the effect of a treatment
or exposure on an outcome that could be truncated by or missing due
to death \cite{frangakis2002principal,zhang2003estimation,robins1986new}.
For individuals who die before the end of study, the outcome of interest
remains undefined. Such issues occur both in clinical trials and in
observational studies. Many researchers note that a comparison of
risk differences of those populations that survive at least until
the end of the study do not offer causal conclusions \cite{rubin2006causal,tchetgen2014identification,robins1986new}.
Studies based upon principal stratification, in which comparisons
are made of average causal effects in the `always survivor' group,
attempt to provide causal conclusions in the presence of truncation
by death. 

We refer the interested reader to Wang et al \cite{wang2017identification},
who provide a detailed description of the different approaches with
regard to survivor average causal effects. Our approach differs from
these previous approaches \cite{ding2011identifiability,wang2017identification,frangakis2002principal,zhang2003estimation,wang2017causal}
in that our causal estimand is not the survivor average causal effect
estimand used in these previous papers. Instead, we derive a null
hypothesis that when falsified informs us that the `always survivor'
principal stratum must exist and that there are individuals in this
principal stratum of `always survivors' for whom the treatment has
an effect of the post-survival outcome of interest and allowing us
to construct lower bounds on the size of this subgroup. Our approach
embeds testing for these causal effects firmly within the Neyman-Pearson
paradigm. The identification assumptions used to detect such effects
are exactly those assumptions made to identify the average treatment
effect, that is, we make no identification assumption other than randomization
of treatment. 

\section{Notation and assumptions}

Suppose a binary treatment $X\in\{0,1\}$ is randomized at baseline.
We assume data has been collected over time on survival and on a binary
outcome of interest defined only among those who survive. We assume
that censoring may take place either for survival or for the binary
outcome. The individuals of our study, denoted by symbol $\omega,$
compose a finite population $\Omega.$ At pre-specified times $T=\{1,\ldots,t_{f}\}$
over the course of the study duration $t_{f}$, we define variable
$S(\omega,t)\in\{0,1,2\}$ of an individual $\omega$ for each $t\in T.$
Here, $S(\omega,t)=2$ denotes that the individual's survival response
was censored or missing at time $t.$ If the individual's response
is not censored at time $t,$ then $S(\omega,t)=1$ denotes that the
individual survived at least until time $t,$ and $S(\omega,t)=0$
denotes that the individual did not survive until time $t.$ A binary
outcome of primary interest is also measured at each of the pre-specified
times $t\in T,$ though at each $t\in T,$ the outcome might be truncated
through death, that is, undefined because the individual died. 

If the individual's response is censored by time $t,$ then let $Y(\omega,t)=3,$
and if the individual was not censored but did not survive until $t\in T,$
then let $Y(\omega,t)=2.$ If the individual was not censored and
did not die before time $t$ then let $Y(\omega,t)=1$ if the binary
outcome is present and $Y(\omega,t)=0$ if the binary outcome is absent.
Consequently the ordinarily binary outcome is transformed to a categorical
variable. For this paper, it is important to note that censoring by
time $t$, denoted as $Y(\omega,t)=3,$ refers to either drop-out
of the individual from the trial or missing data, and that death of
an uncensored individual is treated as a different event and is denoted
as $Y(\omega,t)=2.$ If the individual $\omega$ survives at least
until $t\in T$, then $Y(\omega,t)=0$ denotes that the primary outcome
of interest is 0 at follow up period $t$, and $Y(\omega,t)=1$ denotes
that the primary outcome of interest is 1 at follow up period $t$. 

We use this re-coding of the outcome variables $Y$ and $S$ for notational
simplicity. However, we also offer an alternative proof of our results
in the Generalizations and Extensions section that uses a different
coding of the variables that avoids specifying a new level for the
missing responses. The reason we first provide this first coding of
the outcome variables is to draw the connection between the statistical
units in the randomized experiment and the counterfactual question
under investigation. We first use deterministic counterfactuals or
potential outcomes in order for readers to understand how an adaptation
of Neyman's seminal contribution to the use of potential outcomes
in agricultural field experiments \cite{splawa1990application,rubin1990application}
provides a design based understanding of causal effects in the presence
of truncation by death and missing outcome data. The framework can
be extended to use stochastic counterfactuals for continuous outcomes
as developed in the Generalizations and Extensions section.

Define potential outcomes $Y_{x}(\omega,t)$ and $S_{x}(\omega,t)$
to be the value of $Y(\omega,t)$ and $S(\omega,t)$ respectively
at follow up time $t$ had we set the value of the treatment $X$
to $x$ for individual $\omega.$ Similarly, should the individual's
counterfactual $S_{x}(\omega,t)=0$ for $t\in T,$ then $Y_{x}(\omega,t)=2,$
and if $S_{x}(\omega,t)=1$ for $t\in T,$ then $Y_{x}(\omega,t)$
is the counterfactual $Y(\omega,t)$ had we forced individual to take
treatment $x\in\{0,1\}.$ Also, for some $t\in T,$ let $Y_{x}(\omega,t)=3$
denote that had we forced individual to take treatment $x\in\{0,1\}$
their counterfactual of the primary outcome $Y(\omega,t)$ would be
censored or missing to the experimenter in the counterfactual world.
Similarly, let $S_{x}(\omega,t)=2$ denote that had we forced individual
to take treatment $x\in\{0,1\}$ their counterfactual of the survival
status would be censored or missing to the experimenter. This setting
of potential outcomes or counterfactuals allowing for the possibility
of censoring or missing data at a given time $t$ in the counterfactual
world provides us with a theoretical framework to examine causal effects
in the presence of censoring and truncation by death. 

Consider now the setting of a randomized trial. With treatment randomized
at baseline, we can make the `weak ignorability' assumption $(Y_{x}(\omega,t),S_{x}(\omega,t))\amalg X$
for all $t\in T.$ We require the consistency assumption for both
$Y_{x}(\omega,t)$ and $S_{x}(\omega,t),$ which means that when $X(\omega)=x,$
then $Y_{x}(\omega,t)=Y(\omega,t)$ and $S_{x}(\omega,t)=S(\omega,t).$
This assumption states that the value of $Y(\omega,t)$ and $S(\omega,t)$
that would be observed if $X$ had been set to what in fact they were
observed to be is equal respectively to the value of $Y(\omega,t)$
and $S(\omega,t)$ that was observed. The randomization and consistency
assumption are assumed throughout this paper. Additional assumptions,
whenever needed, are explicitly detailed in the relevant section,
theorem and proposition.

For ease of notation, we drop the $\omega$ in $Y(\omega,t)$ and
$S(\omega,t)$ whenever the meaning is clear. Denote for all $t\in T,$
$P^{c(t)}(y_{1},y_{0},s_{1},s_{0})=P(Y_{1}(t)=y_{1},Y_{0}(t)=y_{0},S_{1}(t)=s_{1},S_{0}(t)=s_{0})$,
where $y_{1}$ and $y_{0}$ take values in the set $\{0,1,2,3\}$,
and $s_{1}$ and $s_{0}$ that take values in the set $\{0,1,2\}.$
Also, for all $t\in T,$ we shall use the notation $P_{y,s.x}^{r(t)}=P(Y(t)=y,S(t)=s\mid X=x)$
where $y\in\{0,1,2,3\},$ $s\in\{0,1,2\},$ and $x\in\{0,1\}.$ The
counterfactual probability $P^{c(t)}(y_{1},y_{0},s_{1},s_{0})$ is
an important quantity for understanding the magnitude of different
causal effects, and is used throughout this paper. The observed probability
$P_{y,s.x}^{r(t)}$ is used in empirical conditions throughout this
paper. Under the assumption of randomization and consistency of counterfactuals,
the empirical conditions that only use the observed probability $P_{y,x.s}^{r(t)}$
imply resulting constraints placed upon $P^{c(t)}(y_{1},y_{0},s_{1},s_{0}),$
which, as we will show, enables investigators to evaluate the magnitude
of causal effects for always survivors using only empirical conditions
from the observed data. 

The superscripts are meant to distinguish these probabilities from
one another. The symbol $c(t)$ denotes that this is probability concerns
the counterfactual world, and similarly the symbol $r(t)$ denotes
that this probability concerns our observed data or real world. All
proofs are provided in Online Supplement 1. Let $I(\cdot)$ denote
the usual indicator function. Define $Y_{x}^{[y]}(\omega,t)=I(Y_{x}(\omega,t)=y),$
and $S_{x}^{[s]}(\omega,t)=I(S_{x}(\omega,t)=s)$ for $y\in\{0,1,2,3\},$
$s\in\{0,1,2\},$ $\omega\in\Omega$ and $t\in T.$ We say an individual
is an `always survivor' at time $t$ if $S_{1}(\omega,t)=S_{0}(\omega,t)=1,$
that is the individual survives at least until time $t$ regardless
of treatment. We say that an individual displays an individual level
`always survivor causal effect' at time $t,$ if for such an individual
$Y_{1}(\omega,t)=y,$ $Y_{0}(\omega,t)=1-y,$ $S_{1}(\omega,t)=1,$
and $S_{0}(\omega,t)=1$ for $y\in\{0,1\},$ which means that this
individual would have survived at least until time $t$ regardless
of treatment and there is a causal effect on $Y.$ 

Joint counterfactuals of the form $Y_{x}(\omega,t)=1$ and $S_{x}(\omega,t)=0$
are never considered, as the individual $\omega\in\Omega$ has not
survived until time $t\in T.$ Consequently, any counterfactual probability
$P^{c(t)}(y_{1},y_{0},s_{1},s_{0})$ for which $s_{1}=0$ and $y_{1}\in\{0,1\}$
or $s_{0}=0$ and $y_{0}\in\{0,1\}$ is identically zero. Also $P^{c(t)}(y_{1},y_{0},s_{1},s_{0})=0$
whenever $s_{1}=2$ and $y_{1}\neq3$ or $s_{0}=2$ and $y_{0}\neq3.$
Similarly, $P_{y,s.x}^{r(t)}$ is equal to zero whenever any of the
following conditions hold: (1) $s=2$ and $y\neq3;$ (2) or $y=3$
and $s\neq2$ (3) or $s=0$ and $y\neq2;$ (4) or $y=2$ and $s\neq0.$
Situations where, at time $t,$ the realized survival value $s$ is
not censored or missing but the realized value $y$ is censored or
missing are possible in practice. However, in our application below,
these situations do not occur in our data analysis by design of the
clinical trial. The theoretical framework to accommodate such situations
follows the arguments presented here, and Online Supplement 2 fully
explicates the framework and results in the setting where there exists
individuals $\omega\in\Omega$ for whom $Y(\omega,t)$ is censored
but $S(\omega,t)$ is not censored for any specified time $t$. 

We also provide a section that generalizes our work to tuple valued
$Y(\omega,t)$ that has support on $\mathbf{R}^{n_{y}},$ where $n_{y}$
denotes the number of elements of tuple $Y(\omega,t),$ as well as
multivariate $S(\omega,t).$ This generalization also allows for more
extensive sensitivity analyses for the role of missing data on outcomes.
The simpler setting presented below is provided to make the individual
level interpretation transparent.

\section{Identification of individual level always survivor causal effects}

Here, we only require randomization and consistency for Theorem 1,
Proposition 1, and Corollary 1. In the next section, we will examine
the derivation of similar results in the presence of monotonicity
assumptions. 

\subsection*{Theorem 1}

Suppose $X$ is randomized at baseline. If for some $t\in T$ and
$y\in\{0,1\}$ $P_{y,1.1}^{r(t)}+P_{1-y,1.0}^{r(t)}>1$, then there
exists a non-empty subpopulation $\Omega_{s}\subseteq\Omega$ such
that for every $\omega\in\Omega_{s},$ $S_{1}(\omega,t)=S_{0}(\omega,t)=1$
and $Y_{1}(\omega,t)=y,$ $Y_{0}(\omega,t)=1-y.$ 

Theorem 1 allows for the empirical detection of individual level total
effects within the `always survivor' principal stratum. A difference
between our result and previous methods \cite{ding2011identifiability,wang2017identification,frangakis2002principal,zhang2003estimation,wang2017causal}
is that Theorem 1 effectively only requires randomization and yet
still provides conditions for which the `always survivor' principal
stratum exists and has individuals for whom the treatment changes
the outcome. To the best of our knowledge, this is the first time
an empirical condition to detect causal effects is formulated in the
presence of truncation by death and censoring without any additional
assumptions other than those guaranteed by design of the clinical
trial. Proposition 1 provides further context, comparing the proportion
of individuals randomized at baseline that are always survivors and
who display a positive (or negative) effect of treatment $X$ on an
outcome $Y$ to the proportion of individuals randomized at baseline
that are always survivors and who display a negative (or positive)
effect of treatment $X$ on outcome $Y.$  Corollary 1 provides further
results for this difference for all individuals in the always survivor
principal stratum (observed or censored). Proofs of all results are
given in Online Supplement 1.

\subsection*{Proposition 1}

Suppose $X$ is randomized at baseline. For some $t\in T$ and $y\in\{0,1\},$
the contrast $P_{y,1.1}^{r(t)}+P_{1-y,1.0}^{r(t)}-1$ is equal to 

\begin{eqnarray*}
P^{c(t)}(y,1-y,1,1) & - & \{P^{c(t)}(1-y,y,1,1)+P^{c(t)}(2,y,0,1)+P^{c(t)}(1-y,2,1,0)\\
 &  & +P^{c(t)}(2,2,0,0)+P^{c(t)}(3,3,2,2)+P^{c(t)}(3,y,2,1)\\
 &  & +P^{c(t)}(3,2,2,0)+P^{c(t)}(1-y,3,1,2)+P^{c(t)}(2,3,0,2)\}.
\end{eqnarray*}

\subsection*{Corollary 1}

Suppose $X$ is randomized at baseline. The expression $P_{1,1.1}^{r(t)}+P_{0,1.0}^{r(t)}-1$
is a lower bound on the risk difference between the proportion of
individuals randomized at baseline that are always survivors (observed
or censored) at time $t$ for whom the treatment causes the outcome
and the proportion of individuals randomized at baseline that are
always survivors (observed or censored) at time $t$ for whom the
treatment prevents the outcome of interest. 

In counterfactual notation, the expression $P_{1,1.1}^{r(t)}+P_{0,1.0}^{r(t)}-1$
is a lower bound on $P^{c(t)}(1,0,1,1)-[P^{c(t)}(0,1,1,1)+P^{c(t)}(3,3,2,2)+P^{c(t)}(3,1,2,1)+P^{c(t)}(0,3,1,2)].$
Note $P^{c(t)}(1,0,1,1)$ is a lower bound on the proportion of individuals
randomized at baseline that are always survivors for whom the treatment
causes the outcome $Y$ as it does not count any censored individual
that could be an always survivor for whom the treatment causes the
outcome $Y.$ The term in the square brackets is an upper bound on
the proportion of individuals randomized at baseline that are always
survivors for whom the treatment prevents the outcome $Y,$ because
any censored individual that could possibly be an always survivor
at time $t$ for whom the treatment prevents the outcome $Y$ is represented
in the square brackets. Note, from Theorem 1 and Proposition 1, if
$P_{1,1.1}^{r(t)}+P_{0,1.0}^{r(t)}-1>0,$ then we have established
that treatment efficacy at time $t$ of population randomized at baseline.
Our result establishes that the proportion of individuals randomized
at baseline that are always survivors at time $t$ and treatment causes
the outcome at time $t$ is necessarily greater by at least $P_{1,1.1}^{r(t)}+P_{0,1.0}^{r(t)}-1$
than the proportion of individuals that are always survivors at time
$t$ and the treatment prevents the outcome at time $t.$ 

It follows from Corollary 1, the expression $\max\{P_{1,1.1}^{r(t)}+P_{0,1.0}^{r(t)}-1,0\}$
is thus a lower bound on the proportion of individuals randomized
at baseline that are always survivors (observed or censored) at time
$t$ for whom the treatment causes the outcome: $P^{c(t)}(1,0,1,1)$.
Similarly, the expression $P_{0,1.1}^{r(t)}+P_{1,1.0}^{r(t)}-1$ is
a lower bound on the risk difference between the proportion of individuals
randomized at baseline that are always survivors (observed or censored)
at time $t$ for whom the treatment prevents the outcome and the proportion
of individuals randomized at baseline that are always survivors (observed
or censored) at time $t$ for whom the treatment causes the outcome
of interest. In counterfactual notation, the expression $P_{0,1.1}^{r(t)}+P_{1,1.0}^{r(t)}-1$
is a lower bound on $P^{c(t)}(0,1,1,1)-[P^{c(t)}(1,0,1,1)+P^{c(t)}(3,3,2,2)+P^{c(t)}(3,0,2,1)+P^{c(t)}(1,3,1,2)]$.
The expression $\max\{P_{0,1.1}^{r(t)}+P_{1,1.0}^{r(t)}-1,0\}$ is
a lower bound on the proportion of individuals randomized at baseline
that are always survivors (observed or censored) at time $t$ for
whom the treatment prevents the outcome: $P^{c(t)}(0,1,1,1).$ 

Thus, without any assumption on the censoring mechanism, we can formulate
lower bounds on the proportion of individuals randomized at baseline
that are always survivors (censored or not) at time $t$ and for whom
the treatment causes (prevents) the outcome minus the proportion of
individuals randomized at baseline that are always survivors (censored
or not) at time $t$ and for whom the treatment prevents (causes respectively)
the outcome. To the best of our knowledge, no other paper provides
empirical conditions to detect causal effects in the presence of censoring
without assumptions on the censoring mechanism. Corollary 1 also gives
us a contrast only involving real-world probabilities corresponding
to the minimum proportion of of individuals randomized at baseline
that are always survivors (censored or not) at time $t$ for whom
treatment has an effect.

Theorem 1, Proposition 1 and Corollary 1 provide scientists with interpretations,
under weaker assumptions, that are not possible in any of the previous
approaches \cite{imai2008sharp,wang2017identification,ding2011identifiability}
to deal with truncation by death. Previous approaches that examine
the survivor average causal effect make more assumptions than randomization
and consistency \cite{ding2011identifiability,imai2008sharp,wang2017identification}.
Bounds on the survivor average causal effect and other causal estimands
are discussed in the Online Supplement 2 as well as the Generalization
and Extensions Section. Other approaches rely on either principal
ignorability or a sufficient covariate to essentially choose the set
of individuals that are always survivors at time $t,$ and then characterize
the average treatment effect for this set of always survivors \cite{ding2011identifiability,wang2017identification}.

Our approach is substantially different in terms of assumptions and
interpretation. Previous approaches, including principal ignorability
or sufficient covariate methods, make assumptions that are not guaranteed
by design, and therefore cannot guarantee the causal interpretation
of their results \cite{tchetgen2014identification,wang2017causal,wang2017identification,zhang2003estimation,ding2011identifiability,frangakis2002principal}.
For example, to identify survivor average causal effects two recent
papers propose a substitution variable denoted $A$ that essentially
amounts to providing the statistician with the relevant units that
form the always survivor principal strata. Some of the assumptions,
for example principal ignorability or sufficient covariate assumptions,
that are used to get these relevant units are not only assumptions
which cannot be guaranteed to hold in any experimental design, but
are also unfalsifiable \cite{ding2011identifiability,wang2017causal}
from the observed data as these assumptions involve conditioning on
joint potential outcomes. Scientists that use these previous methods
\cite{ding2011identifiability,wang2017falsification} will rely on
assumptions that are not only unverifiable, or even unfalsifiable,
but also offer little in the way of formal scientific or statistical
methodology for the statistician to assess whether they are even plausible
given that the observed data distribution cannot be used to assess
assumptions where joint counterfactuals (e.g. $S_{1}(t)=1$ and $S_{0}(t)=1$)
are both in the conditioning statement.

\section{Identification of individual level always survivor causal effects
under monotonicity}

Monotonicity assumptions are often employed in methods to evaluate
principal stratum direct effects. Recall, we denote $Y_{x}^{[y]}(\omega,t)=I(Y_{x}(\omega,t)=y),$
and $S_{x}^{[s]}(\omega,t)=I(S_{x}(\omega,t)=s)$ for $y\in\{0,1,2,3\},$
$s\in\{0,1,2\},$ and $t\in T.$ Here we will consider additional
results under two different monotonicity assumptions. In the first
monotonicity assumption, we will assume $S_{1}^{[0]}(\omega,t)+S_{0}^{[1]}(\omega,t)\leq1$
for all $\omega\in\Omega$ and for a pre-specified $t\in T.$ This
assumption translates to it being the case that no individual exists
who would be alive and not censored at least until time $t$ when
forced to take control, but would be dead and not censored by time
$t$ when forced to take treatment. This means that we do not observe
any individual for whom $S_{1}(\omega,t)=0$ and $S_{0}(\omega,t)=1$
for the specified $t\in T.$ In the absence of censoring, this assumption
is simply that for no individual does the treatment itself cause death.
If treatment is considered to be non-smoking and the control is smoking,
scientific literature might lead us to believe that such an assumption
is reasonable to make. Subject matter expertise could provide guidance
for which $t\in T,$ $S_{1}^{[0]}(\omega,t)+S_{0}^{[1]}(\omega,t)\leq1$
for all $\omega\in\Omega.$ If for all $\omega\in\Omega,$ $S_{1}^{[0]}(\omega,t)+S_{0}^{[1]}(\omega,t)\leq1$
holds for only for a subset of $T_{s}\subset T,$ then Theorem 2A
and Proposition 2A, provided below, applies only for those $t\in T_{s}.$
While such monotonicity constraints are never verifiable, they are
falsifiable, and again can sometimes be justified with subject matter
knowledge. 

For the second monotonicity assumption, we shall assume that $S_{1}^{[2]}(\omega,t)+S_{0}^{[1]}(\omega,t)\leq1$
for all $\omega\in\Omega$ and for a pre-specified $t\in T.$ Heuristically,
this assumption means that there is no individual who would drop out
of the study by time $t$ when forced to take treatment but who would
not drop out of the study and is alive when forced to take control
condition (e.g. a prior or older treatment). This means that we do
not observe any individual for whom $S_{1}(\omega,t)=2$ and $S_{0}(\omega,t)=1$
for the specified $t\in T.$ Again, such monotonicity constraints
are falsifiable and could be judged using subject matter knowledge.
For instance, if clinicians believe that the new treatment is less
toxic than the old treatment, then this could provide some evidence
that patients might be less disposed to drop out if they are forced
to take the less toxic new treatment in comparison to the more toxic
old treatment. 

The monotonicity assumption on censoring is weaker than the non-informative
censoring that is commonly used in statistical analysis. Non-informative
censoring would have that if an individual $\omega\in\Omega$ is censored
under treatment at any fixed time $t\in T,$ denoted $S_{1}(\omega,t)=2$,
then the same individual would be censored under the control condition
at the same fixed time $t\in T$, denoted $S_{0}(\omega,t)=2.$ In
contrast, the monotonicity assumption on censoring, enables statisticians
to examine situations where censoring is likely to occur sooner in
one of the two arms than in the other.

We will consider analogous results to Theorem 1 and Proposition 1
when one or both of these monotonicity assumptions hold. If such monotonicity
assumptions are untenable for the population in question, investigators
can still use the empirical conditions provided earlier that do not
need any monotonicity assumptions. Causally interpretable sensitivity
analyses are also presented in Section 6 when one or both of the monotonicity
assumptions do not hold. Such sensitivity analysis provide researchers
tools to quantify the consequences of violations of these assumptions,
and also directly consider how each individual in their sample could
contribute to the causal conclusions.

\subsection*{Theorem 2A}

Suppose $X$ is randomized at baseline. In addition, suppose that
for some $t\in T,$ $S_{1}^{[0]}(\omega,t)+S_{0}^{[1]}(\omega,t)\leq1$
for all $\omega\in\Omega.$ For some $t\in T$ and $y\in\{0,1\},$
if the empirical condition $P_{1-y,1.0}^{r(t)}-P_{1-y,1.1}^{r(t)}-P_{3,2.1}^{r(t)}>0$
holds, then there exists a non-empty subpopulation $\Omega_{s}\subseteq\Omega$
such that for every $\omega\in\Omega_{s},$ $S_{1}(\omega,t)=S_{0}(\omega,t)=1,$
and $Y_{1}(\omega,t)=y,$ $Y_{0}(\omega,t)=1-y.$

\subsection*{Proposition 2A}

Suppose $X$ is randomized at baseline. In addition, suppose that
for some $t\in T,$ $S_{1}^{[0]}(\omega,t)+S_{0}^{[1]}(\omega,t)\leq1$
for all $\omega\in\Omega.$ For some $t\in T$ and $y\in\{0,1\},$
the contrast $P_{1-y,1.0}^{r(t)}-P_{1-y,1.1}^{r(t)}-P_{3,2.1}^{r(t)}$
is equal to
\begin{eqnarray*}
P^{c(t)}(y,1-y,1,1) & - & \{P^{c(t)}(1-y,y,1,1)+P^{c(t)}(1-y,2,1,0)\\
 &  & +P^{c(t)}(3,3,2,2)+P^{c(t)}(3,y,2,1)\\
 &  & +P^{c(t)}(3,2,2,0)+P^{c(t)}(1-y,3,1,2)\}.
\end{eqnarray*}

\subsection*{Corollary 2A}

Suppose $X$ is randomized at baseline. In addition, suppose that
for some $t\in T,$ $S_{1}^{[0]}(\omega,t)+S_{0}^{[1]}(\omega,t)\leq1$
for all $\omega\in\Omega.$ The expression $P_{0,1.0}^{r(t)}-P_{0,1.1}^{r(t)}-P_{3,2.1}^{r(t)}$
is a lower bound on the risk difference between the proportion of
individuals randomized at baseline that are always survivors (observed
or censored) at time $t$ for whom the treatment causes the outcome
and the proportion of individuals randomized at baseline that are
always survivors (observed or censored) at time $t$ for whom the
treatment prevents the outcome of interest. 

The expression $\max\{P_{0,1.0}^{r(t)}-P_{0,1.1}^{r(t)}-P_{3,2.1}^{r(t)},0\}$
is thus also a lower bound on the proportion of individuals randomized
at baseline that are always survivors (observed or censored) at time
$t$ for whom treatment causes the outcome. Similarly, the expression
$P_{1,1.0}^{r(t)}-P_{1,1.1}^{r(t)}-P_{3,2.1}^{r(t)}$ is a lower bound
on the risk difference between the proportion of individuals randomized
at baseline that are always survivors (observed or censored) at time
$t$ for whom the treatment prevents the outcome and the proportion
of individuals randomized at baseline that always survivors (observed
or censored) at time $t$ for whom the treatment causes the outcome
of interest. The expression $\max\{P_{1,1.0}^{r(t)}-P_{1,1.1}^{r(t)}-P_{3,2.1}^{r(t)},0\}$
is a lower bound on the proportion of individuals randomized at baseline
that are always survivors (observed or censored) at time $t$ for
whom treatment prevents the outcome. 

The interpretations for Theorem 2A and Proposition 2A are similar
to those provided for Theorem 1 and Proposition 1. Specifically, if
we believe that no individual exists that would die and not be censored
by time $t$ when forced to take treatment but would not die and not
be censored by time $t$ when forced to take control and if $P_{0,1.0}^{r(t)}-P_{0,1.1}^{r(t)}-P_{3,2.1}^{r(t)}>0,$
then there exists a set of individuals that would live at least until
time $t$ and for whom the treatment causes outcome. From Corollary
2A, we learn that the proportion of individuals randomized at baseline
that are always survivors (observed or censored) until time $t$ and
for whom the treatment causes (prevents) outcome of interest is at
least greater the proportion of individuals randomized at baseline
that are always survivors (observed or censored) at time $t$ regardless
of treatment assignment and for whom the treatment prevents (causes
respectively) outcome of interest. Also, Corollary 2A provides a contrast
involving only real-world probabilities that is the minimum proportion
of individuals randomized at baseline that are always survivors (observed
or censored) at time $t$ for whom the treatment causes (or prevents)
the outcome. The other monotonicity assumption also provides useful
results and is presented next.

\subsection*{Theorem 2B}

Suppose $X$ is randomized at baseline. In addition, suppose that
for some $t\in T,$ $S_{1}^{[2]}(\omega,t)+S_{0}^{[1]}(\omega,t)\leq1$
for all $\omega\in\Omega.$ For some $t\in T$ and $y\in\{0,1\},$
if the empirical condition $P_{1-y,1.0}^{r(t)}-P_{1-y,1.1}^{r(t)}-P_{2,0.1}^{r(t)}>0$
holds, then there exists a non-empty subpopulation $\Omega_{s}\subseteq\Omega$
such that for every $\omega\in\Omega_{s},$ $S_{1}(\omega,t)=S_{0}(\omega,t)=1,$
and $Y_{1}(\omega,t)=y,$ $Y_{0}(\omega,t)=1-y.$ 

\subsection*{Proposition 2B}

Suppose $X$ is randomized at baseline. In addition, suppose that
for some $t\in T,$ $S_{1}^{[2]}(\omega,t)+S_{0}^{[1]}(\omega,t)\leq1$
for all $\omega\in\Omega.$ For some for some $t\in T$ and $y\in\{0,1\},$
the empirical condition $P_{1-y,1.0}^{r(t)}-P_{1-y,1.1}^{r(t)}-P_{2,0.1}^{r(t)}$
is equal to 
\begin{eqnarray*}
P^{c(t)}(y,1-y,1,1) & - & \{P^{c(t)}(1-y,y,1,1)+P^{c(t)}(1-y,2,1,0)\\
 &  & +P^{c(t)}(2,y,0,1)+P^{c(t)}(2,2,0,0)\\
 &  & +P^{c(t)}(1-y,3,1,2)+P^{c(t)}(2,3,0,2)\}.
\end{eqnarray*}

\subsection*{Corollary 2B}

Suppose $X$ is randomized at baseline. In addition, suppose that
for some $t\in T,$ $S_{1}^{[2]}(\omega,t)+S_{0}^{[1]}(\omega,t)\leq1$
for all $\omega\in\Omega.$ The expression $\max\{P_{0,1.0}^{r(t)}-P_{0,1.1}^{r(t)}-P_{2,0.1}^{r(t)},0\}$
is a lower bound on the proportion of individuals randomized at baseline
that are always survivors (observed or censored) at time $t$ for
whom treatment causes the outcome. 

Similarly, the expression $\max\{P_{1,1.0}^{r(t)}-P_{1,1.1}^{r(t)}-P_{2,0.1}^{r(t)},0\}$
is a lower bound on the proportion of individuals randomized at baseline
that are always survivors (observed or censored) at time $t$ for
whom treatment prevents the outcome. The interpretations for Theorem
2B and Proposition 2B are similar to those provided for Theorem 2A
and Proposition 2A. Finally, if we believe that the two monotonicity
assumptions hold, then we have the following results.

\subsection*{Theorem 2C}

Suppose $X$ is randomized at baseline. In addition, suppose that
for some $t\in T,$ $S_{1}^{[2]}(\omega,t)+S_{0}^{[1]}(\omega,t)\leq1$
and $S_{1}^{[0]}(\omega,t)+S_{0}^{[1]}(\omega,t)\leq1$ for all $\omega\in\Omega.$
For some $t\in T$ and $y\in\{0,1\},$ if the empirical condition
$P_{1-y,1.0}^{r(t)}-P_{1-y,1.1}^{r(t)}>0$ holds, then there exists
a non-empty subpopulation $\Omega_{s}\subseteq\Omega$ such that for
every $\omega\in\Omega_{s},$ $S_{1}(\omega,t)=S_{0}(\omega,t)=1,$
and $Y_{1}(\omega,t)=y,$ $Y_{0}(\omega,t)=1-y.$

\subsection*{Proposition 2C}

Suppose $X$ is randomized at baseline. In addition, suppose that
for some $t\in T,$ $S_{1}^{[2]}(\omega,t)+S_{0}^{[1]}(\omega,t)\leq1$
and $S_{1}^{[0]}(\omega,t)+S_{0}^{[1]}(\omega,t)\leq1$  for all $\omega\in\Omega.$
For some $t\in T$ and $y\in\{0,1\},$ the contrast $P_{1-y,1.0}^{r(t)}-P_{1-y,1.1}^{r(t)}$
is equal to
\begin{eqnarray*}
P^{c(t)}(y,1-y,1,1) & - & \{P^{c(t)}(1-y,y,1,1)+P^{c(t)}(1-y,2,1,0)+P^{c(t)}(1-y,3,1,2)\}.
\end{eqnarray*}

\subsection*{Corollary 2C}

Suppose $X$ is randomized at baseline. In addition, suppose that
for some $t\in T,$ $S_{1}^{[2]}(\omega,t)+S_{0}^{[1]}(\omega,t)\leq1$
and $S_{1}^{[0]}(\omega,t)+S_{0}^{[1]}(\omega,t)\leq1$ for all $\omega\in\Omega.$
The expression $\max\{P_{0,1.0}^{r(t)}-P_{0,1.1}^{r(t)},0\}$ is a
lower bound on the proportion of individuals at randomized at baseline
that are always survivors (observed or censored) at time $t$ for
whom treatment causes the outcome. 

Similarly, the expression $\max\{P_{1,1.0}^{r(t)}-P_{1,1.1}^{r(t)},0\}$
is a lower bound on the proportion of individuals randomized at baseline
that are always survivors (observed or censored) at time $t$ for
whom treatment prevents the outcome. The interpretations of Theorem,
Proposition, and Corollary 2C are similar to those provided for Theorem,
Proposition and Corollary 2B respectively. The conditions are to detect
individual level always survivor causal effects are weaker the more
monotonicity assumptions are made. Again, if such monotonicity assumptions
cannot be justified on scientific grounds, investigators can still
use the results that do not require such assumptions. The sensitivity
analysis that is presented in Section 6 enables researchers to assess
the consequences of violations of such monotonicity assumptions and
also check how violations of these monotonicity assumptions impact
the causal findings.

\section{Inference for individual level always survivor causal effects}

To investigate individual level always survivor causal effects, null
hypotheses can be formulated that when falsified produce the inequalities
associated with Theorem 1, 2A, 2B, or 2C. For Theorem 1, if we reject
$P_{y,1.1}^{r(t)}+P_{1-y,1.0}^{r(t)}\leq1$ for $y\in\{0,1\},$ then
we conclude with a fixed type one error rate that there exists a non-empty
subpopulation $\Omega_{s}\subseteq\Omega$ such that for every $\omega\in\Omega_{s},$
$S_{1}(\omega,t)=S_{0}(\omega,t)=1$ and $Y_{1}(\omega,t)=y,$ $Y_{0}(\omega,t)=1-y.$
Testing $P_{y,1.1}^{r(t)}+P_{1-y,1.0}^{r(t)}\leq1$ is equivalent
to testing $P_{1-y,1.0}^{r(t)}\leq P_{1-y,1.1}^{r(t)}+P_{2,0.1}^{r(t)}+P_{3,2.1}^{r(t)},$
which can also be converted into a one-sided difference of proportions.
Note testing $P_{1-y,1.0}^{r(t)}\leq P_{y,1.1.}^{r(t)}+P_{2,0.1}^{r(t)}+P_{3,2.1}^{r(t)}$
is equivalent to testing $P(Y(t)=1-y,S(t)=1\mid X=0)\leq P(\{Y(t)=1-y,S(t)=1\}\cup\{Y(t)=2,S(t)=0\}\cup\{Y(t)=3,S(t)=2\}\mid X=1)$. 

To use Theorem 2A, if assume $S_{1}^{[0]}(\omega,t)+S_{0}^{[1]}(\omega,t)\leq1$
for all $\omega\in\Omega$ and we reject, $P_{1-y,1.0}^{r(t)}-P_{1-y,1.1}^{r(t)}-P_{3,2.1}^{r(t)}\leq0$
for some $y\in{0,1}$, then we conclude with a fixed type one error
rate that there exists a non-empty subpopulation $\Omega_{s}$ such
that for every $\omega\in\Omega_{s},$ $S_{1}(\omega,t)=S_{0}(\omega,t)=1$
and $Y_{1}(\omega,t)=y,$ $Y_{0}(\omega,t)=1-y.$ Note testing $P_{1-y,1.0}^{r(t)}\leq P_{y,1.1.}^{r(t)}+P_{3,2.1}^{r(t)}$
is equivalent to testing $P(Y(t)=1-y,S=1\mid X=0)\leq P(\{Y(t)=1-y,S(t)=1\}\cup\{Y(t)=3,S(t)=2\}\mid X=1)$.
Similarly, to use Theorem 2B, if assume $S_{1}^{[2]}(\omega,t)+S_{0}^{[1]}(\omega,t)\leq1$
for all $\omega\in\Omega$ and we reject, $P_{1-y,1.0}^{r(t)}-P_{1-y,1.1}^{r(t)}-P_{2,0.1}^{r(t)}\leq0$
for some $y\in\{0,1\}$, then we conclude with a fixed type one error
rate that there exists a non-empty subpopulation $\Omega_{s}\subseteq\Omega$
such that for every $\omega\in\Omega_{s},$ $S_{1}(\omega,t)=1$ $S_{0}(\omega,t)=1$
and $Y_{1}(\omega,t)=y,$ $Y_{0}(\omega,t)=1-y.$ Note testing $P_{1-y,1.0}^{r(t)}\leq P_{y,1.1.}^{r(t)}+P_{2,0.1}^{r(t)}$
is equivalent to testing $P(Y(t)=1-y,S=1\mid X=0)\leq P(\{Y(t)=1-y,S(t)=1\}\cup\{Y(t)=2,S(t)=0\}\mid X=1)$.
Finally, to use Theorem 2C, if assume and $S_{1}^{[2]}(\omega,t)+S_{0}^{[1]}(\omega,t)\leq1$
and $S_{1}^{[0]}(\omega,t)+S_{0}^{[1]}(\omega,t)\leq1$ for all $\omega\in\Omega$
and we reject, $P_{1-y,1.0}^{r(t)}-P_{1-y,1.1}^{r(t)}\leq0$ for some
$y\in{0,1}$, then we conclude with a fixed type one error rate that
there exists a non-empty subpopulation $\Omega_{s}$ such that for
every $\omega\in\Omega_{s},$ $S_{1}(\omega,t)=S_{0}(\omega,t)=1,$
and $Y_{1}(\omega,t)=y,$ $Y_{0}(\omega,t)=1-y.$ 

In our data application below we have seven fixed time periods of
interest and so we use Theorem 1 seven times and apply a Bonferroni
correction. We also apply Theorem 2A, 2B, 2C arguing that the relevant
monotonicity assumptions likely hold, and demonstrate similar conclusions.
Note for the data application, we use the equivalent test expressed
in a one-sided difference of proportions instead of the complement
of the inequality associated with Theorems 1, 2A, 2B, 2C. 

The Online Supplement 2 provides Bayesian and randomization based
inferential methods to assess results associated with Theorem 1, 2A,
2B, 2C. For our data application, we use the standard t-test for a
one-sided difference of proportions to assess always survivor causal
effects, because of the moderately large sample size and ease of implementation
for readers. In the setting of smaller sample sizes, the randomization
or Bayesian approach might be more appropriate. 

\section{Sensitivity analysis for monotonicity assumptions}

As demonstrated in Section 3, we provide methods for identification
of individual level always survivor causal effects with only the assumptions
guaranteed by design of the clinical trial. Theorem 1, Proposition
1 and Corollary 1 effectively only require randomization at baseline
to first detect always survivor causal effects and secondly provide
a population level characterization of such causal effects. Monotonicity
assumptions on survival or censoring enable scientists to detect always
survivor causal effects under assumptions that are justified using
subject matter knowledge, and can assist researchers to detect always
survivor causal effects when the tests without such monotonicity assumptions
are too stringent. Sensitivity analysis can provide scientists and
statisticians with methods to interpret their conclusions when a set
of assumptions is not guaranteed by design.

\subsection*{Theorem 3A}

Suppose $X$ is randomized at baseline. For some $t\in T$ and $y\in\{0,1\},$
the expression $P_{1-y,1.0}^{r(t)}-P_{1-y,1.1}^{r(t)}-P_{3,2.1}^{r(t)}-d_{m}(t)$
for $d_{m}(t)=P^{c(t)}(2,1-y,0,1)-P^{c(t)}(1-y,2,1,0)-P^{c(t)}(3,2,2,0)$
is equal to 

\begin{align*}
 & P^{c(t)}(y,1-y,1,1)-[P^{c(t)}(1-y,y,1,1)+P^{c(t)}(3,3,2,2)+P^{c(t)}(3,y,2,1)+P^{c(t)}(1-y,3,1,2)],
\end{align*}
 and consequently if $P_{1-y,1.0}^{r(t)}-P_{1-y,1.1}^{r(t)}-P_{3,2.1}^{r(t)}>d_{m}(t),$
then 

\begin{align*}
 & P^{c(t)}(y,1-y,1,1)>P^{c(t)}(1-y,y,1,1)+P^{c(t)}(3,3,2,2)+P^{c(t)}(3,y,2,1)+P^{c(t)}(1-y,3,1,2).
\end{align*}

\subsection*{Theorem 3B}

Suppose $X$ is randomized at baseline. If for some $t\in T$ and
$y\in\{0,1\}$ the expression $P_{1-y,1.0}^{r(t)}-P_{1-y,1.1}^{r(t)}-P_{2,0.1}^{r(t)}-a_{m}(t)$
for $a_{m}(t)=P^{c(t)}(3,1-y,2,1)-P^{c(t)}(1-y,2,1,0)-P^{c(t)}(1-y,3,1,2)-P^{c(t)}(2,y,0,1)-P^{c(t)}(2,2,0,0)-P^{c(t)}(2,3,0,2)$
is equal to 
\begin{align*}
P^{c(t)}(y,1-y,1,1)-P^{c(t)}(1-y,y,1,1),
\end{align*}
 and consequently if $P_{1-y,1.0}^{r(t)}-P_{1-y,1.1}^{r(t)}-P_{2,0.1}^{r(t)}>a_{m}(t),$
then $P^{c(t)}(y,1-y,1,1)>P^{c(t)}(1-y,y,1,1).$

\subsection*{Theorem 3C}

Suppose $X$ is randomized at baseline. If for some $t\in T$ and
$y\in\{0,1\}$ the expression $P_{1-y,1.0}^{r(t)}-P_{1-y,1.1}^{r(t)}-k_{m}(t)$
for $k_{m}(t)=P^{c(t)}(2,1-y,0,1)+P^{c(t)}(3,1-y,2,1)-P^{c(t)}(1-y,2,1,0)-P^{c(t)}(1-y,3,1,2)$
is equal to 
\begin{align*}
P^{c(t)}(y,1-y,1,1)-P^{c(t)}(1-y,y,1,1),
\end{align*}
 and consequently if $P_{1-y,1.0}^{r(t)}-P_{1-y,1.1}^{r(t)}>k_{m}(t),$
then $P^{c(t)}(y,1-y,1,1)>P^{c(t)}(1-y,y,1,1).$

The quantities $d_{m}(t),$ $a_{m}(t)$ and $k_{m}(t)$ are not point
identified in randomized studies. A simple, yet effective, approach
for sensitivity analysis would be to first estimate the left hand
side of the inequalities associated with Theorem 3A, Theorem 3B, and
Theorem 3C, and then crudely interpret $d_{m}(t),$ $a_{m}(t),$ $k_{m}(t)$
as the proportion of individuals of a specific counterfactual form
to render the respective inferences associated with Theorem 3A, 3B,
3C to be invalid. As an example, suppose the researcher estimates
$P_{0,1.0}^{r(t)}-P_{0,1.1}^{r(t)}$ for a specific $t\in T$ as 0.2.
Using Theorem 2C, the scientist would conclude that at least 20\%
of her population randomized at baseline would follow the counterfactual
response pattern $Y_{1}(\omega,t)=1,$ $Y_{0}(\omega,t)=0,$ $S_{1}(\omega,t)=1,$
and $S_{0}(\omega,t)=1.$ For the scientist's conclusion to be possibly
incorrect on the existence of such an individual level effect, the
exact 20\% of individuals in her population that she believed to follow
counterfactual response pattern $Y_{1}(\omega,t)=1,$ $Y_{0}(\omega,t)=0,$
and $S_{1}(\omega,t)=S_{0}(\omega,t)=1,$ would need to follow either
one of two counterfactual response types $\{\omega_{1},\omega_{2}\}\in\Omega:$
$Y_{1}(\omega_{1},t)=2,$ $Y_{0}(\omega_{1},t)=1,$ $S_{1}(\omega_{1},t)=0,$
$S_{0}(\omega_{1},t)=1$ or $Y(\omega_{2},t)=3,$ $Y_{0}(\omega_{2},t)=0,$
$S_{1}(\omega_{2},t)=2,$ $S_{0}(\omega_{2},t)=1.$ Additionally,
there has to be at least 20\% (possibly significantly greater than
20\%, if $P^{c(t)}(0,2,1,0)$ and $P^{c(t)}(0,3,1,2)$ are non-zero)
of individuals that follow counterfactual response types $Y_{1}(\omega_{1},t)=2,$
$Y_{0}(\omega_{1},t)=1,$ $S_{1}(\omega_{1},t)=0,$ $S_{0}(\omega_{1},t)=1$
or $Y(\omega_{2},t)=3,$ $Y_{0}(\omega_{2},t)=0,$ $S_{1}(\omega_{2},t)=2,$
$S_{0}(\omega_{2},t)=1$ at time $t$ for there to be no individual
of counterfactual response type $Y_{1}(\omega,t)=1,$ $Y_{0}(\omega,t)=0,$
$S_{1}(\omega,t)=1,$ and $S_{0}(\omega,t)=1$ in our population. 

Theorem 3A, 3B, and 3C provide further context on how violations of
individual level monotonicity assumptions impact the relevant counterfactual
proportions. As an example, consider a researcher that estimates $P_{0,1.0}^{r(t)}-P_{0,1.1}^{r(t)}-P_{3,2.1}^{r(t)}$
as 0.25 at time $t=1.$ If the scientist believes that at most five
percent of her clinical trial subject's randomized at baseline satisfy
the following three conditions, (1) not survive when given the new
treatment by time $t=1$, (2) survive when given the control condition
at least until time $t=1$, and (3) not develop the outcome of interest
under the control condition (denoted $P^{c(t)}(2,0,0,1)\leq0.05)$,
then from Theorem 3A, we know that the proportion of individuals randomized
at baseline that are always survivors at least until time $t=1$ that
would develop outcome $Y$ under the treatment condition but when
prescribed the control condition is at least $0.25-0.05=0.20$. 

Such sensitivity analysis enables researchers to assess either how
many individuals of a specific counterfactual response type need to
exist in a population for the detection of individual principal stratum
causal effects to be invalid if a particular monotonicity assumption
fails, or how given individual level counterfactual response types
attenuate the always survivor causal effect of interest. Additionally,
researchers will develop an understanding of the individual reasons
for trial censoring or death at each particular time in the trial.
If a trial subject dies in a random car crash at time $t=1$, then
a researcher might have strong arguments to support the belief that
the subjects death had nothing to do with treatment assignment. Such
information can be used to choose and refine suitable values of $d_{m}(t),$
$a_{m}(t)$ and $k_{m}(t)$ for different times in the study.

Similarly, if the control condition arm experiences a greater number
of adverse events (for instance higher toxicities) for the duration
the trial and the rate of censoring is at least as large in the control
arm as in the treatment arm, then similarly a researcher might have
grounds to believe that at particular times in the trial that $P^{c(t)}(3,1-y,2,1)$
is less than $P^{c(t)}(1-y,3,1,2).$ In such situations, the researcher
might have solid grounds to believe that $P_{1-y,1.0}^{r(t)}-P_{1-y,1.1}^{r(t)}-P_{2,0.1}^{r(t)}$
is fairly close to $P^{c(t)}(y,1-y,1,1)-P^{c(t)}(1-y,y,1,1).$ After
unblinding the trial, researchers might discover that some toxicities
only occur in one of the two arms, and that such toxicities cause
the patients to drop out of the study. A careful sensitivity analysis
that keeps track of the each individual's reasons for drop out or
death could provide researchers with greater assurance in detecting
always survivor causal effects and also enable them to further refine
the constants $k_{m}(t)$ and $a_{m}(t)$ that are associated with
Theorems and Corollaries 3B and 3C. 

The data analysis presented below illustrates how such sensitivity
analysis for monotonicity assumptions can reassure researchers about
the veracity of their scientific findings. We present a second more
generalized version of sensitivity analysis in the section titled
Generalizations and Extensions as well as some notes on sensitivity
analysis in Online Supplement 2. 

\section{Application to Southwest Oncology Group Trial}

Our results are applied to the data from the Southwest Oncology Group
(SWOG) Trial, which was a phase III trial to compare docetaxel plus
estramustine against mitoxantrone plus prednisone in men with metastatic,
hormone-independent prostate cancer \cite{petrylak2004docetaxel}.
For the trial, the primary end-point was overall survival \cite{petrylak2004docetaxel}.
A total of 338 patients were randomized to the docetaxel plus estramustine
(henceforth referred to as docetaxel), and a total of 336 were randomized
to the mitoxantrone plus prednisone (henceforth referred to as mitoxantrone).
The trial found that docetaxel improved median survival in comparison
to mitoxantrone \cite{petrylak2004docetaxel}. Progression of cancer
is an important secondary outcome of interest and may be subject to
truncation by death. Our hypothesis is that for some pre-specified
times $t\in T_{\text{SWOG}}$, the proportion of individuals randomized
at baseline who would remain alive regardless of treatment at least
until the pre-specified time $t$ and whose cancer would progress
under mitoxantrone but not under docetaxel at time $t$ is greater
than the proportion of individuals randomized at baseline who would
remain alive regardless of treatment at least until the pre-specified
time $t$ and whose cancer would progress under docetaxel but not
under mitoxantrone at time $t$. Such a hypothesis can be evaluated
using our results, but no previous approach could answer such a question
using only the set of assumptions guaranteed by design.

The time periods are 1 month $(t=1)$, 2 months $(t=2)$, 3 months
$(t=3)$, 4 months $(t=4)$, 6 months $(t=5)$, 12 months $(t=6)$,
and 18 months $(t=7)$ post treatment initiation. Here, $X=1$ denotes
that the individual was randomized to the docetaxel arm, and $X=0$
denotes that the individual was randomized to the mitoxantrone arm.
Outcome $Y(t)$ is defined as a categorical variable that takes values
in the set $\{0,1,2,3\}$ for each $t\in T_{\text{SWOG}}=\{1,\ldots,7\},$
where $Y(t)=0$ means that the individual's cancer did not progress
at time point $t,$ $Y(t)=1$ means that the individual's cancer progressed
at time point $t,$ $Y(t)=2$ means that the individual did not survive
until time $t,$ and finally $Y(t)=3$ means that the individual's
observation was censored or missing at time $t.$ Survival status
at time $t,$ denoted $S(t)$ is also defined as a categorical variable
that takes value in the set $\{0,1,2\}$ for each $t\in T_{\text{SWOG}},$
where $S(t)=0$ means that the individual did not survive at time
period $t,$ $S(t)=1$ means that the individual survived to at least
time period $t,$ and $S(t)=2$ means that the individual's survival
status is censored or missing at time period $t.$ We provide the
relevant contingency tables in the appendix and sample R code that
is used to produce all our results in Online Supplement 2.

\subsection{Data Analysis without monotonicity assumptions}

Without any monotonicity assumptions, we need to test $P_{0,1.1}^{r(t)}+P_{1,1.0}^{r(t)}\leq1$
for each $t\in\{1,\ldots,7\}$ to evaluate at each time point whether
there exist individuals for whom $S_{1}(\omega,t)=S_{0}(\omega,t)=1,$
$Y_{1}(\omega,t)=0,$ $Y_{0}(\omega,t)=1,$ that is these, individuals
would survive at least until time $t$ regardless of treatment, but
treatment with docetaxel (vs. mitoxantrone) would prevent their cancer
from progressing. This hypothesis test is equivalent to testing $P_{1,1.0}^{r(t)}\leq P_{1,1.1}^{r(t)}+P_{2,0.1}^{r(t)}+P_{3,2.1}^{r(t)}.$
To conduct this hypothesis test, we provide the following three contingency
tables:

\begin{table}[H]
\begin{centering}
\begin{tabular}{cccccccc}
\toprule 
 & 1 month & 2 months & 3 months & 4 months & 6 months & 12 months & 18 months\tabularnewline
\midrule
\midrule 
$X=1$ & 6 & 30 & 72 & 86 & 116 & 175 & 144\tabularnewline
\midrule 
$X=0$ & 40 & 94 & 146 & 150 & 164 & 147 & 121\tabularnewline
\bottomrule
\end{tabular} 
\par\end{centering}
\caption{Survive and cancer progression at time $t$ $(Y(t)=1,S(t)=1)$}
\end{table}

\begin{table}[H]
\begin{centering}
\begin{tabular}{cccccccc}
\toprule 
 & 1 month & 2 months & 3 months & 4 months & 6 months & 12 months & 18 months\tabularnewline
\midrule
\midrule 
$X=1$ & 3 & 10 & 14 & 22 & 41 & 88 & 139\tabularnewline
\midrule 
$X=0$ & 3 & 8 & 15 & 25 & 47 & 114 & 182\tabularnewline
\bottomrule
\end{tabular} 
\par\end{centering}
\caption{Did not survive until $t,$ $(Y(t)=2,S(t)=0)$}
\end{table}

\begin{table}[H]
\begin{centering}
\begin{tabular}{cccccccc}
\toprule 
 & 1 month & 2 months & 3 months & 4 months & 6 months & 12 months & 18 months\tabularnewline
\midrule
\midrule 
$X=1$ & 9 & 9 & 9 & 9 & 9 & 10 & 11\tabularnewline
\midrule 
$X=0$ & 15 & 15 & 15 & 15 & 16 & 17 & 19\tabularnewline
\bottomrule
\end{tabular} 
\par\end{centering}
\caption{Censored at time $t,$ $(Y(t)=3,S(t)=2)$}
\end{table}

Two-sided confidence intervals for $P(Y(t)=1,S(t)=1\mid X=0)-P(\{Y(t)=1,S(t)=1\}\cup\{Y(t)=3,S(t)=2\}\cup\{Y(t)=2,S(t)=0\}\mid X=1)$
and one-sided p-values for Wald test of the null hypothesis $P(Y(t)=1,S(t)=1\mid X=0)-P(\{Y(t)=1,S(t)=1\}\cup\{Y(t)=3,S(t)=2\}\cup\{Y(t)=2,S(t)=0\}\mid X=1)\leq0$
are provided below:

\begin{table}[H]
\begin{centering}
\begin{tabular}{cccccccc}
\toprule 
 & 1 month & 2 months & 3 months & 4 months & 6 months & 12 months & 18 months\tabularnewline
\midrule
\midrule 
Estimate & $0.07$ & 0.12 & $0.15$ & $0.10$ & 0.00 & -0.37 & -0.51\tabularnewline
\midrule 
95\% CI & $(0.02,0.11)$ & $(0.07,0.20)$ & $(0.08,0.23)$ & $(0.03,0.18)$ & $(-0.08,0.08)$ & $(-0.44,-0.30)$ & $(-0.58,-0.44)$\tabularnewline
\midrule 
99\% CI & $(0.01,0.12)$ & $(0.05,0.22)$ & $(0.06,0.25)$ & $(0.00,0.20)$ & $(-0.11,0.10)$ & $(-0.46,-0.28)$ & $(-0.59,-0.42)$\tabularnewline
\midrule 
p-value & $0.0018$ & $<0.0001$ & $<0.0001$ & $0.004$ & 0.51 & 1 & 1\tabularnewline
\bottomrule
\end{tabular} 
\par\end{centering}
\caption{Results without monotonicity assumptions}
\end{table}

The results from Table 4 demonstrate by Corollary 1 for the Southwest
Oncology Group Trial, that there is statistical evidence suggesting
that the proportion of individuals randomized at baseline that are
always survivors (observed or censored) at one month post treatment
initiation whose cancer would not progress on docetaxel but would
progress on mitoxantrone is greater by 0.07 with a 95 percent confidence
interval of $(0.02,0.11)$ than the proportion of individuals randomized
at baseline that are always survivors (observed or censored) at one
month post treatment initiation whose cancer would progress on docetaxel
but would not progress on mitoxantrone. Our estimate of the number
of individuals randomized at baseline that are always survivors (observed
or censored) at one month post treatment initiation whose cancer would
not progress on docetaxel but would progress on mitoxantrone is greater
by $0.07\cdot674=47$ than the number of individuals randomized at
baseline that are always survivors (observed or censored) at one month
post treatment initiation whose cancer would progress on docetaxel
but would not progress on mitoxantrone. We do not find evidence for
such individual level always survivor causal effects from 6 months
onwards. At three months, statistical evidence (even after Bonferonni
correction for multiple testing with a threshold of $0.05/7=0.007$)
suggests that the proportion individuals randomized at baseline that
are always survivors three months post treatment initiation whose
cancer would not progress on docetaxel but would on mitoxantrone is
at least $0.15$ with a 95 percent confidence interval of $(0.08,0.23)$.
Note that this is a lower bound of such individuals, and with randomization
alone this proportion is not point identified. In terms of the $336+338=674$
individuals that were enrolled in this clinical trial, our estimate
is that there are at least $0.15\cdot674=81$ \emph{more} individuals
randomized at baseline who would survive at least until 3 months regardless
of which treatment arm they were randomized and their cancer would
progress under mitoxantrone but not docetaxel \emph{than} the individuals
randomized at baseline who would survive at at least until three months
regardless of which treatment arm they were randomized and their cancer
would progress under docetaxel but not mitoxantrone. 

The conclusions drawn in this data analysis could not be made with
any of the previous methods that used principal stratification to
deal with truncation by death \cite{ding2011identifiability,wang2017identification}.
Two papers \cite{ding2011identifiability,wang2017identification}
have used similar data and were unable to draw such conclusions \cite{ding2011identifiability,wang2017identification}.
Note also that these other papers, in contrast to ours, made more
assumptions than those guaranteed by randomization alone. We are able
to draw conclusions here even with weaker assumption because of the
focus, not on the survivor average causal effect, but rather detecting
and quantifying individuals who were always survivors for whom treatment
had a causal effect on cancer progression.

\subsection{Data analysis with monotonicity assumption $S_{1}^{[0]}(\omega,t)+S_{0}^{[1]}(\omega,t)\protect\leq1$}

For each $t\in\{1,\ldots,7\},$ we test the null hypothesis $P_{1,1.0}^{r(t)}\leq P_{1,1.1}^{r(t)}+P_{3,2.1}^{r(t)}$
to evaluate for a fixed $t$ whether the proportion of individuals
for whom $S_{1}(\omega,t)=S_{0}(\omega,t)=1,$ and $Y_{1}(\omega,t)=0,$
$Y_{0}(\omega,t)=1$ is greater than the proportion of individuals
for whom $S_{1}(\omega,t)=S_{0}(\omega,t)=1$ and $Y_{1}(\omega,t)=1,$
$Y_{0}(\omega,t)=0$ under the assumption that $S_{1}^{[0]}(\omega,t)+S_{0}^{[1]}(\omega,t)\leq1$
for all $\omega\in\Omega$ and $t\in T_{\text{SWOG}}.$ The following
table provides $95\%$ and $99\%$ confidence intervals for $P_{1,1.0}^{r(t)}-P_{1,1.1}^{r(t)}-P_{3,2.1}^{r(t)}$
for each $t\in T_{\text{SWOG}}$. Notice $P(S(t)=0\mid X=1)+P(S(t)=1\mid X=0)\leq1$
for all $t\in T_{\text{SWOG}}$ and therefore we fail to falsify the
monotonicity assumption, though of course this does not guarantee
that the assumption holds.

Two-sided confidence intervals of $P(Y(t)=1,S(t)=1\mid X=0)-P(\{Y(t)=1,S(t)=1\}\cup\{Y(t)=3,S(t)=2\}\mid X=1)$
and p-values for Wald test of the null hypothesis $P(Y(t)=1,S(t)=1\mid X=0)-P(\{Y(t)=1,S(t)=1\}\cup\{Y(t)=3,S(t)=2\}\mid X=1)\leq0$
are provided below:

\begin{table}[H]
\begin{centering}
\begin{tabular}{cccccccc}
\toprule 
 & 1 month & 2 months & 3 months & 4 months & 6 months & 12 months & 18 months\tabularnewline
\midrule
\midrule 
Estimate & $0.07$ & $0.16$ & $0.19$ & $0.17$ & $0.12$ & $-0.11$ & -0.10\tabularnewline
\midrule 
95\% CI & $(0.03,0.12)$ & $(0.10,0.23)$ & $(0.12,0.27)$ & $(0.10,0.24)$ & $(0.04,0.20)$ & $(-0.19,-0.03)$ & $(-0.18,-0.02)$\tabularnewline
\midrule 
99\% CI & $(0.02,0.13)$ & $(0.08,0.24)$ & $(0.10,0.29)$ & $(0.07,0.27)$ & $(0.02,0.22)$ & $(-0.21,-0.01)$ & $(-0.20,0.00)$\tabularnewline
\midrule 
p-value & $\text{0.0003}$ & $<0.0001$ & $<0.0001$ & $<0.0001$ & $0.0053$ & $0.9972$ & $0.9941$\tabularnewline
\bottomrule
\end{tabular} 
\par\end{centering}
\caption{Results Monotonicity $S_{1}^{[0]}(\omega,t)+S_{0}^{[1]}(\omega,t)\protect\leq1$
for all $\omega\in\Omega$}
\end{table}

For this section, we have assumed that $S_{1}^{[0]}(\omega,t)+S_{0}^{[1]}(\omega,t)\leq1$
for all $t\in T_{\text{SWOG}}.$ Notice with this assumption, we have
the same conclusions as before regarding always survivors between
one to four months, and now also have statistical evidence for individual
level always survivor causal effects at six months. 

By sensitivity analysis Theorem 3A, in order for the inference of
detecting always survivors six months post treatment initiation whose
cancer would not progress on docetaxel but would on mitoxantrone to
be invalid, we would need at least 12 percent of our study population
at baseline to follow the counterfactual response $S_{1}(\omega,t)=0,$
$S_{0}(\omega,t)=1,$ and $Y_{1}(\omega,t)=2,$ $Y_{0}(\omega,t)=0$
for $t=5,$ that is at least 12 percent of our population randomized
at baseline would need to die within six months if they were prescribed
docetaxel but would survive with a cancer that has not progressed
at least until six months post treatment initiation if prescribed
mitoxantrone. 

\subsection{Data analysis with monotonicity assumption $S_{1}^{[2]}(\omega,t)+S_{0}^{[1]}(\omega,t)\protect\leq1$ }

For each $t\in\{1,\ldots,7\},$ we test the null hypothesis $P_{1,1.0}^{r(t)}\leq P_{1,1.1}^{r(t)}+P_{2,0.1}^{r(t)}$
to evaluate for a fixed $t$ whether the proportion of individuals
for whom $S_{1}(\omega,t)=S_{0}(\omega,t)=1,$ $Y_{1}(\omega,t)=0,$
and $Y_{0}(t,\omega)=1$ is greater than the proportion of individuals
for whom $S_{1}(\omega,t)=S_{0}(\omega,t)=1,$ and $Y_{1}(\omega,t)=1,$
$Y_{0}(\omega,t)=0$ under the assumption that $S_{1}^{[2]}(\omega,t)+S_{0}^{[1]}(\omega,t)\leq1$
for all $\omega\in\Omega$ and $t\in T_{\text{SWOG}}.$ The following
table provides $95\%$ and $99\%$ confidence intervals for $P_{1,1.0}^{r(t)}-P_{1,1.1}^{r(t)}-P_{2,0.1}^{r(t)}$
for each $t\in T_{\text{SWOG}}$. Notice that $P(S(t)=2\mid X=1)+P(S(t)=1\mid X=0)\leq1$
for all $t\in T_{\text{SWOG}},$ though of course this does not guarantee
that the assumption holds.

Two-sided confidence intervals of $P(Y(t)=1,S(t)=1\mid X=0)-P(\{Y(t)=1,S(t)=1\}\cup\{Y(t)=2,S(t)=0\}\mid X=1)$
and one-sided p-values for Wald test of the null hypothesis $P(Y(t)=1,S(t)=1\mid X=0)-P(\{Y(t)=1,S(t)=1\}\cup\{Y(t)=2,S(t)=0\}\mid X=1)\leq0$
are provided below:

\begin{table}[H]
\begin{centering}
\begin{tabular}{cccccccc}
\toprule 
 & 1 month & 2 months & 3 months & 4 months & 6 months & 12 months & 18 months\tabularnewline
\midrule
\midrule 
Estimate & $0.09$ & $0.16$ & $0.18$ & $0.13$ & $0.02$ & $-0.34$ & $-0.48$\tabularnewline
\midrule 
95\% CI & $(0.05,0.13)$ & $(0.10,0.22)$ & $(0.11,0.25)$ & $(0.05,0.21)$ & $(-0.05,0.10)$ & $(-0.41,-0.27)$ & $(-0.54,0.41)$\tabularnewline
\midrule 
99\% CI & $(0.04,0.15)$ & $(0.08,0.24)$ & $(0.08,0.28)$ & $(0.03,0.23)$ & $(-0.08,0.13)$ & $(-0.43,-0.25)$ & $(-0.57,-0.39)$\tabularnewline
\midrule 
p-value & $<0.0001$ & $<0.0001$ & $<0.0001$ & $0.0003$ & 0.30 & 1 & 1\tabularnewline
\bottomrule
\end{tabular} 
\par\end{centering}
\caption{Results Monotonicity $S_{1}^{[2]}(\omega,t)+S_{0}^{[1]}(\omega,t)\protect\leq1$
for all $\omega\in\Omega$}
\end{table}

Under the assumption of monotonicity for censoring, we find evidence
for individual principal stratum causal effects up to 4 months but
not beyond.

\subsection{Data analysis with monotonicity assumptions $S_{1}^{[2]}(\omega,t)+S_{0}^{[1]}(\omega,t)\protect\leq1$
and $S_{1}^{[0]}(\omega,t)+S_{0}^{[1]}(\omega,t)\protect\leq1$ }

For each $t\in\{1,\ldots,7\},$ we test the null hypothesis $P_{1,1.0}^{r(t)}\leq P_{1,1.1}^{r(t)}$
to evaluate for a different fixed $t$ whether the proportion of individuals
for whom $Y_{1}(\omega,t)=0,$ $Y_{0}(t,\omega)=1,$ $S_{1}(t,\omega)=1,$
and $S_{0}(t,\omega)=1$ is greater than the proportion of individuals
for whom $Y_{1}(\omega,t)=1,$ $Y_{0}(\omega,t)=0,$ $S_{1}(\omega,t)=1,$
and $S_{0}(\omega,t)=1$ under the assumption that $S_{1}^{[2]}(\omega,t)+S_{0}^{[1]}(\omega,t)\leq1$
and $S_{1}^{[0]}(\omega,t)+S_{0}^{[1]}(\omega,t)\leq1$ for all $\omega\in\Omega$
and $t\in T_{\text{SWOG}}.$ The following table provides $95\%$
and $99\%$ confidence intervals for $P_{1,1.0}^{r(t)}-P_{1,1.1}^{r(t)}$
for each $t\in T_{\text{SWOG}}$. As before, notice $P(S(t)=0\mid X=1)+P(S(t)=1\mid X=0)\leq1$
and $P(S(t)=2\mid X=1)+P(S(t)=1\mid X=0)\leq1$ for all $t\in T_{\text{SWOG}}$
and therefore we fail to falsify the monotonicity assumption, though
of course this does not guarantee that the assumption holds.

Two-sided confidence intervals of $P(Y(t)=1,S(t)=1\mid X=0)-P(Y(t)=1,S(t)=1\mid X=1)$
and one-sided p-values for Wald test of the null hypothesis $P(Y(t)=1,S(t)=1\mid X=0)-P(Y(t)=1,S(t)=1\mid X=1)\leq0$
are provided below:

\begin{table}[H]
\begin{centering}
\begin{tabular}{cccccccc}
\toprule 
 & 1 month & 2 months & 3 months & 4 months & 6 months & 12 months & 18 months\tabularnewline
\midrule
\midrule 
Estimate & $0.10$ & $0.19$ & $0.22$ & 0.19 & 0.14 & $-0.08$ & -0.07\tabularnewline
\midrule 
95\% CI & $(0.06,0.14)$ & $(0.13,0.25)$ & $(0.15,0.29)$ & $(0.12,0.27)$ & $(0.07,0.22)$ & $(-0.16,0.00)$ & $(-0.14,0.01)$\tabularnewline
\midrule 
99\% CI & $(0.05,0.15)$ & $(0.11,0.27)$ & $(0.13,0.31)$ & $(0.10,0.29)$ & $(0.05,0.24)$ & $(-0.18,0.02)$ & $(-0.17,0.03)$\tabularnewline
\midrule 
p-value & $<0.0001$ & $<0.0001$ & $<0.0001$ & $<0.0001$ & $<0.0001$ & $0.978$ & $0.953$\tabularnewline
\bottomrule
\end{tabular} 
\par\end{centering}
\caption{Results Monotonicity $S_{1}^{[2]}(\omega,t)+S_{0}^{[1]}(\omega,t)\protect\leq1$
and $S_{1}^{[0]}(\omega,t)+S_{0}^{[1]}(\omega,t)\protect\leq1$ for
all $\omega\in\Omega$}

\end{table}

Under the assumption of monotonicity for censoring and survival, statistical
evidence suggests that the proportion of individuals randomized at
baseline that are always survivors three months post treatment initiation
whose cancer would not progress on docetaxel but would on mitoxantrone
is at least 0.22 with a 95 percent confidence interval of $(0.15,0.29)$.
Additionally, we again have statistical evidence for individual principal
stratum causal effects up to 6 months but not beyond. Sensitivity
analysis could also be applied to the results using Theorem 3C and
Corollary 3C.

\section{Generalizations and Extensions}

This section generalizes results to allow for outcomes in $\mathbf{R}^{n}$
and extends notation further. Let $T$ the range of the study period
from baseline $t=0$ to the end of study $t_{f}.$ Let $Y_{x}(\omega,t),$
and $S_{x}(\omega,t)$ denote the potential outcomes at time $t$
for individual $\omega,$ and correspondingly $Y(\omega,t)$ and $S(\omega,t)$
are the observed variables at time $t$ for individual $\omega.$
The variables $Y(\omega,t)$ and $Y_{x}(\omega,t)$ have support on
$\mathbf{R}^{n_{y}}$ where $n_{y}$ is the number of elements corresponding
to the tuple $Y(\omega,t)$. The potential outcomes $Y_{x}(\omega,t)$
and $S_{x}(\omega,t),$ and, correspondingly, the observed variables,
$Y(\omega,t)$ and $S(\omega,t),$ could be missing. Instead of denoting
a new level indicating that these observations are missing as presented
in the previous sections, we account for the missingness in a different
way that is presented below. The coarsest handling of missing data
will stem from using variables $R(\omega,t)$ that are defined as
follows:
\[
R(\omega,t)=\begin{cases}
1 & \text{no missingness in \ensuremath{Y} and \ensuremath{S} for unit \ensuremath{\omega} at time \ensuremath{t}}\\
0 & \text{missingness in tuple \ensuremath{Y} and \ensuremath{S} for unit \ensuremath{\omega} at time \ensuremath{t} }
\end{cases}
\]

For more refined analysis of censoring or missing data, consider the
following variables:

\[
RS(\omega,t)=\begin{cases}
1 & \text{no missingness in outcome \ensuremath{S} for unit \ensuremath{\omega}}\text{ at time t}\\
0 & \text{missingness if any element in outcome \ensuremath{S} for unit \ensuremath{\omega\text{ at time \ensuremath{t}},} }
\end{cases}
\]

and

\[
RY(\omega,t)=\begin{cases}
1 & \text{no missingness in tuples \ensuremath{Y} for unit \ensuremath{\omega}\text{ at time \ensuremath{t}}}\\
0 & \text{missingness if any element in tuples \ensuremath{Y} for unit \ensuremath{\omega}\text{ at time \ensuremath{t}} }
\end{cases}
\]
 Clearly, $R(\omega,t)=RS(\omega,t)\cdot RY(\omega,t).$ We will use
$(S_{x}(t),RY_{x}(t),RS_{x}(t))\in\{(s,a,b)\}$ to denote that $S_{x}(\omega,t)=s$,
$RY_{x}(\omega,t)=a,$ $RM_{x}(\omega,t)=b.$ Let $Y(\omega,t)=\star$
denote that the individual has died by time $t$ which is denoted
$S(\omega,t)=0.$ Similarly, for the potential outcomes if the individual
$\omega$ has died by time $t,$ then $S_{x}(\omega,t)=0$ and $Y_{x}(\omega,t)=\star.$
The random tuple $(Y_{1}(t),Y_{0}(t),S_{1}(t),S_{0}(t),RY_{1}(t),RY_{0}(t),RS_{1}(t),RS_{0}(t))$
is defined for each unit of the randomized experiment, and all the
results provided below retain the individual level interpretations
that are provided in the previous sections. We allow $(Y_{1}(t),Y_{0}(t),S_{1}(t),S_{0}(t),RY_{1}(t),RY_{0}(t),RS_{1}(t),RS_{0}(t))$
to have a distribution function $P(Y_{1}(t)\in y_{1},Y_{0}(t)\in y_{0},S_{1}(t)=s_{1},S_{0}(t)=s_{0},RY_{1}(t)=ry_{1},RY_{0}(t)=ry_{0},RS_{1}(t)=rs_{1},RS_{0}(t)=rs_{0}),$
where $y_{1}$ and $y_{0}$ are subsets of $\mathbf{R}^{n_{y}}$ (or
$y_{1}$ and $y_{0}$ are equal to $\star$ whenever $s_{1}=0$ or
$s_{0}=0$ respectively) and $s_{1},s_{0},ry_{1},ry_{0}$ all take
values in the set $\{0,1\}.$ An important consideration here is that
the distribution function $P(Y_{1}(t)\in y_{1},Y_{0}(t)\in y_{0},S_{1}(t)=s_{1},S_{0}(t)=s_{0},RY_{1}(t)=ry_{1},RY_{0}(t)=ry_{0},RS_{1}(t)=rs_{1},RS_{0}(t)=rs_{0})$
retains all the axiomatic properties of probability theory as provided
by Kolmogorov. In contrast, an causal estimand of the form $P\left(Y_{x}(\omega,t)\in y_{a}\right)$
that is restricted to some $y_{a}$ in $\mathbf{R}^{n_{y}}$ might
not be defined on $y_{a}$ and is undefined when $S_{x}(\omega,t)=0$
at time $t$. 

For our second data application, $S$ still denotes survival, but
a scientist could take $S$ to denote survival and any other outcome
of interest to get at a more refined principal strata. For example,
$S$ could denote whether the individual has survived is not HIV infected.
In this setting $S_{1}(\omega,t)=S_{0}(\omega,t)=1$ denotes that
the individual is an always uninfected, always survivor. Depending
on the scientific question under investigation, these more refined
individual level causal effects might be important. We will also look
at quantifying individual level causal effects at different time points
in the sensitivity analysis section provided below and further in
Online Supplement 2. For instance we will examine whether we can detect
an individual that remains alive at least until time $t_{U}$ with
treatment, $t_{L}$ with control, and will have the outcome in a range
$y_{a}$ at time $t_{U}$ (denoted $Y_{1}(\omega,t_{U})\in y_{a}$
,$S_{1}(\omega,t_{U})=1$) and will not have the outcome in a range
$y_{b}$ at time $t_{L}$ (denoted $Y_{1}(\omega,t_{L})\notin y_{b},S_{1}(\omega,t)=1$).
We will see that these type of causal questions are useful to understand
the life course of disease trajectory. 

\subsection*{Theorem Generalized Always Survivor Causal Effect}

Assume $X$ is randomized at baseline. For some $t\in T,$ $y_{a}\subset\mathbf{R}^{n_{y}},$
and $y_{b}\subset\mathbf{R}^{n_{y}},$ the contrast 
\begin{align*}
 & P\left(Y(t)\in y_{a},S(t)=1,R(t)=1\mid X=1\right)+P(Y(t)\not\in y_{b},S(t)=1,R(t)=1\mid X=0)-1
\end{align*}
 is a lower bound on 
\begin{align*}
 & P\left(Y_{1}(t)\in y_{a},Y_{0}(t)\not\in y_{b},S_{1}(t)=1,R_{1}(t)=1,S_{0}(t)=1,R_{0}(t)=1\right)\\
 & \hspace{1em}-\sum_{(r_{1},r_{0})\in\{0,1\}^{2}}P\left(Y_{1}(t)\not\in y_{a},Y_{0}(t)\in y_{b},S_{1}(t)=1,R_{1}(t)=r_{1},S_{0}(t)=1,R_{0}(t)=r_{0}\right),
\end{align*}
 and is therefore also a lower bound on 
\begin{align*}
 & P\left(Y_{1}(t)\in y_{a},Y_{0}(t)\not\in y_{b},S_{1}(t)=1,S_{0}(t)=1\right)\\
 & \hspace{1em}-P\left(Y_{1}(t)\not\in y_{a},Y_{0}(t)\in y_{b},S_{1}(t)=1,S_{0}(t)=1\right).
\end{align*}

For some $t\in T,$ $y_{a}\subset\mathbf{R}^{n_{y}}$ and $y_{b}\subset\mathbf{R}^{n_{y}},$
the expression
\[
\frac{P(Y(t)\in y_{a},S(t)=1,R(t)=1\mid X=1)+P(Y(t)\not\in y_{b},S(t)=1,R(t)=1\mid X=0)-1}{P\left(S(t)=1,R(t)=1\mid X=1\right)-P\left(R(t)=0\mid X=0\right)-P\left(S(t)=0,R(t)=1\mid X=0\right)}
\]
 is a lower bound on 
\[
\frac{\left(\begin{array}{c}
P\left(Y_{1}(t)\in y_{a},Y_{0}(t)\not\in y_{b},S_{1}(t)=1,R_{1}(t)=1,S_{0}(t)=1,R_{0}(t)=1\right)\\
-\sum_{(r_{1},r_{0})\in\{0,1\}^{2}}P\left(Y_{1}(t)\not\in y_{a},Y_{0}(t)\in y_{b},S_{1}(t)=1,R_{1}(t)=r_{1},S_{0}(t)=1,R_{0}(t)=r_{0}\right)
\end{array}\right)}{P(S_{1}(t)=1,S_{0}(t)=1,R_{1}(t)=1,R_{0}(t)=1)},
\]
and therefore is also a lower bound on
\begin{align*}
 & P\left(Y_{1}(t)\in y_{a},Y_{0}(t)\not\in y_{b}\mid S_{1}(t)=1,R_{1}(t)=1,S_{0}(t)=1,R_{0}(t)=1\right)\\
 & \hspace{1em}-P\left(Y_{1}(t)\not\in y_{a},Y_{0}(t)\in y_{b}\mid S_{1}(t)=1,R_{1}(t)=1,S_{0}(t)=1,R_{0}(t)=1\right),
\end{align*}

For some $t\in T,$ $y_{a}\subset\mathbf{R}^{n_{y}}$ and $y_{b}\subset\mathbf{R}^{n_{y}},$
when $P(Y(t)\in y_{a},S(t)=1,R(t)=1\mid X=1)+P(Y(t)\not\in y_{b},S(t)=1,R(t)=1\mid X=0)>1,$
the expression 
\[
\frac{P(Y(t)\in y_{a},S(t)=1,R(t)=1\mid X=1)+P(Y(t)\not\in y_{b},S(t)=1,R(t)=1\mid X=0)-1}{P\left(S(t)=1,R(t)=1\mid X=1\right)+P(R(t)=0\mid X=1)-P\left(S(t)=0,R(t)=1\mid X=0\right)}
\]
 is a lower bound on 
\[
\frac{\left(\begin{array}{c}
P\left(Y_{1}(t)\in y_{a},Y_{0}(t)\not\in y_{b},S_{1}(t)=1,S_{0}(t)=1\right)\\
-P\left(Y_{1}(t)\not\in y_{a},Y_{0}(t)\in y_{b},S_{1}(t)=1,S_{0}(t)=1\right)
\end{array}\right)}{P(S_{1}(t)=1,S_{0}(t)=1)},
\]

which is equivalent to 
\begin{align*}
 & P(Y_{1}(t)\in y_{a},Y_{0}(t)\notin y_{b}\mid S_{1}(t)=S_{0}(t)=1)\\
 & -P\left(Y_{1}(t)\notin y_{a},Y_{0}(t)\in y_{b}\mid S_{1}(t)=1,S_{0}(t)=1\right).
\end{align*}
 The result above is a generalization of Proposition 1 and Corollary
1. With baseline randomization, we enable statisticians to quantify
always survivor causal effects. In many instances, the scientist will
take $y_{a}=y_{b},$ but as we will see in our data analysis below,
scientists also might want to have $y_{a}\neq y_{b}$ to quantify
a specific form of individual level effect. As noted earlier, without
monotonicity assumptions, the empirical conditions to detect individual
level always survivor causal effects might be too stringent. In this
setting there are many different monotonicity conditions that a scientist
might be willing to make. We provide one result that makes the full
set of monotonicity assumptions. 

\subsection*{Theorem Generalized Always Survivor Causal Effect Monotonicity}

Assume $X$ is randomized at baseline. If we have the monotonicity
assumption at time $t\in T$ that there is no individual $\omega\in\Omega$
of response type $(S_{1}(t),R_{1}(t))\not\in\{(1,1)\}$ and $(S_{0}(t),R_{0}(t))\in\{(1,1)\}$
, then for some for some $t\in T,$ $y_{a}\subset\mathbf{R}^{n_{y}}$
and $y_{b}\subset\mathbf{R}^{n_{y}}$ contrast

\begin{align*}
 & P(Y(t)\not\in y_{b},(S(t),R(t))\in\{(1,1)\}\mid X=0)-P(Y(t)\not\in y_{a},(S(t),R(t))\in\{(1,1)\}\mid X=1)
\end{align*}

is a lower bound on 

\begin{align*}
 & P(Y_{1}(t)\in y_{a},Y_{0}(t)\not\in y_{b},(S_{1}(t),R_{1}(t))\in\{(1,1)\},(S_{0}(t),R_{0}(t))\in\{(1,1)\})\\
 & \hspace{1em}-P(Y_{1}(t)\not\in y_{a},Y_{0}(t)\in y_{b},(S_{1}(t),R_{1}(t))\in\{(1,1)\},(S_{0}(t),R_{0}(t))\in\{(1,1)\})\\
 & \hspace{1em}-P(Y_{1}(t)\not\in y_{a},Y_{0}(t)\in y_{b},(S_{1}(t),R_{1}(t))\in\{(1,1)\},(S_{0}(t),R_{0}(t))\in\{(1,0)\}).
\end{align*}

When there is no individual $\omega\in\Omega$ of response type $(S_{1}(\omega,t),R_{1}(\omega,t))\not\in\{(1,1)\}$
and $(S_{0}(\omega,t),R{}_{0}(\omega,t))\in\{(1,1)\}$, then the expression
\begin{align*}
\frac{P(Y(t)\not\in y_{b},(S(t),R(t))\in\{(1,1)\}\mid X=0)-P(Y(t)\not\in y_{a},(S(t),R(t))\in\{(1,1)\}\mid X=1)}{P(S(t)=1,R(t)=1\mid X=0)}
\end{align*}
 and also is a lower bound on

\begin{align*}
\frac{\left(\begin{array}{c}
P(Y_{1}(t)\in y_{a},Y_{0}(t)\not\in y_{b},S_{1}(t)=1,S_{0}(t)=1,R{}_{1}(t)=R_{0}(t)=1)\\
-P(Y_{1}(t)\not\in y_{a},Y_{0}(t)\in y_{b},S_{1}(t)=1,S_{0}(t)=1,R_{1}(t)=R{}_{0}(t)=1)
\end{array}\right)}{P(S_{1}(t)=1,S_{0}(t)=1,R{}_{1}(t)=R_{0}(t)=1)},
\end{align*}

which is equal to 
\[
\begin{array}{c}
\begin{array}{c}
P(Y_{1}(t)\in y_{a},Y_{0}(t)\not\in y_{b}\mid R_{1}(t)=R_{0}(t)=1,S_{1}(t)=S_{0}(t)=1)\\
-P(Y(t)\not\in y_{a},Y_{0}(t)\in y_{b}\mid R_{1}(t)=R_{0}(t)=1,S_{1}(t)=S_{0}(t)=1).
\end{array}\end{array}
\]

The results to detect and quantify individual level always survivor
causal effects under truncation by death with missing data under partial
monotonicity assumptions are provided in Online Supplement 2. All
estimands provided above can be estimated from the observed data the
associated parameters do not involve the joint of $R(t)=0$ and $S(t)$
or $Y(t)$. 

\subsection*{Generalized Sensitivity Analysis}

We now provided methods to conduct sensitivity analysis with missing
data. The sensitivity analysis provided herein differs from the sensitivity
analysis provided in Section 4 for monotonicity assumptions. The sensitivity
analysis in Section 4 is substantially easier to implement than the
sensitivity analysis provided in this section. However, the methods
provided in this section enable statisticians the ability to understand
how missing responses as well as monotonicity assumptions impact the
quantification of individual level always survivor causal effects.
Statisticians should use both sensitivity analysis methods to assess
the impact of violations of their assumptions on the resulting conclusions.
They should first apply the sensitivity analysis methods provided
in Section 4, because those methods are immediate. The following results
require a little more deliberation from both the statistician and
scientist, but such sensitivity analyses offers different insights
on what needs to be assumed from the observed data to quantify or
point identify $P(Y_{1}(t)\in y_{a},Y_{0}(t)\notin y_{b},S_{1}(t)=1,S_{0}(t)=1)-P(Y_{1}(t)\notin y_{a},Y_{0}(t)\in y_{b},S_{0}(t)=1,S_{0}(t)=1).$
In contrast, the simpler sensitivity analysis will not enable statisticians
to quantify or point identify $P(Y_{1}(t)\in y_{a},Y_{0}(t)\notin y_{b},S_{1}(t)=1,S_{0}(t)=1)-P(Y_{1}(t)\notin y_{a},Y_{0}(t)\in y_{b},S_{0}(t)=1,S_{0}(t)=1).$ 

\subsection*{Theorem Generalized Missing Data Sensitivity Analysis}

Suppose $X$ is randomized at baseline. For some $t\in T,$ $y_{a}\subset\mathbf{R}^{n_{y}}$
and $y_{b}\subset\mathbf{R}^{n_{y}}$ the expression

\begin{align*}
 & P(Y(t)\in y_{a},S(t)=1,R(t)=1\mid X=1)+P(Y(t)\notin y_{b},S(t)=1,R(t)=1\mid X=0)\\
 & +P(Y(t)\in y_{a},S(t)=1,R(t)=0\mid X=1)+P(Y(t)\notin y_{b},S(t)=1,R(t)=0\mid X=0)\\
 & -1
\end{align*}
is a lower bound on 

\begin{align*}
 & \sum_{(r_{1},r_{0})\in\{0,1\}^{2}}P(Y_{1}(t)\in y_{a},Y_{0}(t)\notin y_{b},S_{1}(t)=1,S_{0}(t)=1,R_{1}(t)=r_{1},R_{0}(t)=r_{0})\\
 & -\sum_{(r_{1},r_{0})\in\{0,1\}^{2}}P\left(Y_{1}(t)\notin y_{a},Y_{0}(t)\in y_{b},S_{1}(t)=1,S_{0}(t)=1,R_{1}(t)=r_{1},R_{0}(t)=r_{0}\right),
\end{align*}
 which is equivalent to 

\begin{align*}
 & P(Y_{1}(t)\in y_{a},Y_{0}(t)\notin y_{b},S_{1}(t)=1,S_{0}(t)=1)\\
 & -P\left(Y_{1}(t)\notin y_{a},Y_{0}(t)\in y_{b},S_{1}(t)=1,S_{0}(t)=1\right).
\end{align*}
 When 
\begin{align*}
 & P(Y(t)\in y_{a},S(t)=1,R(t)=1\mid X=1)+P(Y(t)\notin y_{b},S(t)=1,R(t)=1\mid X=0)\\
 & +P(Y(t)\in y_{a},S(t)=1,R(t)=0\mid X=1)+P(Y(t)\notin y_{b},S(t)=1,R(t)=0\mid X=0)\\
 & -1
\end{align*}
 is greater than equal to zero, the expression 
\[
\frac{\left(\begin{array}{c}
P(Y(t)\in y_{a},S(t)=1,R(t)=1\mid X=1)+P(Y(t)\notin y_{b},S(t)=1,R(t)=1\mid X=0)\\
+P(Y(t)\in y_{a},S(t)=1,R(t)=0\mid X=1)+P(Y(t)\notin y_{b},S(t)=1,R(t)=0\mid X=0)\\
-1
\end{array}\right)}{\left(\begin{array}{c}
P(S(t)=1,R(t)=1\mid X=1)+P(S(t)=1,R(t)=0\mid X=1)\\
-P\left(S(t)=0,R(t)=1\mid X=0\right)-P\left(S(t)=0,R(t)=0\mid X=0\right)
\end{array}\right)}
\]
 is a lower bound on 
\begin{align*}
 & P(Y_{1}(t)\in y_{a},Y_{0}(t)\notin y_{b}\mid S_{1}(t)=1,S_{0}(t)=1)\\
 & -P\left(Y_{1}(t)\notin y_{a},Y_{0}(t)\in y_{b}\mid S_{1}(t)=1,S_{0}(t)=1\right).
\end{align*}

The parameters $P(Y(t)\in y_{a},S(t)=1,R(t)=0\mid X=1),$ $P(Y(t)\notin y_{b},S(t)=1,R(t)=0\mid X=0)$
and $P\left(S(t)=0,R(t)=1\mid X=1\right)-P\left(S(t)=0,R(t)=0\mid X=0\right)$
cannot be estimated from the observed data, because when the data
is missing the realized values of $Y(t)$ and $S(t)$ are no longer
known at time $t.$ Therefore, the statistician will need to make
some assumption about the values of these parameters given the observed
data as well as subject matter knowledge about each individual's reason
for dropping out of the study. A second sensitivity analysis method
is provided below. For the second sensitivity analysis, we will have
two time points, $t_{U}$ indexing all $X=1$ (treatment variables)
and $t_{L}$ indexing all $X=0$ (control variables). In most applications,
we will take $t_{L}=t_{U}=t,$ but in one very important example below
$t_{L}$ and $t_{U}$ are distinct time points. If the statistician
is interested in always survivor causal effects at time $t$ and wishes
to use the sensitivity analysis approach from the next theorem, they
should replace all instances where $t_{L}$ and $t_{U}$ appear with
$t.$

\subsection*{Theorem Generalized Missing Data and Monotonicity Sensitivity Analysis}

Assume $X$ is randomized at baseline. For $y_{a}\subset\mathbf{R}^{n_{y}},$
$y_{b}\subset\mathbf{R}^{n_{y}},$ $t_{L}\in T$ and $t_{U}\in T,$
the expression
\begin{align*}
 & P(Y(t_{L})\notin y_{b},S(t_{L})=1,R(t_{L})=1\mid X=0)+P\left(Y(t_{L})\notin y_{b},S(t_{L})=1,R(t_{L})=0\mid X=0\right)\\
 & -P\left(Y(t_{U})\notin y_{a},S(t_{U})=1,R(t_{U})=1\mid X=1\right)-P\left(Y(t_{U})\notin y_{a},S(t_{U})=1,R(t_{U})=0\mid X=1\right)\\
 & -r(t_{L},t_{U}),
\end{align*}
 where
\begin{align*}
r(t_{L},t_{U}) & =P(Y_{1}(t_{U})\in\star,Y_{0}(t_{L})\notin y_{b},S_{1}(t_{U})=0,S_{0}(t_{L})=1)\\
 & \hspace{1em}-P\left(Y_{1}(t_{U})\notin y_{a},Y_{0}(t_{L})\in\star,S_{1}(t_{U})=1,S_{0}(t_{L})=0\right)
\end{align*}
is equal to 
\begin{align*}
 & \sum_{(r_{1},r_{0})\in\{0,1\}^{2}}P(Y_{1}(t_{U})\in y_{a},Y_{0}(t_{L})\notin y_{b},S_{1}(t_{U})=1,S_{0}(t_{L})=1,R_{1}(t_{U})=r_{1},R_{0}(t_{L})=r_{0})\\
 & -\sum_{(r_{1},r_{0})\in\{0,1\}^{2}}P\left(Y_{1}(t_{U})\notin y_{a},Y_{0}(t_{L})\in y_{b},S_{1}(t_{U})=1,S_{0}(t_{L})=1,R_{1}(t_{U})=r_{1},R_{0}(t_{L})=r_{0}\right)
\end{align*}
 or equivalently 
\begin{align*}
 & P(Y_{1}(t_{U})\in y_{a},Y_{0}(t_{L})\notin y_{b},S_{1}(t_{U})=1,S_{0}(t_{L})=1)\\
 & -P\left(Y_{1}(t_{U})\notin y_{a},Y_{0}(t_{L})\in y_{b},S_{1}(t_{U})=1,S_{0}(t_{L})=1\right).
\end{align*}

Consequently, the expression 
\[
\frac{\left(\begin{array}{c}
P(Y(t_{L})\notin y_{b},S(t_{L})=1,R(t_{L})=1\mid X=0)+P\left(Y(t_{L})\notin y_{b},S(t_{L})=1,R(t_{L})=0\mid X=0\right)\\
-P\left(Y(t_{U})\notin y_{a},S(t_{U})=1,R(t_{U})=1\mid X=1\right)-P\left(Y(t_{U})\notin y_{a},S(t_{U})=1,R(t_{U})=0\mid X=1\right)\\
-r(t_{L},t_{U}),
\end{array}\right)}{P(S(t_{L})=1,R(t_{L})=1\mid X=0)+P\left(S(t_{L})=1,R(t_{L})=0\mid X=0\right)-P\left(S_{1}(t_{U})\not=1,S_{0}(t_{L})=1\right)}
\]
 is equal to the normalized principal stratum direct effect 
\[
\frac{P(Y_{1}(t_{U})\in y_{a},Y_{0}(t_{L})\notin y_{b},S_{1}(t_{U})=1,S_{0}(t_{L})=1)-P(Y_{1}(t_{U})\notin y_{a},Y_{0}(t_{L})\in y_{b},S_{1}(t_{U})=1,S_{0}(t_{L})=1)}{P(S_{1}(t_{L})=1,S_{0}(t_{L})=1)}
\]
 which is equivalent to 
\[
P(Y_{1}(t_{U})\in y_{a},Y_{0}(t_{L})\notin y_{b}\mid S_{1}(t_{U})=1,S_{0}(t_{L})=1)-P(Y_{1}(t_{U})\notin y_{a},Y_{0}(t_{L})\in y_{b}\mid S_{1}(t_{U})=1,S_{0}(t_{L})=1).
\]

Sensitivity analysis parameters $r(t_{L},t_{U}),$ $P\left(S(t_{L})=1,R(t_{L})=0\mid X=0\right)$
and $P\left(S_{1}(t_{U})=1,S_{0}(t_{L})\neq1\right)$ cannot be estimated
from the observed data. However, researchers can deliberate on reasonable
values for these sensitivity analysis parameters to detect and quantify
individual level effects. Also, note $r(t_{L},t_{U})$ is equal $P(Y_{1}(t_{U})\in\star,Y_{0}(t_{L})\notin y_{b},S_{1}(t_{U})=0,S_{0}(t_{L})=1)-P\left(Y_{1}(t_{U})\notin y_{a},Y_{0}(t_{L})\in\star,S_{1}(t_{U})=1,S_{0}(t_{L})=0\right),$
and depending on the scientific question under investigation, these
sensitivity analysis parameters can be meaningfully intuited. The
advantage of such type of sensitivity analyses is that they are tied
intrinsically to the actual units, which enables statisticians to
choose these parameters carefully with conversations with the study
investigators. In the presence of truncation by death, statisticians
can fairly attempt to target 
\begin{align*}
 & \sum_{\{r_{1},r_{0}\}\in\{0,1\}^{2}}P(Y_{1}(t)\in y_{a},Y_{0}(t)\notin y_{b},S_{1}(t)=1,S_{0}(t)=1,R_{1}(t)=r_{1},R_{0}(t)=r_{0})\\
 & -\sum_{\{r_{1},r_{0}\}\in\{0,1\}^{2}}P\left(Y_{1}(t)\notin y_{a},Y_{0}(t)\in y_{b},S_{1}(t)=1,S_{0}(t)=1,R_{1}(t)=r_{1},R_{0}(t)=r_{0}\right),
\end{align*}
 and not $P(Y_{1}(t)\in y_{a},Y_{0}(t)\notin y_{b})-P(Y_{0}(t)\notin y_{a},Y_{0}(t)\in y_{b})$
as the counterfactuals $Y_{x}(\omega,t)$ are not meaningfully defined
on $\mathbf{R}^{n_{y}}$ for individuals for whom $S_{x}(\omega,t)=0.$
To identify the counterfactual risk difference $P(Y_{1}(t)\in y_{a},Y_{0}(t)\notin y_{b},S_{1}(t)=1,S_{0}(t)=1)-P\left(Y_{1}(t)\notin y_{a},Y_{0}(t)\in y_{b},S_{1}(t)=1,S_{0}(t)=1\right)$
all causal models are making assumptions of baseline randomization
and an implicit assumption regarding sensitivity analysis parameters
$c(t_{L}=t,t_{U}=t),$ $P\left(Y(t_{L})\notin y_{b},S(t_{L})=1,R(t_{L})=0\mid X=0\right)$
and $P\left(Y(t_{U})\notin y_{a},S(t_{U})=1,R(t_{U})=0\mid X=1\right).$ 

A Bayesian approach to using the above sensitivity analysis would
place a Dirichlet prior $\alpha$ on the Multinomial counts associated
on the joint $Z(t)=(Y(t),S(t),R(t)).$ Counts associated with $(Y(t),S(t),R(t)=0)$
are not observed, and the statistician can choose a range of suitable
values in collaboration with the scientific investigator. 

\section{Second Application to Southwest Oncology Group Trial}

Consider variable $Q,$ which is the quality of life index at 12 weeks.
Given that we have a single time point, we drop the indexing on $t,$
and, similarly, sensitivity analysis parameter $c(t_{L}=t,t_{U}=t)$
is measured at a single time point $t_{L}=t_{U}=t$ of 12 weeks. We
have the following contingency table

\begin{table}[H]
\begin{centering}
\begin{tabular}{cccccc}
\toprule 
 & $Q>70,S=1,R=1$ & $Q\leq70,S=1,R=1$ & $S=0$ & $R=0$ & Total\tabularnewline
\midrule
\midrule 
$X=1$ & 73 & 136 & 13 & $23+93=116$ & 338\tabularnewline
\midrule 
$X=0$ & 89 & 89 & 11 & $30+117=147$ & 336\tabularnewline
\midrule 
Total & 162 & 225 & 24 & 263 & 674\tabularnewline
\bottomrule
\end{tabular} 
\par\end{centering}
\caption{Quality of Life at 12 weeks}
\end{table}

Using Theorem Generalized Always Survivor Causal Effect Monotonicity,
to test the hypothesis that there is an individual that would have
survived regardless of treatment and their quality of life is below
$70$ units with docetaxel and above 70 units with mitoxantrone, the
scientist could examine the null hypothesis $P(Q>70,S=1,R=1\mid X=0)\leq P(DQ>70,S=1,R=1\mid X=1).$
The required monotonicity assumptions ($S_{1}(\omega)\geq S_{0}(\omega)$
and $R_{1}(\omega)\geq R_{0}(\omega)$ for all units in our population)
appear reasonable in this setting. Using Theorem Generalized Always
Survivor Causal Effect Monotonicity, rejecting $P(Q>70,S=1,R=1\mid X=0)\leq P(Q>70,S=1,R=1\mid X=1)$
tells the scientist that there must exist at least one individual
$\omega\in\Omega$ such that $Q_{1}(\omega)\leq70,$ $Q_{0}(\omega)>70,$
$S_{1}(\omega)=S_{0}(\omega)=R_{1}(\omega)=R_{0}(\omega)=1,$ and
the same theorem also tells us that $P(Q>70,S=1,R=1\mid X=0)\leq P(Q>70,S=1,R=1\mid X=1)$
is a lower bound on

\begin{align*}
 & P\left(Q_{1}(\omega)\leq70,Q_{0}(\omega)>70,S_{1}(\omega)=S_{0}(\omega)=1,R_{1}(\omega)=1,R_{0}(\omega)=1\right)\\
 & -P\left(Q_{1}(\omega)>70,Q_{0}(\omega)\leq70,S_{1}(\omega)=S_{0}(\omega)=1,R_{1}(\omega)=1,R_{0}(\omega)=1\right)\\
 & -P\left(Q_{1}(\omega)>70,Q_{0}(\omega)\leq70,S_{1}(\omega)=S_{0}(\omega)=1,R_{1}(\omega)=1,R_{0}(\omega)=0\right).
\end{align*}
 The risk difference $P(Q>70,S=1,R=1\mid X=0)-P(Q>70,S=1,R=1\mid X=1)$
is estimated as $\frac{89}{336}-\frac{73}{338}=0.05$ with a 95\%
confidence interval of $(-0.02,0.12).$ Assuming the monotonicity
assumptions hold true, the counterfactual interpretation of $P(Q>70,S=1,R=1\mid X=0)-P(Q>70,S=1,R=1\mid X=1)=0.05$
is that on average we should expect at least $674\cdot0.05=34$ more
individuals to survive regardless of treatment to twelve weeks and
their quality of life measurement is above 70 with mitoxantrone and
below 70 with docetaxel than the individuals who also survive regardless
of treatment to twelve weeks and whose quality of life is below 70
with mitoxantrone and above 70 with docetaxel. Using the second sensitivity
analysis in the Generalization and Extensions section, a statistician
might entertain values for $P(Q>70,S=1,R=0\mid X=0)-P(Q>70,S=1,R=0\mid X=1).$
Given that $P(Q>70,S=1,R=1\mid X=0)-P(Q>70,S=1,R=1\mid X=1)$ is positive,
a scientist might speculate that $P(Q>70,S=1,R=0\mid X=0)-P(Q>70,S=1,R=0\mid X=1)$
is also positive and similarly around 0.05. Given the results of the
clinical trial where it was established that docetaxel was protective
for survival in comparison to mitoxantrone, a statistician might reasonable
assume that $r(t_{L},t_{U})$ is negative (the number of failures
should be higher for the mitoxantrone arm in comparison to the docetaxel
arm). In such a setting, the counterfactual interpretation of 
\begin{align*}
 & P(Q>70,S=1,R=1\mid X=0)-P(Q>70,S=1,R=1\mid X=1)\\
 & +P(Q>70,S=1,R=0\mid X=0)-P(Q>70,S=1,R=0\mid X=1)\\
 & -r(t_{L}=t,t_{U}=t)>0.10
\end{align*}
 enables scientists to make the following conclusion: on average we
should expect at least 10\% of our population ($647\cdot0.10\approx65)$
randomized at baseline to survive regardless of treatment and their
quality of life is above 70 with mitoxantrone but below 70 with docetaxel.
While such sensitivity analysis require careful consideration, the
advantage of such analyses is that they are intrinsically tied to
the actual units of the statistical experiment, which enables scientists
and statisticians deliberate together to come to a consensus decision
regarding the results of their actual experiment. Quantifying that
10\% of the population displays such individual level effects is informative,
but it might not be that useful from a decision-theoretic or regulatory
perspective if all 65 of such individuals are of the counterfactual
response type $Q_{1}(\omega)=69,$ $Q_{0}(\omega)=71,$ $S_{1}(\omega)=S_{0}(\omega)=1.$
Instead, we might be interested in a meaningful change in quality
of life. 

In such a setting, we might be more interested to discover whether
there are a large number of individuals that follow the counterfactual
response type $Y_{1}(\omega)\leq70,$ $Y_{0}(\omega)>75,$ $S_{1}(\omega)=S_{0}(\omega)=R_{1}(\omega)=R_{0}(\omega)=1.$
To answer this scientific question, we would look at rejecting the
null hypothesis $P\left(Y>75,S=1,R=1\mid X=0\right)\leq P\left(Y>70,S=1,R=1\mid X=1\right)$.
Use the following contingency table, along the the previous contingency
table:

\begin{table}[H]
\begin{centering}
\begin{tabular}{cccccc}
\toprule 
 & $Q>75,S=1,R=1$ & $Q\leq75,S=1,R=1$ & $S=0$ & $R=0$ & Total\tabularnewline
\midrule
\midrule 
$X=1$ & 63 & 146 & 13 & $23+93=116$ & 338\tabularnewline
\midrule 
$X=0$ & 71 & 107 & 11 & $30+117=147$ & 336\tabularnewline
\midrule 
Total & 134 & 253 & 24 & 263 & 674\tabularnewline
\bottomrule
\end{tabular} 
\par\end{centering}
\caption{Quality of Life at 12 weeks}
\end{table}

The estimated risk difference $P\left(Y>75,S=1,R=1\mid X=0\right)-P\left(Y>70,S=1,R=1\mid X=1\right)=\frac{71}{336}-\frac{73}{388}$
is approximately zero with a 95 percent confidence interval of $(-0.07,0.06).$
We therefore find no statistical evidence for these counterfactual
response types of individuals in our population. Our analysis reveals
that scientists can deliberate on continuous outcomes, and decide
on meaningful hypothesis testing or estimation for informed decision
making. In cancer clinical trials where quality of life is an extremely
important consideration in end of life decisions, individuals might
be willing to sacrifice some gain in individual level life expectancy
for an improved quality of life. A difference of one point on a scale
from 0 to 100 might not be meaningful enough for the individual to
contemplate such a sacrifice, but if the difference is 10 points,
then the doctor and patient might well wish to deliberate. 

\section{Survival Average Causal Effects}

Online Supplement 1 provides a fully Bayesian approach using the Dirichlet-Multinomial
distribution to provide lower bounds on $P(Y_{1}(t)\in y_{a},Y_{0}(t)\not\in y_{b}\mid S_{1}(t)=S_{0}(t)=1)-P(Y_{1}(t)\not\in y_{a},Y_{0}(t)\in y_{b}\mid S_{1}(t)=S_{0}(t)=1).$
This approach is applied to the first and second data analysis. The
causal estimand $P(Y_{1}(t)\in y_{a},Y_{0}(t)\not\in y_{b}\mid S_{1}(t)=S_{0}(t)=1)-P(Y_{1}(t)\not\in y_{a},Y_{0}(t)\in y_{b}\mid S_{1}(t)=S_{0}(t)=1)$
is arguably less important than $P(Y_{1}(t)\in y_{a},Y_{0}(t)\not\in y_{b},S_{1}(t)=1,S_{0}(t)=1)-P(Y_{1}(t)\not\in y_{a},Y_{0}(t)\in y_{b},S_{1}(t)=1,S_{0}(t)=1)$
given that the more traditional survivor average causal effect will
never inform us how many people are in the always-survivor principal
stratum. In contrast, $P(Y_{1}(t)\in y_{a},Y_{0}(t)\not\in y_{b},S_{1}(t)=1,S_{0}(t)=1)-P(Y_{1}(t)\not\in y_{a},Y_{0}(t)\in y_{b},S_{1}(t)=1,S_{0}(t)=1)$
provides information on how many people remain in the always survivor
principal strata as well as the treatment effect on $Y(t).$ Our Theorems
demonstrated that bounds or senstivity analysis for the survivor average
causal effect is just rescaling of$P(Y_{1}(t)\in y_{a},Y_{0}(t)\not\in y_{b},S_{1}(t)=1,S_{0}(t)=1)-P(Y_{1}(t)\not\in y_{a},Y_{0}(t)\in y_{b},S_{1}(t)=1,S_{0}(t)=1).$ 

If scientists want life course trajectories, take  $Y_{x}(\omega,t),$
and $S_{x}(\omega,t)$ (also factuals $Y(\omega,t),$ and $S(\omega,t)$)
to be multivariate tuples each defined on $\mathbf{R}^{n_{v}\times n_{y}}\cup\{\star,\ldots,\star\}_{n_{v}\times n_{y}}$
and $\mathbf{R}^{n_{v}\times n_{s}}\cup\{\star,\ldots,\star\}_{n_{v}\times n_{s}}$
respectively, where $n_{v}$ denotes the number of time points of
interest, $n_{y}$ denotes the number of elements associated with
tuple $Y_{x}(\omega,t_{i})$ and $n_{s}$ denote the number of elements
associated with tuple $S_{x}(\omega,t_{i})$. The tuple denoted as
$\{\star,\ldots,\star\}_{n_{v}\times n}$ denotes a tuple with elements
$\star$ of length $n_{v}\times n$ that is used to notationally allow
for undefined variable status whenever needed. In such a setting,
use variables $Y_{x}(\omega,t)=(Y_{x}(\omega,t_{1}),\ldots,Y_{x}(\omega,t_{n_{v}}))$
and $S_{x}(\omega,t)=(S_{x}(\omega,t_{1}),\ldots,S_{x}(\omega,t_{n_{v}}))$
in all the Theorems below. Each $Y_{x}(\omega,t_{i})$ and $S_{x}(\omega,t_{i})$
is defined on $\mathbf{R}^{n_{y}}\cup\{\star,\ldots,\star\}_{n_{y}}$
and $\mathbf{R}^{n_{s}}\cup\{\star,\ldots,\star\}_{n_{s}}$ respectively.
The full generality of defining $Y_{x}(\omega,t)$ and $S_{x}(\omega,t)$
to be multivariate tuples each defined on $\mathbf{R}^{n_{v}\times n_{y}}\{\star,\ldots,\star\}_{n_{v}\times n_{y}}$
and $\mathbf{R}^{n_{v}\times n_{s}}\cup\{\star,\ldots,\star\}_{n_{v}\times n_{s}}$
respectively was not necessary for our applciations, but we provide
this notation in order or scientists to consider as refined scientific
questions they deem important.

\section{Conclusion}

This work demonstrates that testing and estimating individual causal
effects within `always survivor' principal stratum can be conducted
with simply the identification assumptions for a non-zero total effect.
The identifiability assumptions we use to identify these individual
level always survivor causal effects are no stronger than randomization.
Consequently, our results embeds testing for individual level always
survivor causal effects firmly within the Neyman-Pearson paradigm.
We also enable statistician to assess the existence of such individual
level always survivor causal effects using randomization based and
Bayesian inference as described in the Online Supplement 2. Previous
literature required stronger assumptions to identify or provide bounds
for survivor average causal effect, which is a different causal estimand
of interest \cite{cai2008bounds,imai2008sharp}. Our results are generally
applicable to clinical trials in the situation where an outcome of
interest is potentially not observed when death of the individual
occurs prior to the end of study. Our identifiability assumptions
are guaranteed by design in a randomized trial. 

Our proposed methodology provides a procedure to evaluate whether
a new treatment would delay cancer progression. The effect of a new
treatment on overall survival is a different scientific question than
whether new treatment delays cancer progression. Our data analysis
demonstrates the usefulness of our methods in this setting. In oncology
trials, determining treatment effectiveness sooner rather than later
can extend and improve patient lives as the progression to mortality
for many type III cancers can be rapid. Our methodology is the first
and, as of yet, only proposed procedure that enables detecting and
quantifying causal effects in the presence of truncation by death
and censoring using only the assumptions that are guaranteed by design
of the clinical trial. 

\section*{Acknowledgements}

The authors thank Linbo Wang and Peng Ding for providing them with
the actual data from the Southwest Oncology Group Trial that is used
in all the analysis presented in this article. This research was funded
by the National Institutes of Health. 

\section*{Online Supplement 1}

Let $I(\cdot)$ denote the usual indicator function. Define $Y_{x}^{[y]}(\omega,t)=I(Y_{x}(\omega,t)=y),$
and $S_{x}^{[s]}(\omega,t)=I(S_{x}(\omega,t)=s)$ for $y\in\{0,1,2,3\},$
$s\in\{0,1,2\},$ and $t\in T.$ Also note $Y_{x}^{[y]}(\omega,t)S_{x}^{[s]}(\omega,t)=I(Y_{x}(\omega,t)=y,S_{x}(\omega,t)=s)$
as a property of indicator functions. For ease of notation and space
considerations, we drop the $\omega$ in $Y(\omega,t)$, $S(\omega,t)$,
$Y_{x}(\omega,t)$, $S_{x}(\omega,t)$, whenever the meaning is clear.
The main paper and Online Supplement 1 assumes deterministic counterfactuals.
The Online Supplement 2 provides the same results with stochastic
counterfactuals and derives the same results. The Online Supplement
2 also provides results for the situation where there exists individuals
$\omega$ whose variable $S(\omega,t)$ observed at time $t,$ but
the outcome $Y(\omega,t)$ is censored. Table 8 presented at the end
of this paper is used for all our proofs. 

\subsection*{Proofs of identification of individual level always survivor causal
effects}

We first present the proofs of identification of individual level
always survivor causal effects.

\subsection*{Proof of Theorem 1}

We prove the contrapositive. Assume that no individual $\omega$ of
response type $Y_{1}(\omega,t)=y,$ $Y_{0}(\omega,t)=1-y,$ $S_{1}(\omega,t)=1$
and $S_{0}(\omega,t)=1$ exists in our population for a fixed $t\in T$
and $y\in\{0,1\}.$ Then for all individuals $\omega$ in our population
$\Omega,$ $I(Y_{1}(\omega,t)=y,S_{1}(\omega,t)=1)+I(Y_{0}(\omega,t)=1-y,S_{0}(\omega,t)=1)\leq1.$
Taking expectations, 

\begin{eqnarray*}
P(Y_{1}(\omega,t)=y,S_{1}(\omega,t)=1)+P(Y_{0}(\omega,t)=1-y,S_{0}(\omega,t)=1) & \leq & 1\iff\\
P(Y_{1}(\omega,t)=y,S_{1}(\omega,t)=1\mid X=0)+P(Y_{0}(\omega,t)=1-y,S_{0}(\omega,t)=1\mid X=1) & \leq & 1\iff\\
P(Y(t)=y,S(t)=1\mid X=0)-P(Y(t)=1-y,S(t)=1\mid X=1) & \leq & 1.
\end{eqnarray*}
 The first to second line follows from $(Y_{x}(t),S_{x}(t))\amalg X$
for all $t$ and the second to third line follows from consistency
of counterfactuals. The same proof applies for any arbitrary $t\in T.$

\subsection*{Proof of Proposition 1}

For any fixed $t\in T,$ taking expectation of $Y_{1}^{[1]}(\omega,t)S_{1}^{[1]}(\omega,t)+Y_{0}^{[0]}(\omega,t)S_{0}^{[1]}(\omega,t)-1$
and $Y_{1}^{[0]}(\omega,t)S_{1}^{[1]}(\omega,t)+Y_{0}^{[1]}(\omega,t)S_{1}^{[1]}(\omega,t)-1$
gives the result. A table with the relevant frequencies of the counterfactuals
is provided below in Table 8 (see also Tables 1.3s in Online Supplement
2 to avoid computation). The same proof applies for any arbitrary
$t\in T.$

\subsection*{Proof of Corollary 1}

For any fixed $t\in T,$ consider from arguments presented in Theorem
1 and Proposition 1 (see Tables 1.3s and 1.4s in Online Supplement
2),
\begin{align*}
P_{y,1.1}^{r(t)}+P_{1-y,1.0}^{r(t)}-1 & =E\left[I(Y_{1}(\omega,t)=y,S_{1}(\omega,t)=1)+I(Y_{0}(\omega,t)=1-y,S_{0}(\omega,t)=1)-1\right]\\
 & =P^{c(t)}(y,1-y,1,1)-P^{c(t)}(1-y,y,1,1)-P^{c(t)}(2,y,0,1)\\
 & \hspace{1em}-P^{c(t)}(1-y,2,1,0)-P^{c(t)}(2,2,0,0)-P^{c(t)}(3,3,2,2)\\
 & \hspace{1em}-P^{c(t)}(3,y,2,1)-P^{c(t)}(3,2,2,0)-P^{c(t)}(1-y,3,1,2)\\
 & \hspace{1em}-P^{c(t)}(2,3,0,2).\\
\end{align*}
 Therefore, 
\begin{align*}
 & P^{c(t)}(y,1-y,1,1)-P^{c(t)}(1-y,y,1,1)-P^{c(t)}(3,y,2,1)\\
 & \hspace{1em}-P^{c(t)}(1-y,3,1,2)-P^{c(t)}(3,3,2,2)\\
 & =\\
 & \hspace{1em}P_{y,1.1}^{r(t)}+P_{1-y,1.0}^{r(t)}-1\\
 & \hspace{1em}+P^{c(t)}(2,y,0,1)+P^{c(t)}(1-y,2,1,0)\\
 & \hspace{1em}+P^{c(t)}(2,2,0,0)\\
 & \hspace{1em}+P^{c(t)}(3,2,2,0)+P^{c(t)}(2,3,0,2).
\end{align*}
Consequently, 
\begin{align*}
P_{y,1.1}^{r(t)}+P_{1-y,1.0}^{r(t)}-1 & <P^{c(t)}(y,1-y,1,1)-P^{c(t)}(1-y,y,1,1)\\
 & \hspace{1em}-P^{c(t)}(3,y,2,1)-P^{c(t)}(1-y,3,1,2)\\
 & \hspace{1em}-P^{c(t)}(3,3,2,2).
\end{align*}
This demonstrates $P_{1,1.1}^{r(t)}+P_{0,1.0}^{r(t)}-1$ serves as
a lower bound for the counterfactual contrast 
\begin{align*}
P^{c(t)}(1,0,1,1)-\left[P^{c(t)}(0,1,1,1)+P^{c(t)}(3,1,2,1)+P^{c(t)}(3,3,2,2)+P^{c(t)}(0,3,1,2)\right].
\end{align*}
 The first term in this counterfactual contrast, $P^{c(t)}(1,0,1,1),$
is a lower bound of the proportion of individuals randomized at baseline
that are always survivors (observed or censored) and for whom the
treatment causes the outcome, because it does not include any censored
individuals that are also always survivors and for whom the treatment
causes the outcome. The term in the square brackets is an upper bound
on the proportion of individual randomized at baseline that are always
survivors (observed or censored) for whom the treatment prevents the
outcome, because it effectively treats any censored individual that
could possibly be an always survivor and for whom the treatment prevents
the outcome is actually an always survivor and for whom the treatment
prevents the outcome. To see that $P_{1,1.1}^{r(t)}+P_{0,1.0}^{r(t)}-1$
serves as a lower bound to $P^{c(t)}(1,0,1,1),$ note 
\begin{align*}
P_{1,1.1}^{r(t)}+P_{0,1.0}^{r(t)}-1 & \leq P^{c(t)}(1,0,1,1)-\left[P^{c(t)}(0,1,1,1)+P^{c(t)}(3,1,2,1)+P^{c(t)}(3,3,2,2)+P^{c(t)}(0,3,1,2)\right]\implies\\
P_{1,1.1}^{r(t)}+P_{0,1.0}^{r(t)}-1 & \leq P^{c(t)}(1,0,1,1).
\end{align*}
 The proofs of the other bounds are similar. The same proof applies
for any arbitrary $t\in T.$ This completes the proof.

\subsection*{Proof of Theorem 2A}

We prove the contrapositive. For a fixed $t\in T,$ assume that no
individual $\omega$ of response type $Y_{1}(\omega,t)=1,$ $Y_{0}(\omega,t)=0,$
$S_{1}(\omega,t)=1$ and $S_{0}(\omega,t)=1$ exists in our population.
Then for all individuals $\omega$ in our population $\Omega,$ $I(Y_{0}(\omega,t)=0,S_{0}(\omega,t)=1)-I(Y_{1}(\omega,t)=0,S_{1}(\omega,t)=1)-I(Y_{1}(\omega,t)=3,S_{1}(\omega,t)=2)\leq0.$
This last assertion is true after examining the counterfactual table
provided in tables 2.1s-2.3s in Online Supplement 2. Taking expectations,
we have
\begin{eqnarray*}
P(Y_{0}(\omega,t)=0,S_{0}(\omega,t)=1)-P(Y_{1}(\omega,t)=0,S_{1}(\omega,t)=1)-P(Y_{1}(\omega,t)=3,S_{1}(\omega,t)=2) & \leq & 0\iff\\
P(Y_{0}(t)=0,S_{0}(t)=1\mid X=0)-P(Y_{1}(t)=0,S_{1}(t)=1\mid X=1)-P(Y_{1}(t)=3,S_{1}(t)=2\mid X=1) & \leq & 0\iff\\
P(Y(t)=0,S(t)=1\mid X=0)-P(Y(t)=0,S(t)=1\mid X=1)-P(Y(t)=3,S(t)=2\mid X=1) & \leq & 0.
\end{eqnarray*}
 The first to second line follows from $(Y_{x}(t),S_{x}(t))\amalg X$
and the second to third line follows from consistency of counterfactuals.
The same proof applies for any arbitrary $t\in T.$ The other inequality
associated with Theorem 2A can be derived similarly. This completes
the proof.

The proofs of Theorem 2B and 2C can be derived similarly. 

\subsection*{Proof of Proposition 2A}

For a fixed integer $t\in T,$ taking expectation of $Y_{0}^{[1]}(\omega,t)S_{0}^{[1]}(\omega,t)-Y_{1}^{[1]}(\omega,t)S_{1}^{[1]}(\omega,t)-Y_{1}^{[3]}(\omega,t)S_{1}^{[2]}(\omega,t)$
and $Y_{0}^{[0]}(\omega,t)S_{0}^{[1]}(\omega,t)-Y_{1}^{[0]}(\omega,t)S_{1}^{[1]}(\omega,t)-Y_{1}^{[3]}(\omega,t)S_{1}^{[2]}(\omega,t)$
gives the result. A table with the relevant frequencies of the counterfactuals
is provided below. In Table 8, set any probability that does not satisfy
the monotonicity assumption $S_{1}^{[1]}(\omega,t)+S_{0}^{[0]}(\omega,t)\leq1$
to zero. The same proof applies for any arbitrary $t\in T.$ This
completes the proof.

The proofs of Proposition 2B and 2C can be derived similarly. 

\subsection*{Proof of Corollary 2A}

For a fixed $t\in T,$ from Table 8 (see also Table 2.3s in Online
Supplement 2 to avoid computation), baseline randomization and consistency
of counterfactuals, we have
\begin{align*}
 & P_{1-y,1.0}^{r(t)}-P_{1-y,1.1}^{r(t)}-P_{3,2.1}^{r(t)}\\
 & =E\left[I(Y_{0}(\omega,t)=1-y,S_{0}(\omega,t)=1)-I(Y_{1}(\omega,t)=1-y,S_{1}(\omega,t)=1)-I(Y_{1}(\omega,t)=3,S_{1}(\omega,t)=2)\right]\\
 & =P^{c(t)}(y,1-y,1,1)-P^{c(t)}(1-y,y,1,1)\\
 & \hspace{1em}-P^{c(t)}(1-y,2,1,0)-P^{c(t)}(3,3,2,2)\\
 & \hspace{1em}-P^{c(t)}(3,y,2,1)-P^{c(t)}(3,2,2,0)-P^{c(t)}(1-y,3,1,2)
\end{align*}
 Therefore, 
\begin{align*}
 & P^{c(t)}(y,1-y,1,1)-P^{c(t)}(1-y,y,1,1)\\
 & \hspace{1em}-P^{c(t)}(3,y,2,1)-P^{c(t)}(1-y,3,1,2)+P^{c(t)}(3,3,2,2)\\
 & =P_{1-y,1.0}^{r(t)}-P_{1-y,1.1}^{r(t)}-P_{3,2.1}^{r(t)}\\
 & \hspace{1em}+P^{c(t)}(1-y,2,1,0)+P^{c(t)}(3,2,2,0).
\end{align*}
Consequently, 
\begin{align*}
P_{1-y,1.0}^{r(t)}-P_{1-y,1.1}^{r(t)}-P_{3,2.1}^{r(t)} & <P^{c(t)}(y,1-y,1,1)-P^{c(t)}(1-y,y,1,1)\\
 & \hspace{1em}-P^{c(t)}(3,y,2,1)-P^{c(t)}(1-y,3,1,2)\\
 & \hspace{1em}-P^{c(t)}(3,3,2,2).
\end{align*}
This demonstrates $P_{0,1.0}^{r(t)}-P_{0,1.1}^{r(t)}-P_{3,2.1}^{r(t)}$
serves as a lower bound for the counterfactual contrast 
\begin{align*}
P^{c(t)}(1,0,1,1)-\left[P^{c(t)}(0,1,1,1)+P^{c(t)}(3,1,2,1)+P^{c(t)}(3,3,2,2)+P^{c(t)}(0,3,1,2)\right].
\end{align*}
 The first term in this counterfactual contrast, $P^{c(t)}(1,0,1,1),$
is a lower bound of the proportion of individuals that are always
survivors (observed or censored) and for whom the treatment causes
the outcome, because it does not include any censored individuals
that are also always survivors and for whom the treatment causes the
outcome. The term in the square brackets is an upper bound on the
always survivors (observed or censored) for whom the treatment prevents
the outcome, because it effectively treats any censored individual
that could possibly be an always survivor and for whom the treatment
prevents the outcome is actually an always survivor and for whom the
treatment prevents the outcome. To see that $P_{0,1.0}^{r(t)}-P_{0,1.1}^{r(t)}-P_{3,2.1}^{r(t)}$
serves as a lower bound to $P^{c(t)}(1,0,1,1),$ note 
\begin{align*}
P_{0,1.0}^{r(t)}-P_{0,1.1}^{r(t)}-P_{3,2.1}^{r(t)} & \leq P^{c(t)}(1,0,1,1)-\left[P^{c(t)}(0,1,1,1)+P^{c(t)}(3,1,2,1)+P^{c(t)}(3,3,2,2)+P^{c(t)}(0,3,1,2)\right]\implies\\
P_{0,1.0}^{r(t)}-P_{0,1.1}^{r(t)}-P_{3,2.1}^{r(t)} & \leq P^{c(t)}(1,0,1,1).
\end{align*}
 The proofs of the other bounds are similar. The same proof applies
for any arbitrary $t\in T.$ This completes the proof. 

The proofs of Corollary 2B and 2C can be derived similarly. 

\subsection*{Proofs involving sensitivity analysis}

We now present the proofs for the sensitivity analysis of monotonicity
assumptions.

\subsection*{Proof of Theorem 3A}

For a fixed $t\in T,$ from Table 8 (see also Table 2.3s in Online
Supplement 2 to avoid computation), we have under consistency of counterfactuals
and baseline randomization,
\begin{align*}
 & P_{1-y,1.0}^{r(t)}-P_{1-y,1.1}^{r(t)}-P_{3,2.1}^{r(t)}\\
 & =E\left[I(Y_{0}(\omega,t)=1-y,S_{0}(\omega,t)=1)-I(Y_{1}(\omega,t)=1-y,S_{1}(\omega,t)=1)-I(Y_{1}(\omega,t)=3,S_{1}(\omega,t)=2)\right]\\
 & =P^{c(t)}(y,1-y,1,1)-P^{c(t)}(1-y,y,1,1)\\
 & \hspace{1em}-P^{c(t)}(1-y,2,1,0)-P^{c(t)}(3,3,2,2)\\
 & \hspace{1em}-P^{c(t)}(3,y,2,1)-P^{c(t)}(3,2,2,0)\\
 & \hspace{1em}-P^{c(t)}(1-y,3,1,2)+P^{c(t)}(2,1-y,0,1).
\end{align*}

Therefore, 
\begin{align*}
 & P^{c(t)}(y,1-y,1,1)-P^{c(t)}(1-y,y,1,1)\\
 & \hspace{1em}-P^{c(t)}(3,y,2,1)-P^{c(t)}(1-y,3,1,2)+P^{c(t)}(3,3,2,2)\\
 & =P_{1-y,1.0}^{r(t)}-P_{1-y,1.1}^{r(t)}-P_{3,2.1}^{r(t)}\\
 & \hspace{1em}+P^{c(t)}(1-y,2,1,0)+P^{c(t)}(3,2,2,0)-P^{c(t)}(2,1-y,0,1).
\end{align*}

Take $d_{m}(t)=P^{c(t)}(2,1-y,0,1)-P^{c(t)}(1-y,2,1,0)-P^{c(t)}(3,2,2,0),$
then, we have 
\begin{align*}
 & P_{1-y,1.0}^{r(t)}-P_{1-y,1.1}^{r(t)}-P_{3,2.1}-d_{m}(t)\\
 & =P^{c(t)}(y,1-y,1,1)-P^{c(t)}(1-y,y,1,1)\\
 & \hspace{1em}-P^{c(t)}(3,y,2,1)-P^{c(t)}(1-y,3,1,2)+P^{c(t)}(3,3,2,2).
\end{align*}

Consequently, if 
\[
P_{1-y,1.0}^{r(t)}-P_{1-y,1.1}^{r(t)}-P_{3,2.1}^{r(t)}>d_{m}(t).
\]
 where 
\[
d_{m}(t)=P^{c(t)}(2,1-y,0,1)-P^{c(t)}(1-y,2,1,0)-P^{c(t)}(3,2,2,0),
\]
 then, 
\begin{align*}
 & P^{c(t)}(y,1-y,1,1)-P^{c(t)}(1-y,y,1,1)\\
 & \hspace{1em}-P^{c(t)}(3,y,2,1)-P^{c(t)}(1-y,3,1,2)-P^{c(t)}(3,3,2,2)\\
 & >0
\end{align*}
The same proof applies for any arbitrary $t\in T.$ This completes
the proof.

An alternative proof would examine 
\begin{align*}
 & E\left[I(Y_{0}(\omega,t)=1-y,S_{0}(\omega,t)=1)-I(Y_{1}(\omega,t)=1-y,S_{1}(\omega,t)=1)-I(Y_{1}(\omega,t)=3,S_{1}(\omega,t)=2)\right]\\
 & =P(Y_{0}(\omega,t)=1-y,S_{0}(\omega,t)=1)-P(Y_{1}(\omega,t)=1-y,S_{1}(\omega,t)=1)-P(Y_{1}(\omega,t)=3,S_{1}(\omega,t)=2),
\end{align*}
 and then expand $P(Y_{0}(\omega,t)=1-y,S_{0}(\omega,t)=1),$ $P(Y_{1}(\omega,t)=1-y,S_{1}(\omega,t)=1),$
and $P(Y_{1}(\omega,t)=3,S_{1}(\omega,t)=2)$ over the remaining counterfactual
variables. For example, 
\begin{align*}
P(Y_{0}(\omega,t) & =1-y,S_{0}(\omega,t)=1)=\sum_{(y_{1},s_{1})\in\{(Y_{1},S_{1})\}}P(Y_{1}(\omega,t)=y_{1},Y_{0}(\omega,t)=1-y,S_{1}(\omega,t)=s_{1},S_{0}(\omega,t)=1).
\end{align*}
 Here, $\{(Y_{1},S_{1})\}$ denotes the joint compatible state space
of the joint variables $Y_{1}(\omega,t)$ and $S_{1}(\omega,t).$
Explicitly, in our context, $\{(Y_{1},S_{1})\}=\{(0,1),(1,1),(2,0),(3,2)\}$
is the state of compatible realizations of the joint counterfactual
random variable $(Y_{1},S_{1}).$ Section 5 and 8 of Online Supplement
2 provides this proof using stochastic counterfactuals. The proofs
of Theorems 3B and 3C are provided in Section 12 of Online Supplement
2.

\subsection*{Proof of Theorem Generalized Always Survivor Causal Effect}

Begin with the equality 
\begin{align*}
 & P\left(Y(t)\in y_{a},S(t)=1,R(t)=1\mid X=1\right)+P(Y(t)\not\in y_{b},S(t)=1,R(t)=1\mid X=0)-1\\
 & =P\left(Y(t)\in y_{a},S(t)\cdot R(t)=1\mid X=1\right)+P(Y(t)\not\in y_{b},S(t)\cdot R(t)=1\mid X=0)-1\\
 & =P\left(Y(t)\in y_{a},S(t)\cdot R(t)=1\mid X=1\right)-P\left(Y(t)\in y_{b}\text{ or }S(t)\cdot R(t)=0\mid X=0\right)\\
 & =P\left(Y(t)\in y_{a},S(t)\cdot R(t)=1\mid X=1\right)-P\left(Y(t)\in y_{b},S(t)\cdot R(t)=1\mid X=0\right)\\
 & \hspace{1em}-P\left(Y(t)\not\in y_{a},S(t)=1,R(t)=0\mid X=0\right)-P\left(Y(t)\in y_{b},S(t)=1,R(t)=0\mid X=0\right)\\
 & \hspace{1em}-P\left(Y\in\star,S(t)=0,R(t)=1\mid X=0\right)-P\left(Y(t)\in\star,S(t)=0,R(t)=0\mid X=0\right)
\end{align*}
 Under randomization and consistency of counterfactuals, the last
expression in the above equality is equal to 
\begin{align*}
 & P\left(Y_{1}(t)\in y_{a},S_{1}(t)\cdot R_{1}(t)=1\right)-P\left(Y_{0}(t)\in y_{b},S_{0}(t)\cdot R_{0}(t)=1\right)\\
 & \hspace{1em}-P\left(Y_{0}(t)\not\in y_{b},S_{0}(t)=1,R_{0}(t)=0\right)-P\left(Y_{0}(t)\in y_{b},S_{0}(t)=1,R_{0}(t)=0\right)\\
 & \hspace{1em}-P\left(Y_{0}(t)\in\star,S_{0}(t)=0,R_{0}(t)=1\right)-P\left(Y_{0}(t)\in\star,S_{0}(t)=0,R_{0}(t)=0\right)
\end{align*}
Now, use the law of total probability $P(A)=P(A,B)+P(A,B^{c}),$ where
$B^{c}$ denotes the complement of $B.$ Consequently, this last expression
is equal to
\begin{align*}
 & P\left(Y_{1}(t)\in y_{a},S_{1}(t)\cdot R_{1}(t)=1,S_{0}(t)\cdot R_{0}(t)=1\right)+P\left(Y_{1}(t)\in y_{a},S_{1}(t)\cdot R_{1}(t)=1,S_{0}(t)\cdot R_{0}(t)=0\right)\\
 & \hspace{1em}-P\left(Y_{0}(t)\in y_{b},S_{1}(t)R_{1}(t)=1,S_{0}(t)\cdot R_{0}(t)=1\right)-P\left(Y_{0}(t)\in y_{b},S_{1}(t)R_{1}(t)=0,S_{0}(t)\cdot R_{0}(t)=1\right)\\
 & \hspace{1em}-P\left(Y_{0}(t)\not\in y_{b},S_{1}(t)\cdot R_{1}(t)=1,S_{0}(t)=1,R_{0}(t)=0\right)-P\left(Y_{0}(t)\not\in y_{b},S_{1}(t)\cdot R_{1}(t)=0,S_{0}(t)=1,R_{0}(t)=0\right)\\
 & \hspace{1em}-P\left(Y_{0}(t)\in y_{b},S_{1}(t)\cdot R_{1}(t)=1,S_{0}(t)=1,R_{0}(t)=0\right)-P\left(Y_{0}(t)\in y_{b},S_{1}(t)\cdot R_{1}(t)=0,S_{0}(t)=1,R_{0}(t)=0\right)\\
 & \hspace{1em}-P\left(Y_{0}(t)\in\star,S_{1}(t)\cdot R_{1}(t)=1,S_{0}(t)=0,R_{0}(t)=1\right)-P\left(Y_{0}(t)\in\star,S_{1}(t)\cdot R_{1}(t)=0,S_{0}(t)=0,R_{0}(t)=1\right)\\
 & \hspace{1em}-P\left(Y_{0}(t)\in\star,S_{1}(t)\cdot R_{1}(t)=1,S_{0}(t)=0,R_{0}(t)=0\right)-P\left(Y_{0}(t)\in\star,S_{1}(t)\cdot R_{1}(t)=0,S_{0}(t)=0,R_{0}(t)=0\right)
\end{align*}
 Again, use the law of total probability $P(A)=P(A,B)+P(A,B^{c}),$
where $B^{c}$ denotes the complement of $B.$ Consequently, this
last expression  is equal to 
\begin{align*}
 & P\left(Y_{1}(t)\in y_{a},Y_{0}(t)\in y_{b},S_{1}(t)\cdot R_{1}(t)=1,S_{0}(t)\cdot R_{0}(t)=1\right)\\
 & +P\left(Y_{1}(t)\in y_{a},Y_{0}(t)\not\in y_{b},S_{1}(t)\cdot R_{1}(t)=1,S_{0}(t)\cdot R_{0}(t)=1\right)\\
 & +P\left(Y_{1}(t)\in y_{a},Y_{0}(t)\in y_{b},S_{1}(t)\cdot R_{1}(t)=1,S_{0}(t)=1,R_{0}(t)=0\right)\\
 & +P\left(Y_{1}(t)\in y_{a},Y_{0}(t)\not\in y_{b},S_{1}(t)\cdot R_{1}(t)=1,S_{0}(t)=1,R_{0}(t)=0\right)\\
 & +P\left(Y_{1}(t)\in y_{a},Y_{0}(t)\in\star,S_{1}(t)\cdot R_{1}(t)=1,S_{0}(t)=0,R_{0}(t)=1\right)\\
 & +P\left(Y_{1}(t)\in y_{a},Y_{0}(t)\in\star,S_{1}(t)\cdot R_{1}(t)=1,S_{0}(t)=0,R_{0}(t)=0\right)\\
 & \hspace{1em}-P\left(Y_{1}(t)\in y_{a},Y_{0}(t)\in y_{b},S_{1}(t)R_{1}(t)=1,S_{0}(t)\cdot R_{0}(t)=1\right)\\
 & \hspace{1em}-P\left(Y_{1}(t)\not\in y_{a},Y_{0}(t)\in y_{b},S_{1}(t)R_{1}(t)=1,S_{0}(t)\cdot R_{0}(t)=1\right)\\
 & \hspace{1em}-P\left(Y_{1}(t)\in y_{a},Y_{0}(t)\in y_{b},S_{1}(t)=1,R_{1}(t)=0,S_{0}(t)\cdot R_{0}(t)=1\right)\\
 & \hspace{1em}-P\left(Y_{1}(t)\not\in y_{a},Y_{0}(t)\in y_{b},S_{1}(t)=1,R_{1}(t)=0,S_{0}(t)\cdot R_{0}(t)=1\right)\\
 & \hspace{1em}-P\left(Y_{1}(t)\in\star,Y_{0}(t)\in y_{b},S_{1}(t)=0,R_{1}(t)=1,S_{0}(t)\cdot R_{0}(t)=1\right)\\
 & \hspace{1em}-P\left(Y_{1}(t)\in\star,Y_{0}(t)\in y_{b},S_{1}(t)=0,R_{1}(t)=0,S_{0}(t)\cdot R_{0}(t)=1\right)\\
 & \hspace{1em}-P\left(Y_{1}(t)\in y_{a},Y_{0}(t)\not\in y_{b},S_{1}(t)\cdot R_{1}(t)=1,S_{0}(t)=1,R_{0}(t)=0\right)\\
 & \hspace{1em}-P\left(Y_{1}(t)\not\in y_{a},Y_{0}(t)\not\in y_{b},S_{1}(t)\cdot R_{1}(t)=1,S_{0}(t)=1,R_{0}(t)=0\right)\\
 & \hspace{1em}-P\left(Y_{1}(t)\in y_{a},Y_{0}(t)\not\in y_{b},S_{1}(t)=1,R_{1}(t)=0,S_{0}(t)=1,R_{0}(t)=0\right)\\
 & \hspace{1em}-P\left(Y_{1}(t)\not\in y_{a},Y_{0}(t)\not\in y_{b},S_{1}(t)=1,R_{1}(t)=0,S_{0}(t)=1,R_{0}(t)=0\right)\\
 & \hspace{1em}-P\left(Y_{1}(t)\in\star,Y_{0}(t)\not\in y_{b},S_{1}(t)=0,R_{1}(t)=1,S_{0}(t)=1,R_{0}(t)=0\right)\\
 & \hspace{1em}-P\left(Y_{1}(t)\in\star,Y_{0}(t)\not\in y_{b},S_{1}(t)=0,R_{1}(t)=0,S_{0}(t)=1,R_{0}(t)=0\right)\\
 & \hspace{1em}-P\left(Y_{1}(t)\in y_{a},Y_{0}(t)\in y_{b},S_{1}(t)\cdot R_{1}(t)=1,S_{0}(t)=1,R_{0}(t)=0\right)\\
 & \hspace{1em}-P\left(Y_{1}(t)\not\in y_{a},Y_{0}(t)\in y_{b},S_{1}(t)\cdot R_{1}(t)=1,S_{0}(t)=1,R_{0}(t)=0\right)\\
 & \hspace{1em}-P\left(Y_{1}(t)\in y_{a},Y_{0}(t)\in y_{b},S_{1}(t)=1,R_{1}(t)=0,S_{0}(t)=1,R_{0}(t)=0\right)\\
 & \hspace{1em}-P\left(Y_{1}(t)\not\in y_{a},Y_{0}(t)\in y_{b},S_{1}(t)=1,R_{1}(t)=0,S_{0}(t)=1,R_{0}(t)=0\right)\\
 & \hspace{1em}-P\left(Y_{1}\in\star,Y_{0}(t)\in y_{b},S_{1}(t)=0,R_{1}(t)=1,S_{0}(t)=1,R_{0}(t)=0\right)\\
 & \hspace{1em}-P\left(Y_{1}\in\star,Y_{0}(t)\in y_{b},S_{1}(t)=0,R_{1}(t)=0,S_{0}(t)=1,R_{0}(t)=0\right)\\
 & \hspace{1em}-P\left(Y_{1}(t)\in y_{a},Y_{0}(t)\in\star,S_{1}(t)\cdot R_{1}(t)=1,S_{0}(t)=0,R_{0}(t)=1\right)\\
 & \hspace{1em}-P\left(Y_{1}(t)\not\in y_{a},Y_{0}(t)\in\star,S_{1}(t)\cdot R_{1}(t)=1,S_{0}(t)=0,R_{0}(t)=1\right)\\
 & \hspace{1em}-P\left(Y_{1}(t)\in y_{a},Y_{0}(t)\in\star,S_{1}(t)=1,R_{1}(t)=0,S_{0}(t)=0,R_{0}(t)=1\right)\\
 & \hspace{1em}-P\left(Y_{1}(t)\not\in y_{a},Y_{0}(t)\in\star,S_{1}(t)=1,R_{1}(t)=0,S_{0}(t)=0,R_{0}(t)=1\right)\\
 & \hspace{1em}-P\left(Y_{1}(t)\in\star,Y_{0}(t)\in\star,S_{1}(t)=0,R_{1}(t)=1,S_{0}(t)=0,R_{0}(t)=1\right)\\
 & \hspace{1em}-P\left(Y_{1}(t)\in\star,Y_{0}(t)\in\star,S_{1}(t)=0,R_{1}(t)=0,S_{0}(t)=0,R_{0}(t)=1\right)\\
 & \hspace{1em}-P\left(Y_{1}(t)\in y_{a},Y_{0}(t)\in\star,S_{1}(t)\cdot R_{1}(t)=1,S_{0}(t)=0,R_{0}(t)=0\right)\\
 & \hspace{1em}-P\left(Y_{1}(t)\not\in y_{a},Y_{0}(t)\in\star,S_{1}(t)\cdot R_{1}(t)=1,S_{0}(t)=0,R_{0}(t)=0\right)\\
 & \hspace{1em}-P\left(Y_{1}(t)\in y_{a},Y_{0}(t)\in\star,S_{1}(t)=1,R_{1}(t)=0,S_{0}(t)=0,R_{0}(t)=0\right)\\
 & \hspace{1em}-P\left(Y_{1}(t)\not\in y_{a},Y_{0}(t)\in\star,S_{1}(t)=1,R_{1}(t)=0,S_{0}(t)=0,R_{0}(t)=0\right)\\
 & \hspace{1em}-P\left(Y_{1}(t)\in\star,Y_{0}(t)\in\star,S_{1}(t)=0,R_{1}(t)=1,S_{0}(t)=0,R_{0}(t)=0\right)\\
 & \hspace{1em}-P\left(Y_{1}(t)\in\star,Y_{0}(t)\in\star,S_{1}(t)=0,R_{1}(t)=0,S_{0}(t)=0,R_{0}(t)=0\right)
\end{align*}
 Simplifying this last expression, we have that $P\left(Y(t)\in y_{a},S(t)=1\mid X=1\right)+P(Y(t)\not\in y_{b},S(t)=1\mid X=0)-1$
is equal to
\begin{align*}
 & P\left(Y_{1}(t)\in y_{a},Y_{0}(t)\not\in y_{b},S_{1}(t)\cdot R_{1}(t)=1,S_{0}(t)\cdot R_{0}(t)=1\right)\\
 & \hspace{1em}-P\left(Y_{1}(t)\not\in y_{a},Y_{0}(t)\in y_{b},S_{1}(t)R_{1}(t)=1,S_{0}(t)\cdot R_{0}(t)=1\right)\\
 & \hspace{1em}-P\left(Y_{1}(t)\in y_{a},Y_{0}(t)\in y_{b},S_{1}(t)=1,R_{1}(t)=0,S_{0}(t)\cdot R_{0}(t)=1\right)\\
 & \hspace{1em}-P\left(Y_{1}(t)\not\in y_{a},Y_{0}(t)\in y_{b},S_{1}(t)=1,R_{1}(t)=0,S_{0}(t)\cdot R_{0}(t)=1\right)\\
 & \hspace{1em}-P\left(Y_{1}(t)\in\star,Y_{0}(t)\in y_{b},S_{1}(t)=0,R_{1}(t)=1,S_{0}(t)\cdot R_{0}(t)=1\right)\\
 & \hspace{1em}-P\left(Y_{1}(t)\in\star,Y_{0}(t)\in y_{b},S_{1}(t)=0,R_{1}(t)=0,S_{0}(t)\cdot R_{0}(t)=1\right)\\
 & \hspace{1em}-P\left(Y_{1}(t)\not\in y_{a},Y_{0}(t)\not\in y_{b},S_{1}(t)\cdot R_{1}(t)=1,S_{0}(t)=1,R_{0}(t)=0\right)\\
 & \hspace{1em}-P\left(Y_{1}(t)\in y_{a},Y_{0}(t)\not\in y_{b},S_{1}(t)=1,R_{1}(t)=0,S_{0}(t)=1,R_{0}(t)=0\right)\\
 & \hspace{1em}-P\left(Y_{1}(t)\not\in y_{a},Y_{0}(t)\not\in y_{b},S_{1}(t)=1,R_{1}(t)=0,S_{0}(t)=1,R_{0}(t)=0\right)\\
 & \hspace{1em}-P\left(Y_{1}(t)\in\star,Y_{0}(t)\not\in y_{b},S_{1}(t)=0,R_{1}(t)=1,S_{0}(t)=1,R_{0}(t)=0\right)\\
 & \hspace{1em}-P\left(Y_{1}(t)\in\star,Y_{0}(t)\not\in y_{b},S_{1}(t)=0,R_{1}(t)=0,S_{0}(t)=1,R_{0}(t)=0\right)\\
 & \hspace{1em}-P\left(Y_{1}(t)\not\in y_{a},Y_{0}(t)\in y_{b},S_{1}(t)\cdot R_{1}(t)=1,S_{0}(t)=1,R_{0}(t)=0\right)\\
 & \hspace{1em}-P\left(Y_{1}(t)\in y_{a},Y_{0}(t)\in y_{b},S_{1}(t)=1,R_{1}(t)=0,S_{0}(t)=1,R_{0}(t)=0\right)\\
 & \hspace{1em}-P\left(Y_{1}(t)\not\in y_{a},Y_{0}(t)\in y_{b},S_{1}(t)=1,R_{1}(t)=0,S_{0}(t)=1,R_{0}(t)=0\right)\\
 & \hspace{1em}-P\left(Y_{1}\in\star,Y_{0}(t)\in y_{b},S_{1}(t)=0,R_{1}(t)=1,S_{0}(t)=1,R_{0}(t)=0\right)\\
 & \hspace{1em}-P\left(Y_{1}\in\star,Y_{0}(t)\in y_{b},S_{1}(t)=0,R_{1}(t)=0,S_{0}(t)=1,R_{0}(t)=0\right)\\
 & \hspace{1em}-P\left(Y_{1}(t)\not\in y_{a},Y_{0}(t)\in\star,S_{1}(t)\cdot R_{1}(t)=1,S_{0}(t)=0,R_{0}(t)=1\right)\\
 & \hspace{1em}-P\left(Y_{1}(t)\in y_{a},Y_{0}(t)\in\star,S_{1}(t)=1,R_{1}(t)=0,S_{0}(t)=0,R_{0}(t)=1\right)\\
 & \hspace{1em}-P\left(Y_{1}(t)\not\in y_{a},Y_{0}(t)\in\star,S_{1}(t)=1,R_{1}(t)=0,S_{0}(t)=0,R_{0}(t)=1\right)\\
 & \hspace{1em}-P\left(Y_{1}(t)\in\star,Y_{0}(t)\in\star,S_{1}(t)=0,R_{1}(t)=1,S_{0}(t)=0,R_{0}(t)=1\right)\\
 & \hspace{1em}-P\left(Y_{1}(t)\in\star,Y_{0}(t)\in\star,S_{1}(t)=0,R_{1}(t)=0,S_{0}(t)=0,R_{0}(t)=1\right)\\
 & \hspace{1em}-P\left(Y_{1}(t)\not\in y_{a},Y_{0}(t)\in\star,S_{1}(t)\cdot R_{1}(t)=1,S_{0}(t)=0,R_{0}(t)=0\right)\\
 & \hspace{1em}-P\left(Y_{1}(t)\in y_{a},Y_{0}(t)\in\star,S_{1}(t)=1,R_{1}(t)=0,S_{0}(t)=0,R_{0}(t)=0\right)\\
 & \hspace{1em}-P\left(Y_{1}(t)\not\in y_{a},Y_{0}(t)\in\star,S_{1}(t)=1,R_{1}(t)=0,S_{0}(t)=0,R_{0}(t)=0\right)\\
 & \hspace{1em}-P\left(Y_{1}(t)\in\star,Y_{0}(t)\in\star,S_{1}(t)=0,R_{1}(t)=1,S_{0}(t)=0,R_{0}(t)=0\right)\\
 & \hspace{1em}-P\left(Y_{1}(t)\in\star,Y_{0}(t)\in\star,S_{1}(t)=0,R_{1}(t)=0,S_{0}(t)=0,R_{0}(t)=0\right).
\end{align*}
 We will label this last expression involving counterfactuals as expression
D. Given that probability is greater than equal to zero and less than
equal to 1, $P\left(Y(t)\in y_{a},S(t)=1,R(t)=1\mid X=1\right)+P(Y(t)\not\in y_{b},S(t)=1,R(t)=1\mid X=0)-1$
serves as a lower bound on 
\begin{align*}
 & P\left(Y_{1}(t)\in y_{a},Y_{0}(t)\not\in y_{b},S_{1}(t)\cdot R_{1}(t)=1,S_{0}(t)\cdot R_{1}(t)=1\right)\\
 & -P\left(Y_{1}(t)\not\in y_{a},Y_{0}(t)\in y_{b},S_{1}(t)=1,R_{1}(t)=1,S_{0}(t)=1,R_{0}(t)=1\right)\\
 & -P\left(Y_{1}(t)\not\in y_{a},Y_{0}(t)\in y_{b},S_{1}(t)=1,R_{1}(t)=0,S_{0}(t)=1,R_{0}(t)=1\right)\\
 & -P\left(Y_{1}(t)\not\in y_{a},Y_{0}(t)\in y_{b},S_{1}(t)=1,R_{1}(t)=1,S_{0}(t)=1,R_{0}(t)=0\right)\\
 & -P\left(Y_{1}(t)\not\in y_{a},Y_{0}(t)\in y_{b},S_{1}(t)=1,R_{1}(t)=0,S_{0}(t)=1,R_{0}(t)=0\right).
\end{align*}
 Consequently, the expression $P\left(Y(t)\in y_{a},S(t)=1,R(t)=1\mid X=1\right)+P(Y(t)\not\in y_{b},S(t)=1,R(t)=1\mid X=0)-1$
also serves as a lower bound on 
\begin{align*}
 & P\left(Y_{1}(t)\in y_{a},Y_{0}(t)\not\in y_{b},S_{1}(t)=1,S_{0}(t)=1\right)\\
 & -P\left(Y_{1}(t)\not\in y_{a},Y_{0}(t)\in y_{b},S_{1}(t)=1,S_{0}(t)=1\right)
\end{align*}
 from an application of the theorem of total probability.

This completes the proof for the the first part of the Theorem. 

From the first part of the Theorem 

\begin{align*}
 & P(Y(t)\in y_{a},S(t)=1\mid X=1)+P(Y(t)\not\in y_{b},S(t)=1\mid X=0)-1
\end{align*}
is equal to expression D, and to simplify expression D, we will use
identities: 

Identity one:

\begin{align*}
 & P(S_{1}(t)=0,S_{0}(t)=0)\\
 & =P\left(Y_{1}(t)\in\star,Y_{0}(t)\in\star,S_{1}(t)=0,R_{1}(t)=0,S_{0}(t)=0,R_{0}(t)=0\right)\\
 & \hspace{1em}+P\left(Y_{1}(t)\in\star,Y_{0}(t)\in\star,S_{1}(t)=0,R_{1}(t)=1,S_{0}(t)=0,R_{0}(t)=0\right)\\
 & \hspace{1em}+P\left(Y_{1}(t)\in\star,Y_{0}(t)\in\star,S_{1}(t)=0,R_{1}(t)=0,S_{0}(t)=0,R_{0}(t)=1\right)\\
 & \hspace{1em}+P\left(Y_{1}(t)\in\star,Y_{0}(t)\in\star,S_{1}(t)=0,R_{1}(t)=1,S_{0}(t)=0,R_{0}(t)=1\right).
\end{align*}

Identity two : 
\begin{align*}
 & P(Y_{1}(t)\in\star,Y_{0}(t)\in y_{b},S_{1}(t)=0,S_{0}(t)=1)\\
 & =P\left(Y_{1}(t)\in\star,Y_{0}(t)\in y_{b},S_{1}(t)=0,R_{1}(t)=1,S_{0}(t)\cdot R_{0}(t)=1\right)\\
 & +P\left(Y_{1}(t)\in\star,Y_{0}(t)\in y_{b},S_{1}(t)=0,R_{1}(t)=0,S_{0}(t)\cdot R_{0}(t)=1\right)\\
 & +P\left(Y_{1}\in\star,Y_{0}(t)\in y_{b},S_{1}(t)=0,R_{1}(t)=1,S_{0}(t)=1,R_{0}(t)=0\right)\\
 & +P\left(Y_{1}\in\star,Y_{0}(t)\in y_{b},S_{1}(t)=0,R_{1}(t)=0,S_{0}(t)=1,R_{0}(t)=0\right)
\end{align*}

Identity three:
\begin{align*}
 & P(Y_{1}(t)\not\in y_{a},Y_{0}(t)\in\star,S_{1}(t)=1,S_{0}(t)=0)\\
 & =P\left(Y_{1}(t)\not\in y_{a},Y_{0}(t)\in\star,S_{1}(t)=1,R_{1}(t)=0,S_{0}(t)=0,R_{0}(t)=0\right)\\
 & +P\left(Y_{1}(t)\not\in y_{a},Y_{0}(t)\in\star,S_{1}(t)\cdot R_{1}(t)=1,S_{0}(t)=0,R_{0}(t)=0\right)\\
 & +P\left(Y_{1}(t)\not\in y_{a},Y_{0}(t)\in\star,S_{1}(t)=1,R_{1}(t)=0,S_{0}(t)=0,R_{0}(t)=1\right)\\
 & +P\left(Y_{1}(t)\not\in y_{a},Y_{0}(t)\in\star,S_{1}(t)\cdot R_{1}(t)=1,S_{0}(t)=0,R_{0}(t)=0\right)
\end{align*}

Denote

\begin{align*}
 & x=P\left(Y_{1}(t)\in y_{a},Y_{0}(t)\in y_{b},S_{1}(t)=1,R_{1}(t)=0,S_{0}(t)\cdot R_{0}(t)=1\right)\\
 & \hspace{1em}+P\left(Y_{1}(t)\not\in y_{a},Y_{0}(t)\not\in y_{b},S_{1}(t)\cdot R_{1}(t)=1,S_{0}(t)=1,R_{0}(t)=0\right)\\
 & \hspace{1em}+P\left(Y_{1}(t)\in y_{a},Y_{0}(t)\not\in y_{b},S_{1}(t)=1,R_{1}(t)=0,S_{0}(t)=1,R_{0}(t)=0\right)\\
 & \hspace{1em}+P\left(Y_{1}(t)\not\in y_{a},Y_{0}(t)\not\in y_{b},S_{1}(t)=1,R_{1}(t)=0,S_{0}(t)=1,R_{0}(t)=0\right)\\
 & \hspace{1em}+P\left(Y_{1}(t)\in\star,Y_{0}(t)\not\in y_{b},S_{1}(t)=0,R_{1}(t)=1,S_{0}(t)=1,R_{0}(t)=0\right)\\
 & \hspace{1em}+P\left(Y_{1}(t)\in\star,Y_{0}(t)\not\in y_{b},S_{1}(t)=0,R_{1}(t)=0,S_{0}(t)=1,R_{0}(t)=0\right)\\
 & \hspace{1em}+P\left(Y_{1}(t)\in y_{a},Y_{0}(t)\in y_{b},S_{1}(t)=1,R_{1}(t)=0,S_{0}(t)=1,R_{0}(t)=0\right)\\
 & \hspace{1em}+P\left(Y_{1}(t)\in y_{a},Y_{0}(t)\in\star,S_{1}(t)=1,R_{1}(t)=0,S_{0}(t)=0,R_{0}(t)=1\right)\\
 & \hspace{1em}+P\left(Y_{1}(t)\in y_{a},Y_{0}(t)\in\star,S_{1}(t)=1,R_{1}(t)=0,S_{0}(t)=0,R_{0}(t)=0\right)
\end{align*}

Now, the contrast 
\begin{align*}
 & P(Y(t)\in y_{a},S(t)=1\mid X=1)+P(Y(t)\not\in y_{b},S(t)=1\mid X=0)-1
\end{align*}
 is equal to expression D if and only if the following equality holds

\begin{align*}
 & \frac{\left(\begin{array}{c}
P\left(Y_{1}(t)\in y_{a},Y_{0}(t)\not\in y_{b},S_{1}(t)\cdot R_{1}(t)=1,S_{0}(t)\cdot R_{1}(t)=1\right)\\
-P\left(Y_{1}(t)\not\in y_{a},Y_{0}(t)\in y_{b},S_{1}(t)R_{1}(t)=1,S_{0}(t)\cdot R_{0}(t)=1\right)\\
-P\left(Y_{1}(t)\not\in y_{a},Y_{0}(t)\in y_{b},S_{1}(t)=1,R_{1}(t)=0,S_{0}(t)\cdot R_{0}(t)=1\right)\\
-P\left(Y_{1}(t)\not\in y_{a},Y_{0}(t)\in y_{b},S_{1}(t)\cdot R_{1}(t)=1,S_{0}(t)=1,R_{0}(t)=0\right)\\
-P\left(Y_{1}(t)\not\in y_{a},Y_{0}(t)\in y_{b},S_{1}(t)=1,R_{1}(t)=0,S_{0}(t)=1,R_{0}(t)=0\right)
\end{array}\right)}{P(S_{1}(t)R_{1}(t)=1,S_{0}(t)R_{0}(t)=1)}\\
 & =\frac{P(Y(t)\in y_{a},S(t)=1,R(t)=1\mid X=1)+P(Y(t)\not\in y_{b},S(t)=1,R(t)=1\mid X=0)-1}{P(S_{1}(t)R_{1}(t)=1,S_{0}(t)R_{0}(t)=1)}\\
 & \hspace{1em}+\frac{P(Y_{1}(t)\in\star,Y_{0}(t)\in y_{b},S_{1}(t)=0,S_{0}(t)=1)}{P(S_{1}(t)R_{1}(t)=1,S_{0}(t)R_{0}(t)=1)}\\
 & \hspace{1em}+\frac{P(Y_{1}(t)\not\in y_{a},Y_{0}(t)\in\star,S_{1}(t)=1,S_{0}(t)=0)}{P(S_{1}(t)R_{1}(t)=1,S_{0}(t)R_{0}(t)=1)}\\
 & \hspace{1em}+\frac{P(S_{1}(t)=0,S_{0}(t)=0)}{P(S_{1}(t)R_{1}(t)=1,S_{0}(t)R_{0}(t)=1)}\\
 & \hspace{1em}+\frac{x}{P(S_{1}(t)R_{1}(t)=1,S_{0}(t)R_{1}(t)=1)}
\end{align*}

Denote 
\begin{align*}
y & =P(Y_{1}(t)\in\star,Y_{0}(t)\in y_{b},S_{1}(t)=0,S_{0}(t)=1)\\
 & \hspace{1em}+P(Y_{1}(t)\not\in y_{a},Y_{0}(t)\in\star,S_{1}(t)=1,S_{0}(t)=0)\\
 & \hspace{1em}+x.
\end{align*}
 out of space considerations. Note, $y\in[0,1).$ 

Note, the normalized principal stratum direct effect is
\[
\frac{\left(\begin{array}{c}
\sum_{\{r_{1},r_{0}\}\in\{0,1\}^{2}}P\left(Y_{1}(t)\in y_{a},Y_{0}(t)\not\in y_{b},S_{1}(t)=1,R_{1}(t)=r_{1},S_{0}(t)=1,R_{0}(t)=r_{0}\right)\\
-\sum_{\{r_{1},r_{0}\}\in\{0,1\}^{2}}P\left(Y_{1}(t)\not\in y_{a},Y_{0}(t)\in y_{b},S_{1}(t)=1,R_{1}(t)=r_{1},S_{0}(t)=1\cdot R_{0}(t)=r_{0}\right)
\end{array}\right)}{\sum_{\{r_{1},r_{0}\}\in\{0,1\}^{2}}P(S_{1}(t)=1,S_{0}(t)=1,R_{1}=r_{1},R_{0}=r_{0})}.
\]
Instead, we target
\[
\frac{\left(\begin{array}{c}
P\left(Y_{1}(t)\in y_{a},Y_{0}(t)\not\in y_{b},S_{1}(t)\cdot R_{1}(t)=1,S_{0}(t)\cdot R_{1}(t)=1\right)\\
-P\left(Y_{1}(t)\not\in y_{a},Y_{0}(t)\in y_{b},S_{1}(t)R_{1}(t)=1,S_{0}(t)\cdot R_{0}(t)=1\right)\\
-P\left(Y_{1}(t)\not\in y_{a},Y_{0}(t)\in y_{b},S_{1}(t)=1,R_{1}(t)=0,S_{0}(t)\cdot R_{0}(t)=1\right)\\
-P\left(Y_{1}(t)\not\in y_{a},Y_{0}(t)\in y_{b},S_{1}(t)\cdot R_{1}(t)=1,S_{0}(t)=1,R_{0}(t)=0\right)\\
-P\left(Y_{1}(t)\not\in y_{a},Y_{0}(t)\in y_{b},S_{1}(t)=1,R_{1}(t)=0,S_{0}(t)=1,R_{0}(t)=0\right)
\end{array}\right)}{P(S_{1}(t)R_{1}(t)=1,S_{0}(t)R_{1}(t)=1)}
\]

Now, from

\begin{align*}
 & \frac{\left(\begin{array}{c}
P\left(Y_{1}(t)\in y_{a},Y_{0}(t)\not\in y_{b},S_{1}(t)\cdot R_{1}(t)=1,S_{0}(t)\cdot R_{1}(t)=1\right)\\
-P\left(Y_{1}(t)\not\in y_{a},Y_{0}(t)\in y_{b},S_{1}(t)R_{1}(t)=1,S_{0}(t)\cdot R_{0}(t)=1\right)\\
-P\left(Y_{1}(t)\not\in y_{a},Y_{0}(t)\in y_{b},S_{1}(t)=1,R_{1}(t)=0,S_{0}(t)\cdot R_{0}(t)=1\right)\\
-P\left(Y_{1}(t)\not\in y_{a},Y_{0}(t)\in y_{b},S_{1}(t)\cdot R_{1}(t)=1,S_{0}(t)=1,R_{0}(t)=0\right)\\
-P\left(Y_{1}(t)\not\in y_{a},Y_{0}(t)\in y_{b},S_{1}(t)=1,R_{1}(t)=0,S_{0}(t)=1,R_{0}(t)=0\right)
\end{array}\right)}{P(S_{1}(t)R_{1}(t)=1,S_{0}(t)R_{0}(t)=1)}\\
 & =\frac{P(Y(t)\in y_{a},S(t)=1,R(t)=1\mid X=1)+P(Y(t)\not\in y_{b},S(t)=1,R(t)=1\mid X=0)-1}{P(S_{1}(t)R_{0}(t)=1,S_{0}(t)R_{0}(t)=1)}\\
 & \hspace{1em}+\frac{\left(y+P(S_{1}(t)=0,S_{0}(t)=0)\right)}{P(S_{1}(t)R_{1}(t)=1,S_{0}(t)R_{0}(t)=1)}\\
 & =\frac{P(Y(t)\in y_{a},S(t)=1,R(t)=1\mid X=1)+P(Y(t)\not\in y_{b},S(t)=1,R(t)=1\mid X=0)-1}{P(S_{1}(t)R_{1}(t)=1,S_{0}(t)R_{0}(t)=1)-P(S_{1}(t)R_{1}(t)=0,S_{0}(t)R_{0}(t)=0)}\times\\
\\
 & \left(\frac{P(S_{1}(t)R_{1}(t)=1,S_{0}(t)R_{0}(t)=1)-P(S_{1}(t)R_{1}(t)=0,S_{0}(t)R_{0}(t)=0)}{P(S_{1}(t)R_{1}(t)=1,S_{0}(t)R_{0}(t)=1)}\right.\\
 & \hspace{1em}+\frac{\left(y+P(S_{1}(t)=0,S_{0}(t)=0)\right)}{P(S_{1}(t)R_{1}(t)=1,S_{0}(t)R_{0}(t)=1)}\times\\
 & \left.\times\frac{\left(P(S_{1}(t)R_{1}(t)=1,S_{0}(t)R_{0}(t)=1)-P(S_{1}(t)R_{1}(t)=0,S_{0}(t)R_{0}(t)=0)\right)}{P(Y(t)\in y_{a},S(t)=1,R(t)=1\mid X=1)+P(Y(t)\not\in y_{b},S(t)=1,R(t)=1\mid X=0)-1}\right).
\end{align*}

We want to show that the next expression, which we denote expression
A,
\begin{align*}
\left(\frac{P(S_{1}(t)R_{1}(t)=1,S_{0}(t)R_{0}(t)=1)-P(S_{1}(t)R_{1}(t)=0,S_{0}(t)R_{0}(t)=0)}{P(S_{1}(t)=1,S_{0}(t)=1,R_{1}(t)=1,R_{0}(t)=1)}\right.\\
+\frac{\left(y+P(S_{1}(t)=0,S_{0}(t)=0)\right)}{P(S_{1}(t)R_{1}(t)=1,S_{0}(t)R_{0}(t)=1)}\times\hspace{1em}\\
\left.\times\frac{\left(P(S_{1}(t)R_{1}(t)=1,S_{0}(t)R_{0}(t)=1)-P(S_{1}(t)R_{1}(t)=0,S_{0}(t)R_{0}(t)=0)\right)}{P(Y(t)\in y_{a},S(t)=1,R(t)=1\mid X=1)+P(Y(t)\not\in y_{a},S(t)=1,R(t)=1\mid X=0)-1}\right).
\end{align*}
is greater than or equal to one to show that 
\begin{align*}
\frac{P(Y(t)\in y_{a},S(t)=1,R(t)=1\mid X=1)+P(Y(t)\not\in y_{b},S(t)=1,R(t)=1\mid X=0)-1}{P(S_{1}(t)R_{1}(t)=1,S_{0}(t)R_{0}(t)=1)-P(S_{1}(t)R_{1}(t)=0,S_{0}(t)R_{0}(t)=0)}
\end{align*}
 is a lower bound to 
\[
\frac{\left(\begin{array}{c}
P\left(Y_{1}(t)\in y_{a},Y_{0}(t)\not\in y_{b},S_{1}(t)\cdot R_{1}(t)=1,S_{0}(t)\cdot R_{1}(t)=1\right)\\
-P\left(Y_{1}(t)\not\in y_{a},Y_{0}(t)\in y_{b},S_{1}(t)R_{1}(t)=1,S_{0}(t)\cdot R_{0}(t)=1\right)\\
-P\left(Y_{1}(t)\not\in y_{a},Y_{0}(t)\in y_{b},S_{1}(t)=1,R_{1}(t)=0,S_{0}(t)\cdot R_{0}(t)=1\right)\\
-P\left(Y_{1}(t)\not\in y_{a},Y_{0}(t)\in y_{b},S_{1}(t)\cdot R_{1}(t)=1,S_{0}(t)=1,R_{0}(t)=0\right)\\
-P\left(Y_{1}(t)\not\in y_{a},Y_{0}(t)\in y_{b},S_{1}(t)=1,R_{1}(t)=0,S_{0}(t)=1,R_{0}(t)=0\right)
\end{array}\right)}{P(S_{1}(t)=1,S_{0}(t)=1,R_{1}(t)=1,R_{0}(t)=1)}.
\]
Through inspection, expression A is greater than or equal to one if
and only if 
\[
\frac{\left(P(S_{1}(t)R_{1}(t)=1,S_{0}(t)R_{0}(t)=1)-P(S_{1}(t)R_{1}(t)=0,S_{0}(t)R_{0}(t)=0)\right)}{P(Y(t)\in y_{a},S(t)=1,R(t)=1\mid X=1)+P(Y(t)\not\in y_{a},S(t)=1,R(t)=1\mid X=0)-1}\geq1.
\]
 We denote 
\[
\frac{\left(P(S_{1}(t)R_{1}(t)=1,S_{0}(t)R_{0}(t)=1)-P(S_{1}(t)R_{1}(t)=0,S_{0}(t)R_{0}(t)=0)\right)}{P(Y(t)\in y_{a},S(t)=1,R(t)=1\mid X=1)+P(Y(t)\not\in y_{a},S(t)=1,R(t)=1\mid X=0)-1}
\]
 as expression B.

Now, from the earlier part of the Theorem, we had the denominator
of expression B is equal to expression D. The numerator of expression
B is 
\[
\left(P(S_{1}(t)R_{1}(t)=1,S_{0}(t)R_{0}(t)=1)-P(S_{1}(t)R_{1}(t)=0,S_{0}(t)R_{0}(t)=0)\right).
\]

We have $P(S_{1}(t)R_{1}(t)=1,S_{0}(t)R_{0}(t)=1)\geq P\left(Y_{1}(t)\in y_{a},Y_{0}(t)\not\in y_{a},S_{1}(t)\cdot R_{1}(t)=1,S_{0}(t)\cdot R_{0}(t)=1\right)$
as $P(A)\geq P(A,B).$ 

Now, 

\begin{align*}
 & P(S_{1}(t)R_{1}(t)=0,S_{0}(t)R_{0}(t)=0)\\
 & =P(S_{1}(t)=1,R_{1}(t)=0,S_{0}(t)=0,R_{0}(t)=0)\\
 & \hspace{1em}+P(S_{1}(t)=0,R_{1}(t)=1,S_{0}(t)=0,R_{0}(t)=0)\\
 & \hspace{1em}+P(S_{1}(t)=0,R_{1}(t)=0,S_{0}(t)=0,R_{0}(t)=0)\\
 & \hspace{1em}+P(S_{1}(t)=1,R_{1}(t)=0,S_{0}(t)=0,R_{0}(t)=1)\\
 & \hspace{1em}+P(S_{1}(t)=0,R_{1}(t)=1,S_{0}(t)=0,R_{0}(t)=1)\\
 & \hspace{1em}+P(S_{1}(t)=0,R_{1}(t)=0,S_{0}(t)=0,R_{0}(t)=1)\\
 & \hspace{1em}+P(S_{1}(t)=1,R_{1}(t)=0,S_{0}(t)=1,R_{0}(t)=0)\\
 & \hspace{1em}+P(S_{1}(t)=0,R_{1}(t)=1,S_{0}(t)=1,R_{0}(t)=0))\\
 & \hspace{1em}+P(S_{1}(t)=0,R_{1}(t)=0,S_{0}(t)=1,R_{0}(t)=0)).
\end{align*}
 This in turn is equal to 
\begin{align*}
 & P(S_{1}(t)R_{1}(t)=0,S_{0}(t)R_{0}(t)=0)\\
 & =P(Y_{1}(t)\in y_{a},Y_{0}(t)\in\star,S_{1}(t)=1,R_{1}(t)=0,S_{0}(t)=0,R_{0}(t)=0)\\
 & \hspace{1em}+P(Y_{1}(t)\not\in y_{a},Y_{0}(t)\in\star S_{1}(t)=1,R_{1}(t)=0,S_{0}(t)=0,R_{0}(t)=0)\\
 & \hspace{1em}+P(Y_{1}(t)\in\star,Y_{0}(t)\in\star,S_{1}(t)=0,R_{1}(t)=1,S_{0}(t)=0,R_{0}(t)=0)\\
 & \hspace{1em}+P(Y_{1}(t)\in\star,Y_{0}(t)\in\star,S_{1}(t)=0,R_{1}(t)=0,S_{0}(t)=0,R_{0}(t)=0)\\
 & \hspace{1em}+P(Y_{1}(t)\in y_{a},Y_{0}(t)\in\star S_{1}(t)=1,R_{1}(t)=0,S_{0}(t)=0,R_{0}(t)=1)\\
 & \hspace{1em}+P(Y_{1}(t)\not\in y_{a},Y_{0}(t)\in\star S_{1}(t)=1,R_{1}(t)=0,S_{0}(t)=0,R_{0}(t)=1\\
 & \hspace{1em}+P(Y_{1}(t)\in\star,Y_{0}(t)\in\star,S_{1}(t)=0,R_{1}(t)=1,S_{0}(t)=0,R_{0}(t)=1)\\
 & \hspace{1em}+P(Y_{1}(t)\in\star,Y_{0}(t)\in\star,S_{1}(t)=0,R_{1}(t)=0,S_{0}(t)=0,R_{0}(t)=1)\\
 & \hspace{1em}+P(Y_{1}(t)\in y_{a},Y_{0}(t)\not\in y_{b},S_{1}(t)=1,R_{1}(t)=0,S_{0}(t)=1,R_{0}(t)=0)\\
 & \hspace{1em}+P(Y_{1}(t)\not\in y_{a},Y_{0}(t)\not\in y_{b},S_{1}(t)=1,R_{1}(t)=0,S_{0}(t)=1,R_{0}(t)=0)\\
 & \hspace{1em}+P(Y_{1}(t)\in y_{a},Y_{0}(t)\in y_{b},S_{1}(t)=1,R_{1}(t)=0,S_{0}(t)=1,R_{0}(t)=0)\\
 & \hspace{1em}+P(Y_{1}(t)\not\in y_{a},Y_{0}(t)\in y_{b},S_{1}(t)=1,R_{1}(t)=0,S_{0}(t)=1,R_{0}(t)=0)\\
 & \hspace{1em}+P(Y_{1}(t)\in\star,Y_{0}(t)\in y_{b},S_{1}(t)=0,R_{1}(t)=1,S_{0}(t)=1,R_{0}(t)=0)\\
 & \hspace{1em}+P(Y_{1}(t)\in\star,Y_{0}(t)\not\in y_{b},S_{1}(t)=0,R_{1}(t)=1,S_{0}(t)=1,R_{0}(t)=0)\\
 & \hspace{1em}+P(Y_{1}(t)\in\star,Y_{0}(t)\in y_{b},S_{1}(t)=0,R_{1}(t)=0,S_{0}(t)=1,R_{0}(t)=0)\\
 & \hspace{1em}+P(Y_{1}(t)\in\star,Y_{0}(t)\not\in y_{b},S_{1}(t)=0,R_{1}(t)=0,S_{0}(t)=1,R_{0}(t)=0).
\end{align*}

If we examine 
\[
\frac{\left(P(S_{1}(t)\cdot R_{1}(t)=1,S_{0}(t)\cdot R_{0}(t)=1)-P(S_{1}(t)\cdot R_{1}(t)=0,S_{0}(t)\cdot R_{0}(t)=0)\right)}{P(Y(t)\in y_{a},S(t)=1,R(t)=1\mid X=1)+P(Y(t)\not\in y_{a},S(t)=1,R(t)=1\mid X=0)-1},
\]
 then the numerator is greater than the denominator as every term
that is negative in the numerator is represented in the counterfactual
representation of the denominator, and the only positive term in the
numerator is greater than equal to the only positive term in the counterfactual
representation of the denominator. Since expression B is greater than
equal to 1, expression A is also greater than equal to 1. 

This implies that 
\begin{align*}
\frac{P(Y(t)\in y_{a},S(t)=1,R(t)=1\mid X=1)+P(Y(t)\not\in y_{a},S(t)=1,R(t)=1\mid X=0)-1}{P(S_{1}(t)R_{1}(t)=1,S_{0}(t)R_{0}(t)=1)-P(S_{1}(t)R_{1}(t)=0,S_{0}(t)R_{0}(t)=0)}
\end{align*}
 serves a a lower bound on 
\[
\frac{\left(\begin{array}{c}
P\left(Y_{1}(t)\in y_{a},Y_{0}(t)\not\in y_{b},S_{1}(t)\cdot R_{1}(t)=1,S_{0}(t)\cdot R_{1}(t)=1\right)\\
-P\left(Y_{1}(t)\not\in y_{a},Y_{0}(t)\in y_{b},S_{1}(t)R_{1}(t)=1,S_{0}(t)\cdot R_{0}(t)=1\right)\\
-P\left(Y_{1}(t)\not\in y_{a},Y_{0}(t)\in y_{b},S_{1}(t)=1,R_{1}(t)=0,S_{0}(t)\cdot R_{0}(t)=1\right)\\
-P\left(Y_{1}(t)\not\in y_{a},Y_{0}(t)\in y_{b},S_{1}(t)\cdot R_{1}(t)=1,S_{0}(t)=1,R_{0}(t)=0\right)\\
-P\left(Y_{1}(t)\not\in y_{a},Y_{0}(t)\in y_{b},S_{1}(t)=1,R_{1}(t)=0,S_{0}(t)=1,R_{0}(t)=0\right)
\end{array}\right)}{P(S_{1}(t)=1,S_{0}(t)=1,R_{1}(t)=1,R_{0}(t)=1)}.
\]
 Now, 
\[
\frac{P(Y(t)\in y_{a},S(t)=1,R(t)=1\mid X=1)+P(Y(t)\not\in y_{a},S(t)=1,R(t)=1\mid X=0)-1}{P(S_{1}(t)R_{1}(t)=1,S_{0}(t)R_{0}(t)=1)-P(S_{1}(t)R_{1}(t)=0,S_{0}(t)R_{0}(t)=0)}
\]
 is equal to 
\[
\frac{P(Y(t)\in y_{a},S(t)=1,R(t)=1\mid X=1)+P(Y(t)\not\in y_{a},S(t)=1,R(t)=1\mid X=0)-1}{P\left(S=1,R=1\mid X=1\right)-P\left(R=0\mid X=0\right)-P\left(S=0,R=1\mid X=0\right)}.
\]
 This holds as 
\begin{align*}
 & P(S_{1}(t)R_{1}(t)=1,S_{0}(t)R_{0}(t)=1)-P(S_{1}(t)R_{1}(t)=0,S_{0}(t)R_{0}(t)=0)\\
 & =P(S_{1}(t)R_{1}(t)=1,S_{0}(t)R_{0}(t)=1)+P(S_{1}(t)R_{1}(t)=1,S_{0}(t)R_{0}(t)=0)\\
 & \hspace{1em}-P(S_{1}(t)R_{1}(t)=1,S_{0}(t)R_{0}(t)=0)-P(S_{1}(t)R_{1}(t)=0,S_{0}(t)R_{0}(t)=0)\\
 & =P(S_{1}(t)R_{1}(t)=1)-P(S_{0}(t)R_{0}(t)=0)\\
 & =P(S_{1}(t)R_{1}(t)=1)\\
 & \hspace{1em}-P(S_{0}(t)=0,R_{0}(t)=0)\\
 & \hspace{1em}-P(S_{0}(t)=1,R_{0}(t)=0)\\
 & \hspace{1em}-P(S_{0}(t)=0,R_{0}(t)=1)\\
 & =P(S_{1}(t)=1,R_{1}(t)=1)\\
 & \hspace{1em}-P(R_{0}(t)=0)\\
 & \hspace{1em}-P(S_{0}(t)=0,R_{0}(t)=1)\\
 & =P\left(S(t)=1,R(t)=1\mid X=1\right)-P\left(R(t)=0\mid X=0\right)-P\left(S(t)=0,R=1\mid X=0\right).
\end{align*}

This implies that 
\[
\frac{P(Y(t)\in y_{a},S(t)=1,R(t)=1\mid X=1)+P(Y(t)\not\in y_{a},S(t)=1,R(t)=1\mid X=0)-1}{P\left(S=1,R=1\mid X=1\right)-P\left(R=0\mid X=0\right)-P\left(S=0,R=1\mid X=0\right)}
\]
 is a lower bound on 
\[
\frac{\left(\begin{array}{c}
P\left(Y_{1}(t)\in y_{a},Y_{0}(t)\not\in y_{b},S_{1}(t)\cdot R_{1}(t)=1,S_{0}(t)\cdot R_{1}(t)=1\right)\\
-P\left(Y_{1}(t)\not\in y_{a},Y_{0}(t)\in y_{b},S_{1}(t)R_{1}(t)=1,S_{0}(t)\cdot R_{0}(t)=1\right)\\
-P\left(Y_{1}(t)\not\in y_{a},Y_{0}(t)\in y_{b},S_{1}(t)=1,R_{1}(t)=0,S_{0}(t)\cdot R_{0}(t)=1\right)\\
-P\left(Y_{1}(t)\not\in y_{a},Y_{0}(t)\in y_{b},S_{1}(t)\cdot R_{1}(t)=1,S_{0}(t)=1,R_{0}(t)=0\right)\\
-P\left(Y_{1}(t)\not\in y_{a},Y_{0}(t)\in y_{b},S_{1}(t)=1,R_{1}(t)=0,S_{0}(t)=1,R_{0}(t)=0\right)
\end{array}\right)}{P(S_{1}(t)=1,S_{0}(t)=1,R_{1}(t)=1,R_{0}(t)=1)}.
\]

The expression

\[
\frac{P(Y(t)\in y_{a},S(t)=1,R(t)=1\mid X=1)+P(Y(t)\not\in y_{a},S(t)=1,R(t)=1\mid X=0)-1}{P\left(S(t)=1,R(t)=1\mid X=1\right)-P\left(R(t)=0\mid X=0\right)-P\left(S(t)=0,R(t)=1\mid X=0\right)}
\]

is a function only of the observed data distribution, and has no terms
that involve counterfactuals or potential outcomes.

We could only make this bound sharper if we could find a function
of the observed data that was greater than 1 but less than expression
A. There is no immediate way to find such a function of the observed
data, because expression A is not point identified from randomization
at baseline and no function of the observed data can be found that
is guaranteed to be less than expression A and also greater than one
for any observed data on $(S(t),Y(t),X).$ 

For the last part of the Theorem, we provide a sketch proof. The complete
proof of this final result of the Theorem is provided in the proof
of Theorem Generalized Missing Data Sensitivity Analysis. 

For some $t\in T,$ $y_{a}\subset\mathbf{R}^{n_{y}}$ and $y_{b}\subset\mathbf{R}^{n_{y}}$
the numerator of the expression 
\[
\frac{P(Y(t)\in y_{a},S(t)=1\mid X=1)+P(Y(t)\not\in y_{b},S(t)=1\mid X=0)-1}{P\left(S(t)=1\mid X=1\right)-P(S(t)=0\mid X=0)}
\]
 is a lower bound on $P(Y_{1}(t)\in y_{a},Y_{0}(t)\not\in y_{b},S_{1}(t)=1,S_{0}(t)=1)-P(Y_{1}(t)\notin y_{a},Y_{0}(t)\in y_{a},S_{1}(t)=1,S_{0}(t)=1).$
The expression 

\[
\frac{P(Y(t)\in y_{a},S(t)=1\mid X=1)+P(Y(t)\not\in y_{b},S(t)=1\mid X=0)-1}{P\left(S(t)=1\mid X=1\right)-P(S(t)=0\mid X=0)}
\]
 is equal to 

\[
\frac{\left(\begin{array}{c}
P(Y(t)\in y_{a},S(t)=1,R(t)=1\mid X=1)+P(Y(t)\in y_{a},S(t)=1,R(t)=0\mid X=1)\\
+P(Y(t)\not\in y_{b},S(t)=1,R(t)=1\mid X=0)+P(Y(t)\not\in y_{b},S(t)=1,R(t)=0\mid X=0)\\
-1
\end{array}\right)}{\left(\begin{array}{c}
P\left(S(t)=1,R(t)=1\mid X=1\right)+P(S(t)=1,R(t)=0\mid X=1)\\
-P(S(t)=0,R(t)=1\mid X=0)-P(S(t)=0,R(t)=0\mid X=0)
\end{array}\right)}.
\]
 The above expression is bounded from below with the following expression

\[
\frac{\left(\begin{array}{c}
P(Y(t)\in y_{a},S(t)=1,R(t)=1\mid X=1)\\
+P(Y(t)\not\in y_{b},S(t)=1,R(t)=1\mid X=0)\\
-1
\end{array}\right)}{\left(\begin{array}{c}
P\left(S(t)=1,R(t)=1\mid X=1\right)+P(R(t)=0\mid X=1)\\
-P(S(t)=0,R(t)=1\mid X=0)
\end{array}\right)}.
\]
 Conseqently, when $P(Y(t)\in y_{a},S(t)=1,R(t)=1\mid X=1)+P(Y(t)\not\in y_{b},S(t)=1,R(t)=1\mid X=0)>1,$
the expression 
\[
\frac{\left(\begin{array}{c}
P(Y(t)\in y_{a},S(t)=1,R(t)=1\mid X=1)\\
+P(Y(t)\not\in y_{b},S(t)=1,R(t)=1\mid X=0)\\
-1
\end{array}\right)}{\left(\begin{array}{c}
P\left(S(t)=1,R(t)=1\mid X=1\right)+P(R(t)=0\mid X=1)\\
-P(S(t)=0,R(t)=1\mid X=0)
\end{array}\right)}
\]
 also serves as a lower bound to 
\[
\frac{P(Y_{1}(t)\in y_{a},Y_{0}(t)\not\in y_{b},S_{1}(t)=1,S_{0}(t)=1)-P(Y_{1}(t)\notin y_{a},Y_{0}(t)\in y_{a},S_{1}(t)=1,S_{0}(t)=1)}{P\left(S_{1}(t)=1,S_{0}(t)=1\right)},
\]
 which is equivalent to 
\[
P(Y_{1}(t)\in y_{a},Y_{0}(t)\not\in y_{b}\mid S_{1}(t)=1,S_{0}(t)=1)-P(Y_{1}(t)\notin y_{a},Y_{0}(t)\in y_{a}\mid S_{1}(t)=1,S_{0}(t)=1).
\]

This completes the proof. 

\subsection*{Proof of Generalized Always Survivor Causal Effect Monotonicity}

We begin with the expression 
\[
P(Y(t)\not\in y_{b},(S(t),R(t))\in\{(1,1)\}\mid X=0)-P(Y(t)\not\in y_{a},(S(t),R(t))\in\{(1,1)\}\mid X=1).
\]
 Under consistency of counterfactuals and randomization, this expression
is equal to 
\[
P(Y_{0}(t)\not\in y_{b},(S_{0}(t),R_{0}(t))\in\{(1,1)\})-P(Y_{1}(t)\not\in y_{a},(S_{1}(t),R_{1}(t))\in\{(1,1)\}).
\]
Now apply the law of total probability, $P(A)=\sum P(A,B_{i}).$ 
\begin{align*}
 & P(Y_{0}(t)\not\in y_{b},(S_{0}(t),R_{0}(t))\in\{(1,1)\})\\
 & \hspace{1em}-P(Y_{1}(t)\not\in y_{a},(S_{1}(t),R_{1}(t))\in\{(1,1)\})\\
 & =P(Y_{0}(t)\not\in y_{b},(S_{1}(t),R_{1}(t))\in\{(1,1)\},(S_{0}(t),R_{0}(t))\in\{(1,1)\})\\
 & \hspace{1em}+P(Y_{0}(t)\not\in y_{b},(S_{1}(t),R_{1}(t))\not\in\{(1,1)\},(S_{0}(t),R_{0}(t))\in\{(1,1)\})\\
 & \hspace{1em}-P(Y_{1}(t)\not\in y_{a},(S_{1}(t),R_{1}(t))\in\{(1,1)\},(S_{0}(t),R_{0}(t))\in\{(1,1)\})\\
 & \hspace{1em}-P(Y_{1}(t)\not\in y_{a},(S_{1}(t),R_{1}(t))\in\{(1,1)\},(S_{0}(t),R_{0}(t))\not\in\{(1,1)\}).
\end{align*}
 Now again apply DeMorgan's laws and $P(A\cup B)=P(A)+P(B)$ for disjoint
$A$ and $B.$ 

\begin{align*}
 & P(Y(t)\not\in y_{b},(S(t),R(t))\in\{(1,1)\}\mid X=0)-P(Y(t)\not\in y_{a},(S(t),R(t))\in\{(1,1)\}\mid X=1).\\
 & =P(Y_{0}(t)\not\in y_{b},(S_{1}(t),R_{1}(t))\in\{(1,1)\},(S_{0}(t),R_{0}(t))\in\{(1,1)\})\\
 & \hspace{1em}+P(Y_{0}(t)\not\in y_{b},(S_{1}(t),R_{1}(t))\in\{(1,0)\},(S_{0}(t),R_{0}(t))\in\{(1,1)\})\\
 & \hspace{1em}+P(Y_{0}(t)\not\in y_{b},(S_{1}(t),R_{1}(t))\in\{(0,1)\},(S_{0}(t),R_{0}(t))\in\{(1,1)\})\\
 & \hspace{1em}+P(Y_{0}(t)\not\in y_{b},(S_{1}(t),R_{1}(t))\in\{(0,0)\},(S_{0}(t),R_{0}(t))\in\{(1,1)\})\\
 & \hspace{1em}-P(Y_{1}(t)\not\in y_{a},(S_{1}(t),R_{1}(t))\in\{(1,1)\},(S_{0}(t),R_{0}(t))\in\{(1,1)\})\\
 & \hspace{1em}-P(Y_{1}(t)\not\in y_{a},(S_{1}(t),R_{1}(t))\in\{(1,1)\},(S_{0}(t),R_{0}(t))\in\{(1,0)\})\\
 & \hspace{1em}-P(Y_{1}(t)\not\in y_{a},(S_{1}(t),R_{1}(t))\in\{(1,1)\},(S_{0}(t),R_{0}(t))\in\{(0,1)\})\\
 & \hspace{1em}-P(Y_{1}(t)\not\in y_{a},(S_{1}(t),R_{1}(t))\in\{(1,1)\},(S_{0}(t),R_{0}(t))\in\{(0,0)\})
\end{align*}
 Now, again apply the law of total probability. Also, apply $Y_{1}(t)\in\star$
when $S_{1}(t)=0.$ 

\begin{align*}
 & P(Y(t)\not\in y_{b},(S(t),R(t))\in\{(1,1)\}\mid X=0)\\
 & \hspace{1em}-P(Y(t)\not\in y_{a},(S(t),R(t))\in\{(1,1)\}\mid X=1).\\
 & =P(Y_{1}(t)\in y_{a},Y_{0}(t)\not\in y_{b},(S_{1}(t),R_{1}(t))\in\{(1,1)\},(S_{0}(t),R_{0}(t))\in\{(1,1)\})\\
 & \hspace{1em}+P(Y_{1}(t)\not\in y_{a},Y_{0}(t)\not\in y_{b},(S_{1}(t),R_{1}(t))\in\{(1,1)\},(S_{0}(t),R_{0}(t))\in\{(1,1)\})\\
 & \hspace{1em}+P(Y_{1}(t)\in y_{a},Y_{0}(t)\not\in y_{b},(S_{1}(t),R_{1}(t))\in\{(1,0)\},(S_{0}(t),R_{0}(t))\in\{(1,1)\})\\
 & \hspace{1em}+P(Y_{1}(t)\not\in y_{a},Y_{0}(t)\not\in y_{b},(S_{1}(t),R_{1}(t))\in\{(1,0)\},(S_{0}(t),R_{0}(t))\in\{(1,1)\})\\
 & \hspace{1em}+P(Y_{1}(t)\in\star,Y_{0}(t)\not\in y_{b},(S_{1}(t),R_{1}(t))\in\{(0,1)\},(S_{0}(t),R_{0}(t))\in\{(1,1)\})\\
 & \hspace{1em}+P(Y_{1}(t)\in\star,Y_{0}(t)\not\in y_{b},(S_{1}(t),R_{1}(t))\in\{(0,0)\},(S_{0}(t),R_{0}(t))\in\{(1,1)\})\\
 & \hspace{1em}-P(Y_{1}(t)\not\in y_{a},Y_{0}(t)\in y_{b},(S_{1}(t),R_{1}(t))\in\{(1,1)\},(S_{0}(t),R_{0}(t))\in\{(1,1)\})\\
 & \hspace{1em}-P(Y_{1}(t)\not\in y_{a},Y_{0}(t)\not\in y_{b},(S_{1}(t),R_{1}(t))\in\{(1,1)\},(S_{0}(t),R_{0}(t))\in\{(1,1)\})\\
 & \hspace{1em}-P(Y_{1}(t)\not\in y_{a},Y_{0}(t)\in y_{b},(S_{1}(t),R_{1}(t))\in\{(1,1)\},(S_{0}(t),R_{0}(t))\in\{(1,0)\})\\
 & \hspace{1em}-P(Y_{1}(t)\not\in y_{a},Y_{0}(t)\not\in y_{b},(S_{1}(t),R_{1}(t))\in\{(1,1)\},(S_{0}(t),R_{0}(t))\in\{(1,0)\})\\
 & \hspace{1em}-P(Y_{1}(t)\not\in y_{a},Y_{0}(t)\in\star,(S_{1}(t),R_{1}(t))\in\{(1,1)\},(S_{0}(t),R_{0}(t))\in\{(0,1)\})\\
 & \hspace{1em}-P(Y_{1}(t)\not\in y_{a},Y_{0}(t)\in\star,(S_{1}(t),R_{1}(t))\in\{(1,1)\},(S_{0}(t),R_{0}(t))\in\{(0,0)\}).
\end{align*}
 Simplifying and rearranging terms 
\begin{align*}
 & P(Y(t)\not\in y_{b},(S(t),R(t))\in\{(1,1)\}\mid X=0)\\
 & \hspace{1em}-P(Y(t)\not\in y_{a},(S(t),R(t))\in\{(1,1)\}\mid X=1).\\
 & =P(Y_{1}(t)\in y_{a},Y_{0}(t)\not\in y_{b},(S_{1}(t),R_{1}(t))\in\{(1,1)\},(S_{0}(t),R_{0}(t))\in\{(1,1)\})\\
 & \hspace{1em}-P(Y_{1}(t)\not\in y_{a},Y_{0}(t)\in y_{b},(S_{1}(t),R_{1}(t))\in\{(1,1)\},(S_{0}(t),R_{0}(t))\in\{(1,1)\})\\
 & \hspace{1em}+\left\{ P(Y_{1}(t)\in y_{a},Y_{0}(t)\not\in y_{b},(S_{1}(t),R_{1}(t))\in\{(1,0)\},(S_{0}(t),R_{0}(t))\in\{(1,1)\})\right.\\
 & \hspace{1em}+P(Y_{1}(t)\not\in y_{a},Y_{0}(t)\not\in y_{b},(S_{1}(t),R_{1}(t))\in\{(1,0)\},(S_{0}(t),R_{0}(t))\in\{(1,1)\})\\
 & \hspace{1em}+P(Y_{1}(t)\in\star,Y_{0}(t)\not\in y_{b},(S_{1}(t),R_{1}(t))\in\{(0,1)\},(S_{0}(t),R_{0}(t))\in\{(1,1)\})\\
 & \hspace{1em}+P(Y_{1}(t)\in\star,Y_{0}(t)\not\in y_{b},(S_{1}(t),R_{1}(t))\in\{(0,0)\},(S_{0}(t),R_{0}(t))\in\{(1,1)\})\\
 & \hspace{1em}-P(Y_{1}(t)\not\in y_{a},Y_{0}(t)\in y_{b},(S_{1}(t),R_{1}(t))\in\{(1,1)\},(S_{0}(t),R_{0}(t))\in\{(1,0)\})\\
 & \hspace{1em}-P(Y_{1}(t)\not\in y_{a},Y_{0}(t)\not\in y_{b},(S_{1}(t),R_{1}(t))\in\{(1,1)\},(S_{0}(t),R_{0}(t))\in\{(1,0)\})\\
 & \hspace{1em}-P(Y_{1}(t)\not\in y_{a},Y_{0}(t)\in\star,(S_{1}(t),R_{1}(t))\in\{(1,1)\},(S_{0}(t),R_{0}(t))\in\{(0,1)\})\\
 & \hspace{1em}\left.-P(Y_{1}(t)\not\in y_{a},Y_{0}(t)\in\star,(S_{1}(t),R_{1}(t))\in\{(1,1)\},(S_{0}(t),R_{0}(t))\in\{(0,0)\})\right\} .
\end{align*}
 Under the monotonicity condition that there is no individual $\omega\in\Omega$
such that $(S_{1}(\omega,t),R_{1}(\omega,t))\not\in\{(1,1)\}$ and
$(S_{0}(\omega,t),R_{0}(\omega,t))\in\{(1,1)\}.$ 

\[
P(Y(t)\not\in y_{b},(S(t),R(t))\in\{(1,1)\}\mid X=0)-P(Y(t)\not\in y_{a},(S(t),R(t))\in\{(1,1)\}\mid X=1)
\]
 is a lower bound on 
\begin{align*}
 & P\left(Y_{1}(t)\in y_{a},Y_{0}(t)\not\in y_{b},(S_{1}(t),RS_{1}(t),RY_{1}(t))\in\{(1,1,1)\},(S_{0}(t),RS_{0}(t),RY_{0}(t))\in\{(1,1,1)\}\right)\\
 & \hspace{1em}-P(Y_{1}(t)\not\in y_{a},Y_{0}(t)\in y_{b},(S_{1}(t),RS_{1}(t))\in\{(1,1)\},RY_{1}(t)=1,(S_{0}(t),RS_{0}(t),RY_{0}(t))\in\{(1,1,1)\})\\
 & \hspace{1em}-P(Y_{1}(t)\not\in y_{a},Y_{0}(t)\in y_{b},(S_{1}(t),RS_{1}(t))\in\{(1,1)\},RY_{1}(t)=1,(S_{0}(t),RS_{0}(t))\in\{(1,0)\},RY_{0}(t)\neq1)\\
 & \hspace{1em}-P(Y_{1}(t)\not\in y_{a},Y_{0}(t)\in y_{b},(S_{1}(t),RS_{1}(t))\in\{(1,1)\},RY_{1}(t)=1,(S_{0}(t),RS_{0}(t))\in\{(1,0)\},RY_{0}(t)=1)\\
 & \hspace{1em}-P(Y_{1}(t)\not\in y_{a},Y_{0}(t)\in y_{b},(S_{1}(t),RS_{1}(t))\in\{(1,1)\},RY_{1}(t)=1,(S_{0}(t),RS_{0}(t))\in\{(1,1)\},RY_{0}(t)\neq1)\\
 & \hspace{1em}-P(Y_{1}(t)\not\in y_{a}Y_{0}(t)\in y_{b},(S_{1}(t),RS_{1}(t))\in\{(1,1)\},RY_{1}(t)\not=1,(S_{0}(t),RS_{0}(t),RY_{0}(t))\in\{(1,1,1)\})\\
 & \hspace{1em}-P(Y_{1}(t)\not\in y_{a},Y_{0}(t)\in y_{b},(S_{1}(t),RS_{1}(t))\in\{(1,1)\},RY_{1}(t)\not=1,(S_{0}(t),RS_{0}(t))\in\{(1,0)\},RY_{0}(t)\neq1)\\
 & \hspace{1em}-P(Y_{1}(t)\not\in y_{a},Y_{a}(t)\in y_{b},(S_{1}(t),RS_{1}(t))\in\{(1,1)\},RY_{1}(t)\not=1,(S_{0}(t),RS_{0}(t))\in\{(1,0)\},RY_{0}(t)=1)\\
 & \hspace{1em}-P(Y_{1}(t)\not\in y_{a},Y_{0}(t)\in y_{b},(S_{1}(t),RS_{1}(t))\in\{(1,1)\},RY_{1}(t)\not=1,(S_{0}(t),RS_{0}(t))\in\{(1,1)\},RY_{0}(t)\neq1).
\end{align*}
 When $P(Y(t)\not\in y_{a},(S(t),R(t))\in\{(1,1)\}\mid X=0)>P(Y(t)\not\in y_{a},(S(t),R(t))\in\{(1,1)\}\mid X=1),$
then 
\[
\frac{P(Y(t)\not\in y_{b},(S(t),R(t))\in\{(1,1)\}\mid X=0)-P(Y(t)\not\in y_{a},(S(t),R(t))\in\{(1,1)\}\mid X=1)}{P((S(t),R(t))\in\{(1,1)\}\mid X=0)}
\]
 serves as a lower on 

\[
\frac{\left(\begin{array}{cc}
 & P\left(Y_{1}(t)\in y_{a},Y_{0}(t)\not\in y_{b},(S_{1}(t),RS_{1}(t),RY_{1}(t))\in\{(1,1,1)\},(S_{0}(t),RS_{0}(t),RY_{0}(t))\in\{(1,1,1)\}\right)\\
 & \hspace{1em}-P(Y_{1}(t)\not\in y_{a},Y_{0}(t)\in y_{b},(S_{1}(t),RS_{1}(t))\in\{(1,1)\},RY_{1}(t)=1,(S_{0}(t),RS_{0}(t),RY_{0}(t))\in\{(1,1,1)\})\\
 & \hspace{1em}-P(Y_{1}(t)\not\in y_{a},Y_{0}(t)\in y_{b},(S_{1}(t),RS_{1}(t))\in\{(1,1)\},RY_{1}(t)=1,(S_{0}(t),RS_{0}(t))\in\{(1,0)\},RY_{0}(t)\neq1)\\
 & \hspace{1em}-P(Y_{1}(t)\not\in y_{a},Y_{0}(t)\in y_{b},(S_{1}(t),RS_{1}(t))\in\{(1,1)\},RY_{1}(t)=1,(S_{0}(t),RS_{0}(t))\in\{(1,0)\},RY_{0}(t)=1)\\
 & \hspace{1em}-P(Y_{1}(t)\not\in y_{a},Y_{0}(t)\in y_{b},(S_{1}(t),RS_{1}(t))\in\{(1,1)\},RY_{1}(t)=1,(S_{0}(t),RS_{0}(t))\in\{(1,1)\},RY_{0}(t)\neq1)\\
 & \hspace{1em}-P(Y_{1}(t)\not\in y_{a}Y_{0}(t)\in y_{b},(S_{1}(t),RS_{1}(t))\in\{(1,1)\},RY_{1}(t)\not=1,(S_{0}(t),RS_{0}(t),RY_{0}(t))\in\{(1,1,1)\})\\
 & \hspace{1em}-P(Y_{1}(t)\not\in y_{a},Y_{0}(t)\in y_{b},(S_{1}(t),RS_{1}(t))\in\{(1,1)\},RY_{1}(t)\not=1,(S_{0}(t),RS_{0}(t))\in\{(1,0)\},RY_{0}(t)\neq1)\\
 & \hspace{1em}-P(Y_{1}(t)\not\in y_{a},Y_{a}(t)\in y_{b},(S_{1}(t),RS_{1}(t))\in\{(1,1)\},RY_{1}(t)\not=1,(S_{0}(t),RS_{0}(t))\in\{(1,0)\},RY_{0}(t)=1)\\
 & \hspace{1em}-P(Y_{1}(t)\not\in y_{a},Y_{0}(t)\in y_{b},(S_{1}(t),RS_{1}(t))\in\{(1,1)\},RY_{1}(t)\not=1,(S_{0}(t),RS_{0}(t))\in\{(1,1)\},RY_{0}(t)\neq1)
\end{array}\right)}{P\left((S_{1}(t),R_{1}(t))\in\{(1,1)\},(S_{0}(t),R_{0}(t))\in\{(1,1)\})\right)}
\]
 and also 

\[
\frac{\left(\begin{array}{c}
P(Y_{1}(t)\in y_{a},Y_{0}(t)\not\in y_{b},(S_{1}(t),R_{1}(t))\in\{(1,1)\},(S_{0}(t),R_{0}(t))\in\{(1,1)\})\\
\hspace{1em}-P(Y_{1}(t)\not\in y_{a},Y_{0}(t)\in y_{b},(S_{1}(t),R_{1}(t))\in\{(1,1)\},(S_{0}(t),R_{0}(t))\in\{(1,1)\})
\end{array}\right)}{P\left((S_{1}(t),R_{1}(t))\in\{(1,1)\},(S_{0}(t),R_{0}(t))\in\{(1,1)\})\right)},
\]
 which is equivalent to 
\[
\begin{array}{c}
P(Y_{1}(t)\in y_{a},Y_{0}(t)\not\in y_{b}\mid(S_{1}(t),R_{1}(t))\in\{(1,1)\},(S_{0}(t),R_{0}(t))\in\{(1,1)\})\\
\hspace{1em}-P(Y_{1}(t)\not\in y_{a},Y_{0}(t)\in y_{b}\mid(S_{1}(t),R_{1}(t))\in\{(1,1)\},(S_{0}(t),R_{0}(t))\in\{(1,1)\}).
\end{array}
\]

\subsection*{Proof of Theorem Generalized Missing Data Sensitivity Analysis}

From the law of total probability, we have that 
\begin{align*}
 & P(Y(t)\in y_{a},S(t)=1,R(t)=1\mid X=1)+P(Y(t)\notin y_{b},S(t)=1,R(t)=1\mid X=0)\\
 & +P(Y(t)\in y_{a},S(t)=1,R(t)=0\mid X=1)+P(Y(t)\notin y_{b},S(t)=1,R(t)=0\mid X=0)\\
 & -1
\end{align*}
is equal to
\begin{align*}
 & P(Y_{1}(t)\in y_{a},Y_{0}(t)\in\star,S_{1}(t)=1,S_{0}(t)=0,R_{1}(t)=1,R_{0}(t)=1)\\
 & +P(Y_{1}(t)\in y_{a},Y_{0}(t)\in\star,S_{1}(t)=1,S_{0}(t)=0,R_{1}(t)=1,R_{0}(t)=0)\\
 & +P(Y_{1}(t)\in y_{a},Y_{0}(t)\in y_{a},S_{1}(t)=1,S_{0}(t)=1,R_{1}(t)=1,R_{0}(t)=1)\\
 & +P(Y_{1}(t)\in y_{a},Y_{0}(t)\in y_{a},S_{1}(t)=1,S_{0}(t)=1,R_{1}(t)=1,R_{0}(t)=0)\\
 & +P(Y_{1}(t)\in y_{a},Y_{0}(t)\notin y_{a},S_{1}(t)=1,S_{0}(t)=1,R_{1}(t)=1,R_{0}(t)=1)\\
 & +P(Y_{1}(t)\in y_{a},Y_{0}(t)\notin y_{a},S_{1}(t)=1,S_{0}(t)=1,R_{1}(t)=1,R_{0}(t)=0)\\
 & +P(Y_{1}(t)\in y_{a},Y_{0}(t)\in\star,S_{1}(t)=1,S_{0}(t)=0,R_{1}(t)=0,R_{0}(t)=1)\\
 & +P(Y_{1}(t)\in y_{a},Y_{0}(t)\in\star,S_{1}(t)=1,S_{0}(t)=0,R_{1}(t)=0,R_{0}(t)=0)\\
 & +P(Y_{1}(t)\in y_{a},Y_{0}(t)\in y_{a},S_{1}(t)=1,S_{0}(t)=1,R_{1}(t)=0,R_{0}(t)=1)\\
 & +P(Y_{1}(t)\in y_{a},Y_{0}(t)\in y_{a},S_{1}(t)=1,S_{0}(t)=1,R_{1}(t)=0,R_{0}(t)=0)\\
 & +P(Y_{1}(t)\in y_{a},Y_{0}(t)\notin y_{a},S_{1}(t)=1,S_{0}(t)=1,R_{1}(t)=0,R_{0}(t)=1)\\
 & +P(Y_{1}(t)\in y_{a},Y_{0}(t)\notin y_{a},S_{1}(t)=1,S_{0}(t)=1,R_{1}(t)=0,R_{0}(t)=0)\\
 & -P\left(Y_{1}(t)\in y_{a},Y_{0}(t)\in y_{a},S_{1}(t)=1,S_{0}(t)=1,R_{1}(t)=1,R_{0}(t)=1\right)\\
 & -P\left(Y_{1}(t)\in y_{a},Y_{0}(t)\in y_{a},S_{1}(t)=1,S_{0}(t)=1,R_{1}(t)=0,R_{0}(t)=1\right)\\
 & -P\left(Y_{1}(t)\notin y_{a},Y_{0}(t)\in y_{a},S_{1}(t)=1,S_{0}(t)=1,R_{1}(t)=1,R_{0}(t)=1\right)\\
 & -P\left(Y_{1}(t)\notin y_{a},Y_{0}(t)\in y_{a},S_{1}(t)=1,S_{0}(t)=1,R_{1}(t)=0,R_{0}(t)=1\right)\\
 & -P\left(Y_{1}(t)\in\star,Y_{0}(t)\in y_{a},S_{1}(t)=0,S_{0}(t)=1,R_{1}(t)=1,R_{0}(t)=1\right)\\
 & -P\left(Y_{1}(t)\in\star,Y_{0}(t)\in y_{a},S_{1}(t)=0,S_{0}(t)=1,R_{1}(t)=0,R_{0}(t)=1\right)\\
 & -P\left(Y_{1}(t)\in y_{a},Y_{0}(t)\in y_{a},S_{1}(t)=1,S_{0}(t)=1,R_{1}(t)=1,R_{0}(t)=0\right)\\
 & -P\left(Y_{1}(t)\in y_{a},Y_{0}(t)\in y_{a},S_{1}(t)=1,S_{0}(t)=1,R_{1}(t)=0,R_{0}(t)=0\right)\\
 & -P\left(Y_{1}(t)\notin y_{a},Y_{0}(t)\in y_{a},S_{1}(t)=1,S_{0}(t)=1,R_{1}(t)=1,R_{0}(t)=0\right)\\
 & -P\left(Y_{1}(t)\notin y_{a},Y_{0}(t)\in y_{a},S_{1}(t)=1,S_{0}(t)=1,R_{1}(t)=0,R_{0}(t)=0\right)\\
 & -P\left(Y_{1}(t)\in\star,Y_{0}(t)\in y_{a},S_{1}(t)=0,S_{0}(t)=1,R_{1}(t)=1,R_{0}(t)=0\right)\\
 & -P\left(Y_{1}(t)\in\star,Y_{0}(t)\in y_{a},S_{1}(t)=0,S_{0}(t)=1,R_{1}(t)=0,R_{0}(t)=0\right)\\
 & -P\left(Y_{1}(t)\in y_{a},Y_{0}(t)\in\star,S_{1}(t)=1,S_{0}(t)=0,R_{1}(t)=1,R_{0}(t)=1\right)\\
 & -P\left(Y_{1}(t)\in y_{a},Y_{0}(t)\in\star,S_{1}(t)=1,S_{0}(t)=0,R_{1}(t)=0,R_{0}(t)=1\right)\\
 & -P\left(Y_{1}(t)\notin y_{a},Y_{0}(t)\in\star,S_{1}(t)=1,S_{0}(t)=0,R_{1}(t)=1,R_{0}(t)=1\right)\\
 & -P\left(Y_{1}(t)\notin y_{a},Y_{0}(t)\in\star,S_{1}(t)=1,S_{0}(t)=0,R_{1}(t)=0,R_{0}(t)=1\right)\\
 & -P\left(Y_{1}(t)\in\star,Y_{0}(t)\in\star,S_{1}(t)=0,S_{0}(t)=0,R_{1}(t)=1,R_{0}(t)=1\right)\\
 & -P\left(Y_{1}(t)\in\star,Y_{0}(t)\in\star,S_{1}(t)=0,S_{0}(t)=0,R_{1}(t)=0,R_{0}(t)=1\right)\\
 & -P\left(Y_{1}(t)\in\star,Y_{0}(t)\in\star,S_{1}(t)=1,S_{0}(t)=0,R_{1}(t)=1,R_{0}(t)=0\right)\\
 & -P\left(Y_{1}(t)\in\star,Y_{0}(t)\in\star,S_{1}(t)=1,S_{0}(t)=0,R_{1}(t)=0,R_{0}(t)=0\right)\\
 & -P\left(Y_{1}(t)\in\star,Y_{0}(t)\in\star,S_{1}(t)=0,S_{0}(t)=0,R_{1}(t)=1,R_{0}(t)=0\right)\\
 & -P\left(Y_{1}(t)\in\star,Y_{0}(t)\in\star,S_{1}(t)=0,S_{0}(t)=0,R_{1}(t)=0,R_{0}(t)=0\right).
\end{align*}
and therefore,
\begin{align*}
 & P(Y(t)\in y_{a},S(t)=1,R(t)=1\mid X=1)+P(Y(t)\notin y_{b},S(t)=1,R(t)=1\mid X=0)\\
 & +P(Y(t)\in y_{a},S(t)=1,R(t)=0\mid X=1)+P(Y(t)\notin y_{b},S(t)=1,R(t)=0\mid X=0)\\
 & -1
\end{align*}
 is a lower bound on 
\begin{align*}
 & P(Y_{1}(t)\in y_{a},Y_{0}(t)\notin y_{b},S_{1}(t)=1,S_{0}(t)=1)\\
 & -P\left(Y_{1}(t)\notin y_{a},Y_{0}(t)\in y_{b},S_{1}(t)=1,S_{0}(t)=1\right).
\end{align*}

We label this equality A. Now from the law of total probability, we
have the following identities

\begin{align*}
 & P(Y_{1}(t)\in y_{a},Y_{0}(t)\notin y_{b},S_{1}(t)=1,S_{0}(t)=1)\\
 & =\sum_{(r_{1},r_{2})\in\{0,1\}^{2}}P(Y_{1}(t)\in y_{a},Y_{0}(t)\notin y_{b},S_{1}(t)=1,S_{0}(t)=1,R_{1}(t)=r_{1},R_{0}(t)=r_{2}),
\end{align*}

\begin{align*}
 & P\left(Y_{1}(t)\notin y_{a},Y_{0}(t)\in y_{b},S_{1}(t)=1,S_{0}(t)=1\right)\\
 & =\sum_{(r_{1},r_{2})\in\{0,1\}^{2}}P\left(Y_{1}(t)\notin y_{a},Y_{0}(t)\in y_{b},S_{1}(t)=1,S_{0}(t)=1,R_{1}(t)=r_{1},R_{0}(t)=r_{2}\right),
\end{align*}

\begin{align*}
 & P\left(Y_{1}(t)\in\star,Y_{0}(t)\in y_{b},S_{1}(t)=0,S_{0}(t)=1\right)\\
 & =\sum_{(r_{1},r_{2})\in\{0,1\}^{2}}P\left(Y_{1}(t)\in\star,Y_{0}(t)\in y_{a},S_{1}(t)=0,S_{0}(t)=1,R_{1}(t)=r_{1},R_{0}(t)=r_{2}\right),
\end{align*}

\begin{align*}
 & P\left(Y_{1}(t)\notin y_{a},Y_{0}(t)\in\star,S_{1}(t)=1,S_{0}(t)=0\right)\\
 & =\sum_{(r_{1},r_{2})\in\{0,1\}^{2}}P\left(Y_{1}(t)\notin y_{a},Y_{0}(t)\in\star,S_{1}(t)=1,S_{0}(t)=0,R_{1}(t)=r_{1},R_{0}(t)=r_{2}\right),
\end{align*}
 
\begin{align*}
 & P\left(Y_{1}(t)\in\star,Y_{0}(t)\in\star,S_{1}(t)=0,S_{0}(t)=0\right)\\
 & =\sum_{(r_{1},r_{2})\in\{0,1\}^{2}}P\left(Y_{1}(t)\in\star,Y_{0}(t)\in\star,S_{1}(t)=0,S_{0}(t)=0,R_{1}(t)=r_{1},R_{0}(t)=r_{2}\right).
\end{align*}
 Therefore, we have that 
\begin{align*}
 & P(Y(t)\in y_{a},S(t)=1,R(t)=1\mid X=1)+P(Y(t)\notin y_{b},S(t)=1,R(t)=1\mid X=0)\\
 & +P(Y(t)\in y_{a},S(t)=1,R(t)=0\mid X=1)+P(Y(t)\notin y_{b},S(t)=1,R(t)=0\mid X=0)\\
 & -1
\end{align*}
is equal to
\begin{align*}
 & P(Y_{1}(t)\in y_{a},Y_{0}(t)\notin y_{b},S_{1}(t)=1,S_{0}(t)=1)\\
 & -P\left(Y_{1}(t)\notin y_{a},Y_{0}(t)\in y_{b},S_{1}(t)=1,S_{0}(t)=1\right)\\
 & -P\left(Y_{1}(t)\in\star,Y_{0}(t)\in y_{b},S_{1}(t)=0,S_{0}(t)=1\right)\\
 & -P\left(Y_{1}(t)\notin y_{a},Y_{0}(t)\in\star,S_{1}(t)=1,S_{0}(t)=0\right)\\
 & -P\left(Y_{1}(t)\in\star,Y_{0}(t)\in\star,S_{1}(t)=0,S_{0}(t)=0\right).
\end{align*}
 We label this equality as equality $B.$ 

Provided $P(S_{1}(t)=1,S_{0}(t)=1)>0,$ Equality B only holds if and
only if 

\begin{align*}
 & \frac{P\left(Y_{1}(t)\in y_{a},Y_{0}(t)\not\in y_{b},S_{1}(t)=1,S_{0}(t)=1\right)-P\left(Y_{1}(t)\notin y_{a},Y_{0}(t)\in y_{b},S_{1}(t)=1,S_{0}(t)=1\right)}{P(S_{1}(t)=1,S_{0}(t)=1)}\\
 & =\frac{P(Y(t)\in y_{a},S(t)=1\mid X=1)+P(Y(t)\not\in y_{b},S(t)=1\mid X=0)-1}{P(S_{1}(t)=1,S_{0}(t)=1)}\\
 & \hspace{1em}+\frac{P\left(Y_{1}(t)\in\star,Y_{0}(t)\in y_{b},S_{1}(t)=0,S_{0}(t)=1\right)}{P(S_{1}=1,S_{0}=1)}\\
 & \hspace{1em}+\frac{P\left(Y_{1}(t)\not\in y_{a},Y_{0}(t)\in\star,S_{1}(t)=1,S_{0}(t)=0\right)}{P(S_{1}(t)=1,S_{0}(t)=1)}\\
 & \hspace{1em}+\frac{P\left(Y_{1}(t)\in\star,Y_{0}(t)\in\star,S_{1}(t)=0,S_{0}(t)=0\right)}{P(S_{1}(t)=1,S_{0}(t)=1)}.
\end{align*}

Using the identity, 
\begin{align*}
 & P\left(Y_{1}(t)\in\star,Y_{0}(t)\in\star,S_{1}(t)=0,S_{0}(t)=0\right)\\
 & =P(S_{1}(t)=0,S_{0}(t)=0).
\end{align*}
this equation for the normalized principal stratum direct effect simplifies
to

\begin{align*}
 & \frac{P\left(Y_{1}(t)\in y_{a},Y_{0}(t)\not\in y_{b},S_{1}(t)=1,S_{0}(t)=1\right)-P\left(Y_{1}(t)\notin y_{a},Y_{0}(t)\in y_{b},S_{1}(t)=1,S_{0}(t)=1\right)}{P(S_{1}=1,S_{0}=1)}\\
 & =\frac{P(Y(t)\in y_{a},S((t)=1\mid X=1)+P(Y(t)\not\in y_{b},S(t)=1\mid X=0)-1}{P(S_{1}(t)=1,S_{0}(t)=1)}\\
 & \hspace{1em}+\frac{P\left(Y_{1}(t)\in\star,Y_{0}(t)\in y_{b},S_{1}(t)=0,S_{0}(t)=1\right)}{P(S_{1}(t)=1,S_{0}(t)=1)}\\
 & \hspace{1em}+\frac{P\left(Y_{1}(t)\not\in y_{a},Y_{0}(t)\in\star,S_{1}(t)=1,S_{0}(t)=0\right)}{P(S_{1}=1,S_{0}=1)}\\
 & \hspace{1em}+\frac{P\left(S_{1}(t)=0,S_{0}(t)=0\right)}{P(S_{1}(t)=1,S_{0}(t)=1)}.
\end{align*}

Denote 
\begin{align*}
x & =P\left(Y_{1}(t)\in\star,Y_{0}(t)\in y_{b},S_{1}(t)=0,S_{0}(t)=1\right)\\
 & \hspace{1em}+P\left(Y_{1}(t)\not\in y_{a},Y_{0}(t)\in\star,S_{1}(t)=1,S_{0}(t)=0\right).
\end{align*}
 out of space considerations. Note, $x\in[0,1).$ The normalized principal
stratum direct effect can be formulated to equal:

\begin{align*}
 & \frac{P\left(Y_{1}(t)\in y_{a},Y_{0}(t)\notin y_{b},S_{1}(t)=1,S_{0}(t)=1\right)-P\left(Y_{1}(t)\notin y_{a},Y_{0}(t)\in y_{b},S_{1}(t)=1,S_{0}(t)=1\right)}{P(S_{1}(t)=1,S_{0}(t)=1)}\\
 & =\frac{P(Y(t)\in y_{a},S(t)=1\mid X=1)+P(Y(t)\not\in y_{b},S(t)=1\mid X=0)-1}{P(S_{1}(t)=1,S_{0}(t)=1)}\\
 & \hspace{1em}+\frac{\left(x+P\left(S_{1}(t)=0,S_{0}(t)=0\right)\right)}{P(S_{1}(t)=1,S_{0}(t)=1)}\\
 & =\frac{P(Y(t)\in y_{a},S(t)=1\mid X=1)+P(Y(t)\not\in y_{b},S(t)=1\mid X=0)-1}{P(S_{1}(t)=1,S_{0}(t)=1)-P(S_{1}(t)=0,S_{0}(t)=0)}\times\\
 & \hspace{1em}\left(\frac{P(S_{1}(t)=1,S_{0}(t)=1)-P(S_{1}(t)=0,S_{0}(t)=0)}{P(S_{1}(t)=1,S_{0}(t)=1)}+\right.\\
 & \hspace{1em}\left.+\frac{\left(x+P\left(S_{1}(t)=0,S_{0}(t)=0\right)\right)}{P(S_{1}(t)=1,S_{0}(t)=1)}\frac{\left(P(S_{1}(t)=1,S_{0}(t)=1)-P(S_{1}(t)=0,S_{0}(t)=0)\right)}{P(Y(t)\in y_{a},S(t)=1\mid X=1)+P(Y(t)\not\in y_{a},S(t)=1\mid X=0)-1}\right).
\end{align*}

We want to show that the next expression, which we denote expression
A,
\begin{align*}
\left(\frac{P(S_{1}(t)=1,S_{0}(t)=1)-P(S_{1}(t)=0,S_{0}(t)=0))}{P(S_{1}(t)=1,S_{0}(t)=1)}\right.\\
\hspace{1em}\left.+\frac{\left(x+P\left(S_{1}(t)=0,S_{0}(t)=0\right)\right)}{P(S_{1}(t)=1,S_{0}(t)=1)}\frac{\left(P(S_{1}(t)=1,S_{0}(t)=1)-P(S_{1}(t)=0,S_{0}(t)=0)\right)}{P(Y(t)\in y_{a},S(t)=1\mid X=1)+P(Y(t)\not\in y_{a},S(t)=1\mid X=0)-1}\right).
\end{align*}
is greater than or equal to one to show that 
\begin{align*}
\frac{P(Y(t)\in y_{a},S(t)=1\mid X=1)+P(Y(t)\not\in y_{b},S(t)=1\mid X=0)-1}{P(S_{1}(t)=1,S_{0}(t)=1)-P(S_{1}(t)=0,S_{0}(t)=0)}\\
\end{align*}
 is a lower bound to 
\[
\frac{P(Y_{1}(t)\in y_{a},Y_{0}(t)\not\in y_{b},S_{1}(t)=1,S_{0}(t)=1)-P\left(Y_{1}(t)\not\in y_{a},Y_{0}(t)\in y_{b},S_{1}(t)=1,S_{0}(t)=1\right)}{P(S_{1}(t)=1,S_{0}(t)=1)}.
\]
Through inspection, expression A is greater than or equal to one if
and only if 
\[
\frac{\left(P(S_{1}(t)=1,S_{0}(t)=1)-P(S_{1}(t)=0,S_{0}(t)=0)\right)}{P(Y(t)\in y_{a},S(t)=1\mid X=1)+P(Y(t)\not\in y_{b},S(t)=1\mid X=0)-1}\geq1.
\]

Now use the identity that $P(Y_{1}(t)=\star,Y_{0}(t)=\star,S_{1}(t)=0,S_{0}(t)=0)=P(S_{1}(t)=0,S_{0}(t)=0),$
and therefore, 

\begin{align*}
 & P(Y(t)\in y_{a},S(t)=1\mid X=1)+P(Y(t)\not\in y_{b},S(t)=1\mid X=0)-1
\end{align*}
is equal to
\begin{align*}
 & P\left(Y_{1}(t)\in y_{a},Y_{0}(t)\not\in y_{b},S_{1}(t)=1,S_{0}(t)=1\right)\\
 & -\left\{ P\left(Y_{1}(t)\not\in y_{a},Y_{0}(t)\in y_{b},S_{1}(t)=1,S_{0}(t)=1\right)\right.\\
 & \hspace{1em}+P\left(Y_{1}(t)\in\star,Y_{0}(t)\in y_{b},S_{1}(t)=0,S_{0}(t)=1\right)\\
 & \hspace{1em}+P\left(Y_{1}(t)\not\in y_{a},Y_{0}(t)\in\star,S_{1}(t)=1,S_{0}(t)=0\right)\\
 & \hspace{1em}\left.+P\left(S_{1}(t)=0,S_{0}(t)=0\right)\right\} .
\end{align*}

If we examine 
\[
\frac{\left(P(S_{1}(t)=1,S_{0}(t)=1)-P(S_{1}(t)=0,S_{0}(t)=0)\right)}{P(Y(t)\in y_{a},S(t)=1\mid X=1)+P(Y(t)\not\in y_{b},S(t)=1\mid X=0)-1},
\]
 then the numerator is greater equal to the denominator as 
\begin{align*}
P(S_{1}(t)=1,S_{0}(t)=1) & \geq P\left(Y_{1}(t)\in y_{a},Y_{0}(t)\not\in y_{b},S_{1}(t)=1,S_{0}(t)=1\right),
\end{align*}
 from the counterfactual representation of 
\begin{align*}
 & P(Y(t)\in y_{a},S(t)=1\mid X=1)+P(Y(t)\not\in y_{b},S(t)=1\mid X=0)-1.
\end{align*}
 Also, both the numerator and denominator contain $-P(S_{1}(t)=0,S_{0}(t)=0),$
which means that 

\[
\frac{\left(P(S_{1}(t)=1,S_{0}(t)=1)-P(S_{1}(t)=0,S_{0}(t)=0)\right)}{P(Y(t)\in y_{a},S(t)=1\mid X=1)+P(Y(t)\not\in y_{b},S(t)=1\mid X=0)-1}\geq0.
\]
 This means that expression A is greater than equal to 1. This implies
finally that 

\begin{align*}
 & \frac{P(Y(t)\in y_{a},S(t)=1\mid X=1)+P(Y(t)\not\in y_{b},S(t)=1\mid X=0)-1}{P(S_{1}(t)=1,S_{0}(t)=1)-P(S_{1}(t)=0,S_{0}(t)=0)}\\
 & \leq\frac{P(Y_{1}(t)\in y_{a},Y_{0}(t)\not\in y_{b},S_{1}(t)=1,S_{0}(t)=1)-P\left(Y_{1}(t)\not\in y_{a},Y_{0}(t)\in y_{b},S_{1}(t)=1,S_{0}(t)=1\right)}{P(S_{1}(t)=1,S_{0}(t)=1)}
\end{align*}
 Therefore, 
\begin{align*}
\frac{P(Y(t)\in y_{a},S(t)=1\mid X=1)+P(Y(t)\not\in y_{b},S(t)=1\mid X=0)-1}{P(S_{1}(t)=1,S_{0}=1)-P(S_{1}(t)=0,S_{0}(t)=0)}
\end{align*}
 is a lower bound on
\begin{align*}
\frac{P(Y_{1}(t)\in y_{a},Y_{0}(t)\not\in y_{b},S_{1}(t)=1,S_{0}(t)=1)-P\left(Y_{1}(t)\not\in y_{a},Y_{0}(t)\in y_{b},S_{1}(t)=1,S_{0}(t)=1\right)}{P(S_{1}(t)=1,S_{0}(t)=1)},
\end{align*}

provided $P(Y(t)\in y_{a},S(t)=1\mid X=1)+P(Y(t)\not\in y_{b},S(t)=1\mid X=0)-1>0.$ 

Now, note that
\begin{align*}
 & P(S_{1}(t)=1,S_{0}(t)=1)-P(S_{1}(t)=0,S_{0}(t)=0)\\
 & =P(S_{1}(t)=1,S_{0}(t)=1)+P(S_{1}(t)=1,S_{0}(t)=0)\\
 & \hspace{1em}-P(S_{1}(t)=1,S_{0}(t)=0)-P(S_{1}(t)=0,S_{0}(t)=0)\\
 & =P(S_{1}(t)=1)-P(S_{0}(t)=0)\\
 & =P(S(t)=1\mid X=1)-P(S(t)=0\mid X=0).
\end{align*}
 Consequently, when $P(Y(t)\in y_{a},S(t)=1\mid X=1)+P(Y(t)\not\in y_{b},S(t)=1\mid X=0)-1>0$,
the expression 
\[
\frac{P(Y(t)\in y_{a},S(t)=1\mid X=1)+P(Y(t)\not\in y_{b},S(t)=1\mid X=0)-1}{P(S(t)=1\mid X=1)-P(S(t)=0\mid X=0)}
\]

is a lower bound to 
\begin{align*}
\frac{P(Y_{1}(t)\in y_{a},Y_{0}(t)\not\in y_{b},S_{1}(t)=1,S_{0}(t)=1)-P\left(Y_{1}(t)\not\in y_{a},Y_{0}(t)\in y_{b},S_{1}(t)=1,S_{0}(t)=1\right)}{P(S_{1}(t)=1,S_{0}(t)=1)},
\end{align*}
 which is equivalent to 
\begin{align*}
P(Y_{1}(t)\in y_{a},Y_{0}(t)\not\in y_{b}\mid S_{1}(t)=1,S_{0}(t)=1)-P\left(Y_{1}(t)\not\in y_{a},Y_{0}(t)\in y_{b}\mid S_{1}(t)=1,S_{0}(t)=1\right).
\end{align*}

Note, the expression 
\[
\frac{P(Y(t)\in y_{a},S(t)=1\mid X=1)+P(Y(t)\not\in y_{b},S(t)=1\mid X=0)-1}{P(S(t)=1\mid X=1)-P(S(t)=0\mid X=0)}
\]
 is a function only of the observed data distribution, and has no
terms that involve counterfactuals or potential outcomes. However, 

\[
\frac{P(Y(t)\in y_{a},S(t)=1\mid X=1)+P(Y(t)\not\in y_{b},S(t)=1\mid X=0)-1}{P(S(t)=1\mid X=1)-P(S(t)=0\mid X=0)}
\]
 cannot be calculated from the observed data if there is missing responses.
As before, 

\begin{align*}
 & \frac{P(Y(t)\in y_{a},S(t)=1\mid X=1)+P(Y(t)\not\in y_{b},S(t)=1\mid X=0)-1}{P(S(t)=1\mid X=1)-P(S(t)=0\mid X=0)}
\end{align*}
 is equal to 
\[
\frac{\left(\begin{array}{c}
P(Y(t)\in y_{a},S(t)=1,R(t)=1\mid X=1)+P(Y(t)\notin y_{b},S(t)=1,R(t)=1\mid X=0)\\
+P(Y(t)\in y_{a},S(t)=1,R(t)=0\mid X=1)+P(Y(t)\notin y_{b},S(t)=1,R(t)=0\mid X=0)\\
-1
\end{array}\right)}{\left(\begin{array}{c}
P(S(t)=1,R(t)=1\mid X=1)+P(S(t)=1,R(t)=0\mid X=1)\\
-P\left(S(t)=1,R(t)=1\mid X=1\right)-P\left(S(t)=1,R(t)=0\mid X=0\right)
\end{array}\right)}.
\]
 The parameters $P(Y(t)\in y_{a},S(t)=1,R(t)=0\mid X=1)+P(Y(t)\notin y_{b},S(t)=1,R(t)=0\mid X=0)$
and $P(S(t)=1,R(t)=0\mid X=1)-P\left(S(t)=1,R(t)=0\mid X=0\right)$
are sensitivity analysis parameters, because we cannot calculate any
probability that involves joints of $(R(t)=0,S(t)=s),$ $(R(t)=0,Y(t)=y),$
or $(R(t)=0,S(t)=s,Y(t)=y)$ for $s\in\{0,1\}$ and $y\in\{0,1\}.$
The moment the individual drops out of the clinical trial, $Y(t)$
and $S(t)$ become undefined. We could only make this bound sharper
if we could find a function of the observed data that was greater
than 1 but less than expression A. There is no immediate way to find
such a function of the observed data, because expression A is not
point identified from randomization at baseline and no function of
the observed data can be found that is guaranteed to be less than
expression A and also greater than one for any observed data on $(S(t),Y(t),R(t),X).$
This completes the proof. 

The expression 
\[
\frac{\left(\begin{array}{c}
P(Y(t)\in y_{a},S(t)=1,R(t)=1\mid X=1)+P(Y(t)\notin y_{b},S(t)=1,R(t)=1\mid X=0)\\
+P(Y(t)\in y_{a},S(t)=1,R(t)=0\mid X=1)+P(Y(t)\notin y_{b},S(t)=1,R(t)=0\mid X=0)\\
-1
\end{array}\right)}{\left(\begin{array}{c}
P(S(t)=1,R(t)=1\mid X=1)+P(S(t)=1,R(t)=0\mid X=1)\\
-P\left(S(t)=1,R(t)=1\mid X=1\right)-P\left(S(t)=1,R(t)=0\mid X=0\right)
\end{array}\right)}
\]
 is bounded from below from the expression
\[
\frac{\left(\begin{array}{c}
P(Y(t)\in y_{a},S(t)=1,R(t)=1\mid X=1)+P(Y(t)\notin y_{b},S(t)=1,R(t)=1\mid X=0)\\
-1
\end{array}\right)}{\left(\begin{array}{c}
P(S(t)=1,R(t)=1\mid X=1)+P(R(t)=0\mid X=1)\\
-P\left(S(t)=1,R(t)=1\mid X=1\right)
\end{array}\right)}.
\]
Consequently, when $P(Y(t)\in y_{a},S(t)=1,R(t)=1\mid X=1)+P(Y(t)\notin y_{b},S(t)=1,R(t)=1\mid X=0)>1,$
the expression 
\[
\frac{\left(\begin{array}{c}
P(Y(t)\in y_{a},S(t)=1,R(t)=1\mid X=1)+P(Y(t)\notin y_{b},S(t)=1,R(t)=1\mid X=0)\\
-1
\end{array}\right)}{\left(\begin{array}{c}
P(S(t)=1,R(t)=1\mid X=1)+P(R(t)=0\mid X=1)\\
-P\left(S(t)=1,R(t)=1\mid X=1\right)
\end{array}\right)}
\]
 is a lower bound on 
\begin{align*}
\frac{P(Y_{1}(t)\in y_{a},Y_{0}(t)\not\in y_{b},S_{1}(t)=1,S_{0}(t)=1)-P\left(Y_{1}(t)\not\in y_{a},Y_{0}(t)\in y_{b},S_{1}(t)=1,S_{0}(t)=1\right)}{P(S_{1}(t)=1,S_{0}(t)=1)},
\end{align*}
 which is equivalent to 
\begin{align*}
P(Y_{1}(t)\in y_{a},Y_{0}(t)\not\in y_{b}\mid S_{1}(t)=1,S_{0}(t)=1)-P\left(Y_{1}(t)\not\in y_{a},Y_{0}(t)\in y_{b}\mid S_{1}(t)=1,S_{0}(t)=1\right).
\end{align*}

\subsection*{Proof of Theorem Generalized Missing Data and Monotonicity Sensitivity
Analysis}

Begin with the expression

\begin{align*}
 & P(Y(t_{L})\notin y_{b},S(t_{L})=1,R(t_{L})=1\mid X=0)\\
 & +P\left(Y(t_{L})\notin y_{b},S(t_{L})=1,R(t_{L})=0\mid X=0\right)\\
 & -P\left(Y(t_{U})\notin y_{a},S(t_{U})=1,R(t_{U})=1\mid X=1\right)\\
 & -P\left(Y(t_{U})\notin y_{a},S(t_{U})=1,R(t_{U})=0\mid X=1\right).
\end{align*}

Under randomization and consistency of counterfactuals this expression
is equal to 

\begin{align*}
 & P(Y_{0}(t_{L})\notin y_{a},S_{0}(t_{L})=1,R_{0}(t_{L})=1)\\
 & +P\left(Y_{0}(t_{L})\notin y_{a},S_{0}(t_{L})=1,R_{0}(t_{L})=0\right)\\
 & -P\left(Y_{1}(t_{U})\notin y_{a},S_{1}(t_{U})=1,R_{1}(t_{U})=1\right)\\
 & -P\left(Y_{1}(t_{U})\notin y_{a},S_{1}(t_{U})=1,R_{1}(t_{U})=0\right).
\end{align*}
 We now will use the law of total probability. The above counterfactual
contrast is equal to 

\begin{align*}
 & P(Y_{0}(t_{L})\notin y_{b},S_{1}(t_{U})=1,S_{0}(t_{L})=1,R_{0}(t_{L})=1)\\
 & +P(Y_{0}(t_{L})\notin y_{b},S_{1}(t_{U})=0,S_{0}(t_{L})=1,R_{0}(t_{L})=1)\\
 & +P\left(Y_{0}(t_{L})\notin y_{b},S_{1}(t_{U})=1,S_{0}(t_{L})=1,R_{0}(t_{L})=0\right)\\
 & +P\left(Y_{0}(t_{L})\notin y_{b},S_{1}(t_{U})=0,S_{0}(t_{L})=1,R_{0}(t_{L})=0\right)\\
 & -P\left(Y_{1}(t_{U})\notin y_{a},S_{1}(t_{U})=1,S_{0}(t_{L})=1,R_{1}(t_{U})=1\right)\\
 & -P\left(Y_{1}(t_{U})\notin y_{a},S_{1}(t_{U})=1,S_{0}(t_{L})=0,R_{1}(t_{U})=1\right)\\
 & -P\left(Y_{1}(t_{U})\notin y_{a},S_{1}(t_{U})=1,S_{0}(t_{L})=1,R_{1}(t_{U})=0\right)\\
 & -P\left(Y_{1}(t_{U})\notin y_{a},S_{1}(t_{U})=1,S_{0}(t_{L})=0,R_{1}(t_{U})=0\right).
\end{align*}

We again apply the law of total probability. The above counterfactual
contrast is equal to 

\begin{align*}
 & P(Y_{1}(t_{U})\in y_{a},Y_{0}(t_{L})\notin y_{b},S_{1}(t_{U})=1,S_{0}(t_{L})=1,R_{0}(t_{L})=1)\\
 & +P(Y_{1}(t_{U})\not\in y_{a},Y_{0}(t_{L})\notin y_{b},S_{1}(t_{U})=1,S_{0}(t_{L})=1,R_{0}(t_{L})=1)\\
 & +P(Y_{1}(t_{U})\in\star,Y_{0}(t_{L})\notin y_{b},S_{1}(t_{U})=0,S_{0}(t_{L})=1,R_{0}(t_{L})=1)\\
 & +P\left(Y_{1}(t_{U})\in y_{a},Y_{0}(t_{L})\notin y_{b},S_{1}(t_{U})=1,S_{0}(t_{L})=1,R_{0}(t_{L})=0\right)\\
 & +P\left(Y_{1}(t_{U})\notin y_{a},Y_{0}(t_{L})\notin y_{b},S_{1}(t_{U})=1,S_{0}(t_{L})=1,R_{0}(t_{L})=0\right)\\
 & +P\left(Y_{1}(t_{U})\in\star,Y_{0}(t_{L})\notin y_{b},S_{1}(t_{U})=0,S_{0}(t_{L})=1,R_{0}(t_{L})=0\right)\\
 & -P\left(Y_{1}(t_{U})\notin y_{a},Y_{0}(t_{L})\in y_{b},S_{1}(t_{U})=1,S_{0}(t_{L})=1,R_{1}(t_{U})=1\right)\\
 & -P\left(Y_{1}(t_{U})\notin y_{a},Y_{0}(t_{L})\notin y_{b},S_{1}(t_{U})=1,S_{0}(t_{L})=1,R_{1}(t_{U})=1\right)\\
 & -P\left(Y_{1}(t_{U})\notin y_{a},Y_{0}(t_{L})\in\star,S_{1}(t_{U})=1,S_{0}(t_{L})=0,R_{1}(t_{U})=1\right)\\
 & -P\left(Y_{1}(t_{U})\notin y_{a},Y_{0}(t_{L})\in y_{b},S_{1}(t_{U})=1,S_{0}(t_{L})=1,R_{1}(t_{U})=0\right)\\
 & -P\left(Y_{1}(t_{U})\notin y_{a},Y_{0}(t_{L})\notin y_{b},S_{1}(t_{U})=1,S_{0}(t_{L})=1,R_{1}(t_{U})=0\right)\\
 & -P\left(Y_{1}(t_{U})\notin y_{a},Y_{0}(t_{L})\in\star,S_{1}(t_{U})=1,S_{0}(t_{L})=0,R_{1}(t_{U})=0\right)
\end{align*}
 We again apply the law of total probability. The above counterfactual
contrast is equal to 

\begin{align*}
 & P(Y_{1}(t_{U})\in y_{a},Y_{0}(t_{L})\notin y_{b},S_{1}(t_{U})=1,S_{0}(t_{L})=1,R_{1}(t_{U})=1,R_{0}(t_{L})=1)\\
 & +P(Y_{1}(t_{U})\in y_{a},Y_{0}(t_{L})\notin y_{b},S_{1}(t_{U})=1,S_{0}(t_{L})=1,R_{1}(t_{U})=0,R_{0}(t_{L})=1)\\
 & +P(Y_{1}(t_{U})\not\in y_{a},Y_{0}(t_{L})\notin y_{b},S_{1}(t_{U})=1,S_{0}(t_{L})=1,R_{1}(t_{U})=1,R_{0}(t_{L})=1)\\
 & +P(Y_{1}(t_{U})\not\in y_{a},Y_{0}(t_{L})\notin y_{b},S_{1}(t_{U})=1,S_{0}(t_{L})=1,R_{1}(t_{U})=0,R_{0}(t_{L})=1)\\
 & +P(Y_{1}(t_{U})\in\star,Y_{0}(t_{L})\notin y_{b},S_{1}(t_{U})=0,S_{0}(t_{L})=1,R_{1}(t_{U})=1,R_{0}(t_{L})=1)\\
 & +P(Y_{1}(t_{U})\in\star,Y_{0}(t_{L})\notin y_{b},S_{1}(t_{U})=0,S_{0}(t_{L})=1,R_{1}(t_{U})=0,R_{0}(t_{L})=1)\\
 & +P\left(Y_{1}(t_{U})\in y_{a},Y_{0}(t_{L})\notin y_{b},S_{1}(t_{U})=1,S_{0}(t_{L})=1,R_{1}(t_{U})=1,R_{0}(t_{L})=0\right)\\
 & +P\left(Y_{1}(t_{U})\in y_{a},Y_{0}(t_{L})\notin y_{b},S_{1}(t_{U})=1,S_{0}(t_{L})=1,R_{1}(t_{U})=0,R_{0}(t_{L})=0\right)\\
 & +P\left(Y_{1}(t_{U})\notin y_{a},Y_{0}(t_{L})\notin y_{b},S_{1}(t_{U})=1,S_{0}(t_{L})=1,R_{1}(t_{U})=1,R_{0}(t_{L})=0\right)\\
 & +P\left(Y_{1}(t_{U})\notin y_{a},Y_{0}(t_{L})\notin y_{b},S_{1}(t_{U})=1,S_{0}(t_{L})=1,R_{1}(t_{U})=0,R_{0}(t_{L})=0\right)\\
 & +P\left(Y_{1}(t_{U})\in\star,Y_{0}(t_{L})\notin y_{b},S_{1}(t_{U})=0,S_{0}(t_{L})=1,R_{1}(t_{U})=1,R_{0}(t_{L})=0\right)\\
 & +P\left(Y_{1}(t_{U})\in\star,Y_{0}(t_{L})\notin y_{b},S_{1}(t_{U})=0,S_{0}(t_{L})=1,R_{1}(t_{U})=0,R_{0}(t_{L})=0\right)\\
 & -P\left(Y_{1}(t_{U})\notin y_{a},Y_{0}(t_{L})\in y_{b},S_{1}(t_{U})=1,S_{0}(t_{L})=1,R_{1}(t_{U})=1,R_{0}(t_{L})=1\right)\\
 & -P\left(Y_{1}(t_{U})\notin y_{a},Y_{0}(t_{L})\in y_{b},S_{1}(t_{U})=1,S_{0}(t_{L})=1,R_{1}(t_{U})=1,R_{0}(t_{L})=0\right)\\
 & -P\left(Y_{1}(t_{U})\notin y_{a},Y_{0}(t_{L})\notin y_{b},S_{1}(t_{U})=1,S_{0}(t_{L})=1,R_{1}(t_{U})=1,R_{0}(t_{L})=1\right)\\
 & -P\left(Y_{1}(t_{U})\notin y_{a},Y_{0}(t_{L})\notin y_{b},S_{1}(t_{U})=1,S_{0}(t_{L})=1,R_{1}(t_{U})=1,R_{0}(t_{L})=0\right)\\
 & -P\left(Y_{1}(t_{U})\notin y_{a},Y_{0}(t_{L})\in\star,S_{1}(t_{U})=1,S_{0}(t_{L})=0,R_{1}(t_{U})=1,R_{0}(t_{L})=1\right)\\
 & -P\left(Y_{1}(t_{U})\notin y_{a},Y_{0}(t_{L})\in\star,S_{1}(t_{U})=1,S_{0}(t_{L})=0,R_{1}(t_{U})=1,R_{0}(t_{L})=0\right)\\
 & -P\left(Y_{1}(t_{U})\notin y_{a},Y_{0}(t_{L})\in y_{b},S_{1}(t_{U})=1,S_{0}(t_{L})=1,R_{1}(t_{U})=0,R_{0}(t_{L})=1\right)\\
 & -P\left(Y_{1}(t_{U})\notin y_{a},Y_{0}(t_{L})\in y_{b},S_{1}(t_{U})=1,S_{0}(t_{L})=1,R_{1}(t_{U})=0,R_{0}(t_{L})=0\right)\\
 & -P\left(Y_{1}(t_{U})\notin y_{a},Y_{0}(t_{L})\notin y_{b},S_{1}(t_{U})=1,S_{0}(t_{L})=1,R_{1}(t_{U})=0,R_{0}(t_{L})=1\right)\\
 & -P\left(Y_{1}(t_{U})\notin y_{a},Y_{0}(t_{L})\notin y_{b},S_{1}(t_{U})=1,S_{0}(t_{L})=1,R_{1}(t)=0,R_{0}(t_{L})=0\right)\\
 & -P\left(Y_{1}(t_{U})\notin y_{a},Y_{0}(t_{L})\in\star,S_{1}(t_{U})=1,S_{0}(t_{L})=0,R_{1}(t)=0,R_{0}(t_{L})=1\right)\\
 & -P\left(Y_{1}(t_{U})\notin y_{a},Y_{0}(t_{L})\in\star,S_{1}(t_{U})=1,S_{0}(t_{L})=0,R_{1}(t)=0,R_{0}(t_{L})=0\right).
\end{align*}

Simplifying this expression we have the above expression is equal
to 

\begin{align*}
 & P(Y_{1}(t_{U})\in y_{a},Y_{0}(t_{L})\notin y_{b},S_{1}(t_{U})=1,S_{0}(t_{L})=1,R_{1}(t_{U})=1,R_{0}(t_{L})=1)\\
 & +P(Y_{1}(t_{U})\in y_{a},Y_{0}(t_{L})\notin y_{b},S_{1}(t_{U})=1,S_{0}(t_{L})=1,R_{1}(t_{U})=0,R_{0}(t_{L})=1)\\
 & +P(Y_{1}(t_{U})\in\star,Y_{0}(t_{L})\notin y_{b},S_{1}(t_{U})=0,S_{0}(t_{L})=1,R_{1}(t_{U})=1,R_{0}(t_{L})=1)\\
 & +P(Y_{1}(t_{U})\in\star,Y_{0}(t_{L})\notin y_{b},S_{1}(t_{U})=0,S_{0}(t_{L})=1,R_{1}(t_{U})=0,R_{0}(t_{L})=1)\\
 & +P\left(Y_{1}(t_{U})\in y_{a},Y_{0}(t_{L})\notin y_{b},S_{1}(t_{U})=1,S_{0}(t_{L})=1,R_{1}(t_{U})=1,R_{0}(t_{L})=0\right)\\
 & +P\left(Y_{1}(t_{U})\in y_{a},Y_{0}(t_{L})\notin y_{b},S_{1}(t_{U})=1,S_{0}(t_{L})=1,R_{1}(t_{U})=0,R_{0}(t_{L})=0\right)\\
 & +P\left(Y_{1}(t_{U})\in\star,Y_{0}(t_{L})\notin y_{b},S_{1}(t_{U})=0,S_{0}(t_{L})=1,R_{1}(t_{U})=1,R_{0}(t_{L})=0\right)\\
 & +P\left(Y_{1}(t_{U})\in\star,Y_{0}(t_{L})\notin y_{b},S_{1}(t_{U})=0,S_{0}(t_{L})=1,R_{1}(t_{U})=0,R_{0}(t_{L})=0\right)\\
 & -P\left(Y_{1}(t_{U})\notin y_{a},Y_{0}(t_{L})\in y_{b},S_{1}(t_{U})=1,S_{0}(t_{L})=1,R_{1}(t_{U})=1,R_{0}(t_{L})=1\right)\\
 & -P\left(Y_{1}(t_{U})\notin y_{a},Y_{0}(t_{L})\in y_{b},S_{1}(t_{U})=1,S_{0}(t_{L})=1,R_{1}(t_{U})=1,R_{0}(t_{L})=0\right)\\
 & -P\left(Y_{1}(t_{U})\notin y_{a},Y_{0}(t_{L})\in\star,S_{1}(t_{U})=1,S_{0}(t_{L})=0,R_{1}(t_{U})=1,R_{0}(t_{L})=1\right)\\
 & -P\left(Y_{1}(t_{U})\notin y_{a},Y_{0}(t_{L})\in\star,S_{1}(t_{U})=1,S_{0}(t_{L})=0,R_{1}(t_{U})=1,R_{0}(t_{L})=0\right)\\
 & -P\left(Y_{1}(t_{U})\notin y_{a},Y_{0}(t_{L})\in y_{b},S_{1}(t_{U})=1,S_{0}(t_{L})=1,R_{1}(t_{U})=0,R_{0}(t_{L})=1\right)\\
 & -P\left(Y_{1}(t_{U})\notin y_{a},Y_{0}(t_{L})\in y_{b},S_{1}(t_{U})=1,S_{0}(t_{L})=1,R_{1}(t_{U})=0,R_{0}(t_{L})=0\right)\\
 & -P\left(Y_{1}(t_{U})\notin y_{a},Y_{0}(t_{L})\in\star,S_{1}(t_{U})=1,S_{0}(t_{L})=0,R_{1}(t_{U})=0,R_{0}(t_{L})=1\right)\\
 & -P\left(Y_{1}(t_{U})\notin y_{a},Y_{0}(t_{L})\in\star,S_{1}(t_{U})=1,S_{0}(t_{L})=0,R_{1}(t_{U})=0,R_{0}(t_{L})=0\right),
\end{align*}
 and therefore the expression 
\begin{align*}
 & P(Y(t_{L})\notin y_{b},S(t_{L})=1,R(t_{L})=1\mid X=0)+P\left(Y(t_{L})\notin y_{b},S(t_{L})=1,R(t_{L})=0\mid X=0\right)\\
 & -P\left(Y(t_{U})\notin y_{a},S(t_{U})=1,R(t_{U})=1\mid X=1\right)-P\left(Y(t_{U})\notin y_{a},S(t_{U})=1,R(t_{U})=0\mid X=1\right)\\
 & -r(t_{L},t_{U}),
\end{align*}
 where 
\begin{align*}
r(t_{L},t_{U}) & =P(Y_{1}(t_{U})\in\star,Y_{0}(t_{L})\notin y_{b},S_{1}(t_{U})=0,S_{0}(t_{L})=1,R_{1}(t_{U})=1,R_{0}(t_{L})=1)\\
 & \hspace{1em}+P(Y_{1}(t_{U})\in\star,Y_{0}(t_{L})\notin y_{b},S_{1}(t_{U})=0,S_{0}(t_{L})=1,R_{1}(t_{U})=0,R_{0}(t_{L})=1)\\
 & \hspace{1em}+P\left(Y_{1}(t_{U})\in\star,Y_{0}(t_{L})\notin y_{b},S_{1}(t_{U})=0,S_{0}(t_{L})=1,R_{1}(t_{U})=1,R_{0}(t_{L})=0\right)\\
 & \hspace{1em}+P\left(Y_{1}(t_{U})\in\star,Y_{0}(t_{L})\notin y_{b},S_{1}(t_{U})=0,S_{0}(t_{L})=1,R_{1}(t_{U})=0,R_{0}(t_{L})=0\right)\\
 & \hspace{1em}-P\left(Y_{1}(t_{U})\notin y_{a},Y_{0}(t_{L})\in\star,S_{1}(t_{U})=1,S_{0}(t_{L})=0,R_{1}(t_{U})=1,R_{0}(t_{L})=1\right)\\
 & \hspace{1em}-P\left(Y_{1}(t_{U})\notin y_{a},Y_{0}(t_{L})\in\star,S_{1}(t_{U})=1,S_{0}(t_{L})=0,R_{1}(t_{U})=1,R_{0}(t_{L})=0\right)\\
 & \hspace{1em}-P\left(Y_{1}(t_{U})\notin y_{a},Y_{0}(t_{L})\in\star,S_{1}(t_{U})=1,S_{0}(t_{L})=0,R_{1}(t_{U})=0,R_{0}(t_{L})=1\right)\\
 & \hspace{1em}-P\left(Y_{1}(t_{U})\notin y_{a},Y_{0}(t_{L})\in\star,S_{1}(t_{U})=1,S_{0}(t_{L})=0,R_{1}(t_{U})=0,R_{0}(t_{L})=0\right),
\end{align*}
 is equal to 
\begin{align*}
 & P(Y_{1}(t_{U})\in y_{a},Y_{0}(t_{L})\notin y_{b},S_{1}(t_{U})=1,S_{0}(t_{L})=1,R_{1}(t_{U})=1,R_{0}(t_{L})=1)\\
 & +P(Y_{1}(t_{U})\in y_{a},Y_{0}(t_{L})\notin y_{b},S_{1}(t_{U})=1,S_{0}(t_{L})=1,R_{1}(t_{U})=0,R_{0}(t_{L})=1)\\
 & +P\left(Y_{1}(t_{U})\in y_{a},Y_{0}(t_{L})\notin y_{b},S_{1}(t_{U})=1,S_{0}(t_{L})=1,R_{1}(t_{U})=1,R_{0}(t_{L})=0\right)\\
 & +P\left(Y_{1}(t_{U})\in y_{a},Y_{0}(t_{L})\notin y_{b},S_{1}(t_{U})=1,S_{0}(t_{L})=1,R_{1}(t_{U})=0,R_{0}(t_{L})=0\right)\\
 & -P\left(Y_{1}(t_{U})\notin y_{a},Y_{0}(t_{L})\in y_{b},S_{1}(t_{U})=1,S_{0}(t_{L})=1,R_{1}(t_{U})=1,R_{0}(t_{L})=1\right)\\
 & -P\left(Y_{1}(t_{U})\notin y_{a},Y_{0}(t_{L})\in y_{b},S_{1}(t_{U})=1,S_{0}(t_{L})=1,R_{1}(t_{U})=1,R_{0}(t_{L})=0\right)\\
 & -P\left(Y_{1}(t_{U})\notin y_{a},Y_{0}(t_{L})\in y_{b},S_{1}(t_{U})=1,S_{0}(t_{L})=1,R_{1}(t_{U})=0,R_{0}(t_{L})=1\right)\\
 & -P\left(Y_{1}(t_{U})\notin y_{a},Y_{0}(t_{L})\in y_{b},S_{1}(t_{U})=1,S_{0}(t_{L})=1,R_{1}(t_{U})=0,R_{0}(t_{L})=0\right).
\end{align*}
 Consequently, the expression 
\[
\frac{\left(\begin{array}{c}
P(Y(t_{L})\notin y_{b},S(t_{L})=1,R(t_{L})=1\mid X=0)+P\left(Y(t_{L})\notin y_{b},S(t_{L})=1,R(t_{L})=0\mid X=0\right)\\
-P\left(Y(t_{U})\notin y_{a},S(t_{U})=1,R(t_{U})=1\mid X=1\right)-P\left(Y(t_{U})\notin y_{a},S(t_{U})=1,R(t_{U})=0\mid X=1\right)\\
-r(t_{L},t_{U}),
\end{array}\right)}{P(S(t_{L})=1,R(t_{L})=1\mid X=0)+P\left(S(t_{L})=1,R(t_{L})=0\mid X=0\right)-P\left(S_{1}(t_{U})=1,S_{0}(t_{L})\neq1\right)}
\]
 is equal to the normalized principal stratum direct effect 
\[
\frac{P(Y_{1}(t_{U})\in y_{a},Y_{0}(t_{L})\notin y_{b},S_{1}(t_{U})=1,S_{0}(t_{L})=1)-P(Y_{1}(t_{U})\notin y_{b},Y_{0}(t_{L})\in y_{a},S_{1}(t_{U})=1,S_{0}(t_{L})=1)}{P(S_{1}(t_{U})=1,S_{0}(t_{L})=1)}
\]
 which is equivalent to 
\[
P(Y_{1}(t_{U})\in y_{a},Y_{0}(t_{L})\notin y_{b}\mid S_{1}(t_{U})=1,S_{0}(t_{L})=1)-P(Y_{1}(t_{U})\notin y,Y_{0}(t_{L})\in y_{b}\mid S_{1}(t_{U})=1,S_{0}(t_{L})=1).
\]
 This last part falls out by realizing that 
\[
P(S(t_{U})=1,R(t_{L})=1\mid X=0)+P\left(S(t_{U})=1,R(t_{U})=0\mid X=0\right)-P\left(S_{1}(t_{U})=1,S_{0}(t_{L})\neq1\right)
\]
 is equal to $P(S_{1}(t_{U})=1,S_{0}(t_{L})=1)$ under randomization
and consistency of counterfactuals. This completes the proof.

\subsection*{Associated Tables}

The tables drop the explicit dependence on $\omega$ and integer $t\in T,$
and use the shorthand $Y_{x}(\omega,t)=Y_{x}$ and $S_{x}(\omega,t)=S_{x}$
for $x\in\{0,1\}$ out of space considerations. Recall that we denote
$Y_{x}^{[y]}(\omega,t)=I(Y_{x}(\omega,t)=y),$ and $S_{x}^{[s]}(\omega,t)=I(S_{x}(\omega,t)=s)$
for $y\in\{0,1,2,3\}$ and $s\in\{0,1,2\}.$ Likewise, we drop the
dependence on $\omega$ and $t,$ and use shorthand notation $S_{x}^{[s]}$
and $Y_{x}^{[y]}$ instead of $S_{x}^{[s]}(\omega,t)$ and $Y_{x}^{[y]}(\omega,t)$
for $y\in\{0,1\}$ out of space considerations. The reader should
always assume that all counterfactuals presented in the tables below
are implicitly dependent on $\omega$ and $t$ even if the shorthand
notation does not explicitly indicate such a dependence. The tables
below provide a complete enumeration of the different counterfactual
response types. Without monotonicity assumptions there are 16 different
counterfactual response types, and then with the monotonicity assumption
that $S_{1}^{[0]}(\omega,t)+S_{0}^{[1]}(\omega,t)\leq1$ for all individuals
in our population, the response types $\omega\in\{5,6\}$ are no longer
possible. Under monotonicity assumption that $S_{1}^{[2]}(\omega,t)+S_{0}^{[1]}(\omega,t)\leq1$
response types $\omega\in\{11,12\}$ in Table 8 are no longer possible. 

All of our results can be derived using Table 8 only, but we provide
further tables in Online Supplement 2 that could aid readers. The
table below is for any fixed $t\in T.$ The counterfactual table presented
below is the joint distribution of $(Y_{1}(\omega,t),Y_{0}(\omega,t),S_{1}(\omega,t),S_{0}(\omega,t)),$
where all joint counterfactuals $(Y_{1}(\omega,t),Y_{0}(\omega,t),S_{1}(\omega,t),S_{0}(\omega,t))$
are measured at the same fixed $t\in T.$ 

\begin{table}[H]
\begin{centering}
\begin{tabular}{|c|c|c|c|c||c|}
\hline 
$\omega$ & $Y_{1}$ & $Y_{0}$ & $S_{1}$ & $S_{0}$ & Probability\tabularnewline
\hline 
\hline 
1 & 1 & 1 & 1 & 1 & $P(Y_{1}=1,Y_{0}=1,S_{1}=1,S_{0}=1)$\tabularnewline
\hline 
2 & 0 & 1 & 1 & 1 & $P(Y_{1}=0,Y_{0}=1,S_{1}=1,S_{0}=1)$\tabularnewline
\hline 
3 & 1 & 0 & 1 & 1 & $P(Y_{1}=1,Y_{0}=0,S_{1}=1,S_{0}=1)$\tabularnewline
\hline 
4 & 0 & 0 & 1 & 1 & $P(Y_{1}=0,Y_{0}=0,S_{1}=1,S_{0}=1)$\tabularnewline
\hline 
5 & 2 & 1 & 0 & 1 & $P(Y_{1}=2,Y_{0}=1,S_{1}=0,S_{0}=1)$\tabularnewline
\hline 
6 & 2 & 0 & 0 & 1 & $P(Y_{1}=2,Y_{0}=0,S_{1}=0,S_{0}=1)$\tabularnewline
\hline 
7 & 1 & 2 & 1 & 0 & $P(Y_{1}=1,Y_{0}=2,S_{1}=1,S_{0}=0)$\tabularnewline
\hline 
8 & 0 & 2 & 1 & 0 & $P(Y_{1}=0,Y_{0}=2,S_{1}=1,S_{0}=0)$\tabularnewline
\hline 
9 & 2 & 2 & 0 & 0 & $P(Y_{1}=2,Y_{0}=2,S_{1}=0,S_{0}=0)$\tabularnewline
\hline 
10 & 3 & 3 & 2 & 2 & $P(Y_{1}=3,Y_{0}=3,S_{1}=2,S_{0}=2)$\tabularnewline
\hline 
11 & 3 & 1 & 2 & 1 & $P(Y_{1}=3,Y_{0}=1,S_{1}=2,S_{0}=1)$\tabularnewline
\hline 
12 & 3 & 0 & 2 & 1 & $P(Y_{1}=3,Y_{0}=0,S_{1}=2,S_{0}=1)$\tabularnewline
\hline 
13 & 3 & 2 & 2 & 0 & $P(Y_{1}=3,Y_{0}=2,S_{1}=2,S_{0}=0)$\tabularnewline
\hline 
14 & 1 & 3 & 1 & 2 & $P(Y_{1}=1,Y_{0}=3,S_{1}=1,S_{0}=2)$\tabularnewline
\hline 
15 & 0 & 3 & 1 & 2 & $P(Y_{1}=0,Y_{0}=3,S_{1}=1,S_{0}=2)$\tabularnewline
\hline 
16 & 2 & 3 & 0 & 2 & $P(Y_{1}=2,Y_{0}=3,S_{1}=0,S_{0}=2)$\tabularnewline
\hline 
\end{tabular} 
\par\end{centering}
\caption{Counterfactual Distribution}
\end{table}

\section*{Data}

A total of 338 individuals were randomized to Docetaxel and 336 individuals
were randomized to Mitoxantrone. The full dataset is provided here
to enable reproducibility of our results.

\subsection*{Survive and Cancer Progression}

\begin{table}[H]
\begin{centering}
\begin{tabular}{cccccccc}
\toprule 
 & 1 month & 2 months & 3 months & 4 months & 6 months & 12 months & 18 months\tabularnewline
\midrule
\midrule 
$X=1$ & 6 & 30 & 72 & 86 & 116 & 175 & 144\tabularnewline
\midrule 
$X=0$ & 40 & 94 & 146 & 150 & 164 & 147 & 121\tabularnewline
\bottomrule
\end{tabular} 
\par\end{centering}
\caption{Survive and cancer progression at time $t$ $(Y(t)=1,S(t)=1)$}
\end{table}

\subsection*{Survive and Cancer Did not Progress}

\begin{table}[H]
\begin{centering}
\begin{tabular}{cccccccc}
\toprule 
 & 1 month & 2 months & 3 months & 4 months & 6 months & 12 months & 18 months\tabularnewline
\midrule
\midrule 
$X=1$ & 320 & 289 & 243 & 221 & 172 & 65 & 17\tabularnewline
\midrule 
$X=0$ & 278 & 219 & 160 & 146 & 109 & 58 & 14\tabularnewline
\bottomrule
\end{tabular} 
\par\end{centering}
\caption{Survive and cancer does not progress at time $t,$ $(Y(t)=0,S(t)=1)$}
\end{table}

\subsection*{Did not Survive}

\begin{table}[H]
\begin{centering}
\begin{tabular}{cccccccc}
\toprule 
 & 1 month & 2 months & 3 months & 4 months & 6 months & 12 months & 18 months\tabularnewline
\midrule
\midrule 
$X=1$ & 3 & 10 & 14 & 22 & 41 & 88 & 166\tabularnewline
\midrule 
$X=0$ & 3 & 8 & 15 & 25 & 47 & 114 & 182\tabularnewline
\bottomrule
\end{tabular} 
\par\end{centering}
\caption{Did not survive until $t,$ $(Y(t)=2,S(t)=0)$}

\end{table}

\subsection*{Censored}

\begin{table}[H]
\begin{centering}
\begin{tabular}{cccccccc}
\toprule 
 & 1 month & 2 months & 3 months & 4 months & 6 months & 12 months & 18 months\tabularnewline
\midrule
\midrule 
$X=1$ & 9 & 9 & 9 & 9 & 9 & 10 & 11\tabularnewline
\midrule 
$X=0$ & 15 & 15 & 15 & 15 & 16 & 17 & 19\tabularnewline
\bottomrule
\end{tabular} 
\par\end{centering}
\caption{Censored at time $t,$ $(Y(t)=3,S(t)=2)$}
\end{table}

\bibliographystyle{plain}
\bibliography{bibmediationiv}

\end{document}